\documentclass[sn-basic]{sn-jnl} 

\usepackage{graphicx}%
\usepackage{multirow}%
\usepackage{amsmath,amssymb,amsfonts}%
\usepackage{amsthm}%
\usepackage{mathrsfs}%
\usepackage[title]{appendix}%
\usepackage{xcolor}%
\usepackage{textcomp}%
\usepackage{manyfoot}%
\usepackage{booktabs}%
\usepackage{algorithm}%
\usepackage{algorithmicx}%
\usepackage{algpseudocode}%
\usepackage{listings}%

\begin{document}

\title[Article Title]{Ground-breaking Exoplanet Science with the ANDES spectrograph at the ELT}


\author*[1,2]{\fnm{Enric} \sur{Palle}}\email{epalle@iac.es}



\author[3]{\fnm{Katia} \sur{Biazzo}}
\author[4,5]{\fnm{Emeline} \sur{Bolmont}}
\author[6]{\fnm{Paul} \sur{Molliere}}
\author[7,8]{\fnm{Katja} \sur{Poppenhaeger}}


\author[9]{\fnm{Jayne} \sur{Birkby}}
\author[10,11]{\fnm{Matteo} \sur{Brogi}}
\author[12]{\fnm{Gael} \sur{Chauvin}}
\author[12]{\fnm{Andrea} \sur{Chiavassa}}
\author[13]{\fnm{Jens} \sur{Hoeijmakers}}
\author[14]{\fnm{Emmanuel } \sur{Lellouch}}
\author[15]{\fnm{Christophe} \sur{Lovis}}
\author[16]{\fnm{Roberto} \sur{Maiolino}}
\author[17]{\fnm{Lisa} \sur{Nortmann}}
\author[1,2]{\fnm{Hannu} \sur{Parviainen}}
\author[18]{\fnm{Lorenzo} \sur{Pino}}
\author[19,20]{\fnm{Martin} \sur{Turbet}}
\author[21]{\fnm{Jesse} \sur{Wender}}

\author[22]{\fnm{Simon} \sur{Albrecht}}
\author[3]{\fnm{Simone} \sur{Antoniucci}}
\author[23,24]{\fnm{Susana C.} \sur{Barros}}
\author[25]{\fnm{Andre} \sur{Beaudoin}}
\author[25]{\fnm{Bjorn} \sur{Benneke}}
\author[26]{\fnm{Isabelle} \sur{Boisse}}
\author[27]{\fnm{Aldo S.} \sur{Bonomo}}
\author[28]{\fnm{Francesco} \sur{Borsa}}
\author[29]{\fnm{Alexis} \sur{Brandeker}}
\author[6]{\fnm{Wolfgang} \sur{Brandner}}
\author[30]{\fnm{Lars A.} \sur{Buchhave}}
\author[18]{\fnm{Anne-Laure} \sur{Cheffot}}
\author[16]{\fnm{Robin} \sur{Deborde}}
\author[31]{\fnm{Florian} \sur{Debras}}
\author[25]{\fnm{Rene} \sur{Doyon}}
\author[32]{\fnm{Paolo} \sur{Di Marcantonio}}
\author[11]{\fnm{Paolo} \sur{Giacobbe}}
\author[1,2]{\fnm{Jonay I.} \sur{Gonz\'alez Hern\'andez}}
\author[33]{\fnm{Ravit} \sur{Helled}}
\author[6]{\fnm{Laura} \sur{Kreidberg}}
\author[34,35]{\fnm{Pedro} \sur{Machado}}
\author[36]{\fnm{Jesus} \sur{Maldonado}}
\author[37]{\fnm{Alessandro} \sur{Marconi}}
\author[38]{\fnm{B.L. Canto} \sur{Martins}}
\author[39,17]{\fnm{Adriano} \sur{Miceli}}
\author[40]{\fnm{Christoph} \sur{Mordasini}}
\author[12]{\fnm{Mamadou} \sur{N'Diaye}}
\author[41]{\fnm{Andrezj} \sur{Niedzielski}}
\author[3]{\fnm{Brunella} \sur{Nisini}}
\author[42]{\fnm{Livia} \sur{Origlia}}
\author[43]{\fnm{Celine} \sur{Peroux}}
\author[7]{\fnm{Alex G.M.} \sur{Pietrow}}
\author[18]{\fnm{Enrico} \sur{Pinna}}
\author[44]{\fnm{Emily} \sur{Rauscher}}
\author[45]{\fnm{Sabine} \sur{Reffert}}
\author[46]{\fnm{Philippe} \sur{Rousselot}}
\author[18]{\fnm{Nicoletta} \sur{Sanna}}
\author[12]{\fnm{Adrien} \sur{Simonnin}}
\author[1,2]{\fnm{Alejandro} \sur{Su\'{a}rez Mascare\~{n}o}}
\author[47]{\fnm{Alessio} \sur{Zanutta}}
\author[17]{\fnm{Mathias} \sur{Zechmeister}}

\affil*[1]{Instituto de Astrof\'isica de Canarias (IAC), 38200 La Laguna, Tenerife, Spain}
\affil[2]{Deptartamento de Astrof\'isica, Universidad de La Laguna (ULL), 38206 La Laguna, Tenerife, Spain}


\affil[3]{INAF - Astronomical Observatory of Rome, I-00043 Monte Porzio Catone, Rome, Italy}
\affil[4]{Observatoire de Gen\`eve, Universit\'e de Gen\`eve, Chemin Pegasi 51, 1290, Sauverny, Switzerland}
\affil[5]{Centre sur la Vie dans l'Univers, Universit\'e de Gen\`eve, Geneva, Switzerland}
\affil[6]{Max Planck Insitute for Astronomy, K{\"o}nigstuhl 17, 69117 Heidelberg, Germany}
\affil[7]{Leibniz Institute for Astrophysics Potsdam, An der Sternwarte 16, 14482 Potsdam, Germany}
\affil[8]{Potsdam University, Institute for Physics and Astronomy, Karl-Liebknecht-Str. 24/25, 14476 Potsdam-Golm}

\affil[9]{Astrophysics, Department of Physics, University of Oxford, Denys Wilkinson Building, Keble Road, Oxford, OX1 3RH, UK}
\affil[10]{Dipartimento di Fisica, Universit\'a degli Studi di Torino, via Pietro Giuria 1, I-10125, Torino, Italy}	
\affil[11]{INAF-Osservatorio Astrofisico di Torino, Via Osservatorio 20, I-10025 Pino Torinese, Italy}
\affil[12]{Universit\'e C\^ote d'Azur, Observatoire de la C\^ote d'Azur, CNRS, Lagrange, CS 34229, Nice,France}
\affil[13]{Lund Observatory, Box 43, SE-221 00 Lund, Sweden}
\affil[14]{LESIA, Observatoire de Paris, 92195 Meudon, France}
\affil[15]{D\'epartement d'Astronomie, Universit\'e de Gen\`eve, Chemin Pegasi 51, CH-1290 Versoix, Switzerland}
\affil[16]{Cavendish Laboratory, University of Cambridge, 19 J. J. Thomson Ave., Cambridge CB3 0HE, UK}
\affil[17]{Institut fur Astrophysik und Geophysik, Georg-August-Universitat, D-37077 Gottingen, Germany}
\affil[18]{INAF - Osservatorio Astrofisico di Arcetri, Largo Enrico Fermi 5, 50125 Firenze, Italy}
\affil[19]{Laboratoire de M\'et\'eorologie Dynamique/IPSL, CNRS, Sorbonne Universit\'e, Ecole Normale Sup\'erieure, PSL Research University, Ecole Polytechnique, 75005 Paris, France}
\affil[20]{Laboratoire d'astrophysique de Bordeaux, Univ. Bordeaux, CNRS, B18N, allée Geoffroy Saint-Hilaire, 33615 Pessac, France}
\affil[21]{Physikalisches Institut, Universit{\"a}t Bern,Gesellschaftsstrasse 6, 3012Bern, Switzerland}


\affil[22]{Department of Physics and Astronomy, Aarhus University, Ny Munkegade 120, DK-8000 Aarhus C, Denmark}
\affil[23]{Instituto de Astrof\'isica e Ci\^encias do Espa\c{c}o, Universidade do Porto, CAUP, Rua das Estrelas, PT4150-762 Porto, Portugal}
\affil[24]{Departamento\,de\,Fisica\,e\,Astronomia,\,Faculdade\,de\,Ciencias,\,Universidade\,do\,Porto,\,Rua\,Campo\,Alegre,\,4169-007\,Porto,\,Portugal}
\affil[25]{Department of Physics and Trottier Institute of Research on Exoplanets, Universit\'e de Montr\'eal, Montr\'eal, QC H3C 3J7, Canada}
\affil[26]{Aix Marseille Univ, CNRS, CNES, LAM, Marseille, France}
\affil[27]{INAF - Osservatorio Astrofisico di Torino, via Osservatorio 20, 10025 Pino Torinese, Italy}
\affil[28]{INAF - Osservatorio Astronomico di Brera, Via E. Bianchi, 46, 23807 Merate (LC), Italy}
\affil[29]{Department of Astronomy, Stockholm University, AlbaNova University Center, 10691 Stockholm, Sweden}
\affil[30]{DTU Space, Technical University of Denmark, Elektrovej 328, DK-2800 Kgs. Lyngby, Denmark}
\affil[31]{IRAP, CNRS - UMR 5277, Universite de Toulouse, Toulouse, France}
\affil[32]{INAF - Osservatorio Astronomico di Trieste, Via G.B. Tiepolo, 11 I-34143 Trieste, Italy}
\affil[33]{Institute for Computational Science, Center for Theoretical Astrophysics \& Cosmology, University of Zurich Winterthurerstr. 190, CH-8057 Zurich, Switzerland}
\affil[34]{Institute of Astrophysics and Space Sciences, Observatorio Astronomico de Lisboa, Ed. Leste, Tapada da Ajuda, 1349-018 Lisbon, Portugal}
\affil[35]{Departamento de Fisica, Faculdade de Ciencias da Universidade de Lisboa, Campo Grande, Lisboa Portugal}
\affil[36]{INAF - Osservatorio Astronomico di Palermo, Piazza del Parlamento 1, 90134 Palermo, Italy}
\affil[37]{Dipartimento di Fisica e Astronomia, Universita degli Studi di Firenze, Via G. Sansone 1,I-50019, Sesto Fiorentino, Firenze, Italy}
\affil[38]{Departamento de F\'isica Te\'orica e Experimental, Universidade Federal do Rio Grande do Norte, Campus Universit\'ario, Natal, RN, 59072-970, Brazil}
\affil[39]{Dipartimento di Fisica e Astronomia, Universitaa di Firenze, Via G. Sansone 1, 50019 Sesto Fiorentino, Firenze, Italy}
\affil[40]{Weltraumforschung und Planetologie, Physikalisches Institut, Universitat Bern, Gesellschaftsstrasse 6, 3012Bern, Switzerland}
\affil[41]{Institute of Astronomy, Nicolaus Copernicus University in Toru\'n, Gagarina 11, 87-100 Toru\'n, Poland}
\affil[42]{Osservatorio di Astrofisica e Scienza dello Spazio di Bologna, Via Gobetti 93/3, I-40129 Bologna, Italy}
\affil[43]{European Southern Observatory, Karl-Schwarzschild-Str. 2, 85748 Garching-bei-M\"unchen, Germany}

\affil[44]{Department of Astronomy, University of Michigan, 1085 S University Ave, Ann Arbor, MI 48109, USA}
\affil[45]{Landessternwarte, Zentrum fur Astronomie der Universitat Heidelberg, Konigstuhl 12, 69117 Heidelberg, Germany}
\affil[46]{Institut UTINAM - UMR 6213, CNRS / Univ. de Franche-Comte, OSU THETA, 41 bis Av. de l’Observatoire, BP 1615, F-25010 Besançon Cedex, France}

\affil[47]{INAF - Osservatorio Astronomico di Brera, via E. Bianchi 46, 23807 Merate, Italy}





\abstract{In the past decade the study of exoplanet atmospheres at high-spectral resolution, via transmission/emission spectroscopy and cross-correlation techniques for atomic/molecular mapping, has become a powerful and consolidated methodology. The current limitation is the signal-to-noise ratio that one can obtain during a planetary transit, which is in turn ultimately limited by telescope size. This limitation will be overcome by ANDES, an optical and near-infrared high-resolution spectrograph for the Extremely Large Telescope, which is currently in Phase B development. ANDES will be a powerful transformational instrument for exoplanet science. It will enable the study of giant planet atmospheres, allowing not only an exquisite determination of atmospheric composition, but also the study of isotopic compositions, dynamics and weather patterns, mapping the planetary atmospheres and probing atmospheric formation and evolution models. The unprecedented angular resolution of ANDES, will also
allow us to explore the initial conditions in which planets form
in proto-planetary disks. The main science case of ANDES, however, is the study of small, rocky exoplanet atmospheres, including the potential for biomarker detections, and the ability to reach this science case is driving its instrumental design. Here we discuss our simulations and the observing strategies to achieve this specific science goal. Since ANDES will be operational at the same time as NASA's JWST and ESA's ARIEL missions, it will provide enormous synergies in the characterization of planetary atmospheres at high and low spectral resolution. Moreover, ANDES will be able to probe for the first time the atmospheres of several giant and small planets in reflected light. In particular, we show how ANDES will be able to unlock the reflected light atmospheric signal of a golden sample of nearby non-transiting habitable zone earth-sized planets within a few tenths of nights, a scientific objective that no other currently approved astronomical facility will be able to reach.}




\maketitle

\section{Introduction}
\label{sec:introduction}




Exoplanets exhibit an immense diversity in their mass, size, internal composition, temperature, atmospheric makeup, and orbital configurations. Despite significant advancements over the past two decades, we still need a comprehensive understanding of the formation and evolution of exoplanetary systems, and the factors influencing their composition and surface conditions. Consequently, the investigation of exoplanet atmospheres across a broad spectrum of planetary types, spanning from gas giants to rocky planets and from scorching hot to temperate worlds, remains a primary focus within the field of exoplanetary science. This is because atmospheric metallicity and some ratios of elemental abundances (C/O) are tracers of the planet's formation and evolution history \citep{Madhusudhan2019, Turrini2021, Eistrup2023}. Currently the JWST is opening a new era of exoplanet atmospheres exploration at low spectral resolution \citep{Tsai2023, Coulombe2023, Esparza-Borges2023}, but (perhaps with the exception of the Trappist1 system) it will not be able to reach the atmospheres of habitable zone rocky planets. 

Ground-based exoplanet astronomy will profoundly transform with the construction of extremely large telescopes \citep{Gilmozzi2005}. The community will be able to probe the population of known planets in many different directions due to the high angular resolution and increased sensitivity. The study of ``classical'' targets such as hot Jupiters and directly imaged gas giant planets will see a paradigm shift, and we will directly probe the three-dimensional structure and wind patterns of their atmospheres. This will be possible with the use of high-resolution spectrographs, which are beginning to show their potential on 8\,m-class telescopes today.

With the advent of the Extremely Large Telescope (ELT), a specific emphasis in exoplanetary research will be placed on examining ``habitable'' terrestrial planets \citep{Snellen2010}. While the concept of habitability is still under exploration, in practical terms, this involves the study of solid, rocky celestial bodies whose surface temperatures permit the presence of liquid water. Recent studies of the mass-radius relationship for small exoplanets have shown that rocky planets typically have masses below approximately 5--8 times that of Earth \citep{Fulton17}. Objects exceeding this mass range may be dominated by water or hydrogen envelops, resembling small Neptunes \citep{Owen2016, Venturini2020, Burn2021, Luque2022}. Additionally, the requirement for liquid water implies that the level of irradiation from the host star falls within a factor of approximately two of Earth's irradiation \citep{Kopparapu2017, Turbet2023}. Thus, terrestrial habitable planets occupy a relatively narrow range of parameter space.

Although Earth-like planets are relatively common in the universe \citep{Bryson2021}, with the current telescopes and instrumentation, attempting to study the atmosphere of an Earth twin is still out of reach. The main difficulty to study small exoplanet atmospheres using direct imaging observations is their enormous planet-to-star contrast ratio. The ELT possesses two distinct qualities that will set it apart in its ability to investigate exoplanets: its unparalleled light-collecting capacity and angular resolution. These attributes offer in turn two distinct avenues for ANDES, previously known as HIRES \citep{Maiolino2013}, to scrutinize exoplanet atmospheres: to expand the characterization of exoplanets via transmission/emission spectroscopy down to the rocky regime, and to attempt for the first time the detection of reflected light from the planetary atmosphere. 

Our chances of characterizing small rocky planets in the coming decade greatly rely on the on-going and successful discovery of the rocky planet population around the brightest and closest M dwarfs \citep{Berta-Thompson2015,Dittmann2017,Gillon2017a,Luque2019}. M dwarfs are the most numerous stars in the solar neighbourhood \citep{henry18, reyle21}, and their lower masses and radii favour the detection of less massive planets. However, planets around M stars have been theoretically predicted to be very vulnerable to atmospheric mass loss \citep[e.g.,][]{Luger2015, Wordsworth2015, Dong2018}, and it is still a topic of hot debate whether they are habitable at all, after being exposed to intense radiation from the host stars for sustained periods of time \citep{Schaefer2016, Owen2016}. 

By the time the ELT becomes operational, it is anticipated that the best-suited potentially habitable planets will have been identified and characterized, thanks to ongoing and forthcoming space-based, namely TESS \citep{Ricker2015}, CHEOPS \citep{Benz2021}, and PLATO\citep{Rauer2014}, and ground-based exoplanet photometric and radial velocity surveys such as NGTS, MEarth, SPECULOOS, HARPS, HARPS-N, ESPRESSO, CARMENES, MAROON-X, or SPiROU, among others. The role of ANDES on the ELT will be to delve deeper into the characterization of these worlds, moving beyond basic parameters like mass and radius. Specifically, ANDES aims to explore their atmospheric structure, composition, and surface conditions, including the detection of possible bio-markers \citep{DesMarais2002, Palle2009}. In this detailed atmospheric characterization, there will be space for synergistic observations with facilities such as JWST, ARIEL and other ELT instrumentation that will be contemporaneous to ANDES. 

The unprecedented angular resolution of ANDES, will also allow us to explore the initial conditions in which planets form and the physical mechanism that give rise to the observed exoplanet population. The majority of the known planets are believed to have formed within $\approx15-20$ au from their hosting star \citep{Pollack1996, Morbidelli2016}, and understanding the composition and spatial distribution of atomic and molecular gas in this inner region of young ($<$10 Myr) circumstellar disks is an essential step towards understanding planetary system’s formation and evolution. One of the main scientific objective of ANDES will be to settle the properties of the gas in the inner star-disk region, where different competing mechanisms of disk gas dispersal are at play, namely magnetospheric accretion, jets, photo-evaporated and magnetically driven disk winds \citep{Rigliaco2013, Nisini2018, Banzatti2019}.

In the following sections, we will elaborate on these science cases and observing techniques, and present up to date simulations of the ANDES capabilities, including real or realistic observations that address the primary scientific goals of ANDES. These simulations have helped us establish the top-level requirements for the instrument and decide on a ranking of since case priorities. We have chosen to emphasize the atmospheric characterization of habitable terrestrial planets, considering it the most compelling of our science objectives, encompassing both transmission spectroscopy and the high-contrast/high-resolution method. However, we will also explore a broader array of intriguing scientific scenarios.

\section{ANDES observing modes and capabilities for exoplanet observations}

\subsection{ANDES Top Level Requirements and Observing Modes}

ANDES will have two fundamental observing modes: a seeing-limited $R=100,000$ fibre-fed spectrograph covering the 0.5--1.8\,\textmu m   interval (with the goal to extend the coverage to the blue and to the K band), and an integral field unit (IFU) AO-assisted mode (also at $R=100,000$) that will be operating in the Y, J, and H bands.  

Exploring small rocky planets in the habitable zone of their stars via transmission spectroscopy is the leading science case of ANDES, while rocky exoplanet reflected light detection is the third in priority (Technical Requirements Specification for ANDES, ESO Document ESO-391757, 2022). Thus, these scientific cases drive ANDES's top-level requirements (TLRs), which are:

\begin{itemize}
    \item Spectral resolution: $R \geq 100\,000$
    \item Spectral sampling $>3.0$ pixels per resolution element 
    \item Wavelength range: 0.5--1.8\,\textmu m (requirement), 0.38--2.4\,\textmu m (goal). In general, the case of exoplanet atmospheres requires a range of 550--1800\,nm for two major reasons: 1) To cover all the major molecules H$_2$O, O$_2$, CO$_2$, CH$_4$, NH$_3$, C$_2$H$_2$, HCN, and 2) To include most of the stellar flux for M dwarfs, which are prime targets. Maximizing stellar flux is key to reach sufficient SNR both in transmission and reflected light. Extension towards the blue down to 0.38\,\textmu m and/or to the red towards 2.4~\textmu m are not crucial, but would be highly beneficial to expand the potential for exoplanet research; for gas giant planets K-band (2 to 2.4 \textmu m) has been one of the most important bands to date. These potential extensions will be discussed in detail in the following sections. 
    \item A diffraction-limited IFU mode in the covering the near-IR Y, J and H bands (goal: I band), with several spaxel sizes (10-100 mas), giving a field of view from $\approx10 \times 10$ to $100 \times 100$ mas.
    \item AO performance: contrast enhancement of around 1000 at 3$\lambda/D$.
    \item High PSF stability on daily timescales
    \item High flat-field accuracy
    \item Wavelength calibration precision 1\,m s$^{-1}$  (goal: 20\,cm s$^{-1}$), stable at the time scales of several hours.
\end{itemize}

There are essentially two observational techniques that can be used to infer the surface and atmospheric properties of exoplanets using the high-dispersion and high spatial contrast capabilities of ANDES, namely atmospheric transmission/emission spectroscopy and planetary reflected light detection, which we will discuss in detail in the following sections. We note that the science case of transmission/emission spectroscopy relies only on using ANDES as a seeing-limited spectrograph. However, the reflected light science case relies on adaptive optics systems coupled with smaller spaxel-size IFUs.

\subsection{Seeing limited transmission and emission spectroscopy}

Atmospheres can be probed in transmission during the transit of an exoplanet in front of its host star. This technique involves comparing observations during the transit to those taken when the planet is not transiting, revealing spectral characteristics of the exoplanet and separating them from the star's spectra over time.

The strength of the atmospheric signal in transmission spectroscopy primarily relies on the atmosphere's scale height, calculated as $H = k T/(\mu g)$ under hydrostatic equilibrium using the ideal gas law. Here, $T$ represents temperature, $\mu$ is the mean molecular weight, and $g$ is the acceleration due to gravity, while $k$ is the Boltzmann constant. The apparent size of the host star's disk affects the signal, making smaller host stars, like M dwarfs, advantageous compared to solar-type stars. For instance, an Earth-like planet orbiting an M4 dwarf would produce a signal of approximately 40\,ppm in transmission, while the same signal for the Sun-Earth system would only be 0.2\,ppm.

In these observations, the primary noise source comes from the small number of photons emitted by the host star. Consequently, bright targets are essential even with the ELT because achieving a cumulative signal-to-noise Ratio (S/N) on the order of $10^5$ is necessary during in-transit observations. It is worth noting that some studies have already achieved noise levels of 10--20\, ppm using high-resolution spectroscopy and cross-correlation techniques, as demonstrated by previous research \citep[e.g.][]{Leigh2003, Brogi2013, Scandariato2021}. These precision levels appear to be primarily limited by photon counts, suggesting that further improvements can be made with increased photon accumulation. The ELT's superior light-gathering capabilities during the relatively short duration of exoplanet transits (typically a few hours) are crucial in this context \citep{Hook2005}.

Seeing-limited observations, when the planet and the star are not resolved, can also be used to measure the emission spectrum of a planet by taking advantage of their differential velocities. The differentiation between star and planet spectral features is possible due to the presence of Doppler shifts between both signals. Emission spectroscopy can be performed for both transiting and non-transiting planets. One uses the cross-correlation technique to detect the planetary signal, where one performs the cross-correlation of the stellar-signal corrected spectra with modelled spectra of the expected planetary signals \citep{Snellen2010}. This is useful when looking for atoms and molecules that may originate hundreds to thousands of individual absorption lines in the transmission spectrum, making the cross-correlation technique the most powerful tool for detection. Using this methodology, the contribution of all these lines is combined, reducing the photon noise and permitting the detection of atoms and molecules hidden in the noise when analysed individually. This technique can also be applied to transmission spectroscopy during transit \citep{Nugroho2017, Sanchez-lopez2019}.

The high dispersion observing techniques that we will discuss throughout this manuscript are phase-resolved transmission and emission spectroscopy (the latter also dubbed `high dispersion phase curves'  \citep{Flowers2019, Ehrenreich2020, Beltz2021, Pino2022}; High signal-to-noise transmission spectroscopy of individual lines \citep[e.g., sodium doublet, $\mathrm{H} \alpha$, among others][]{MillerRicci2012, Louden2015, Brogi2016, Seidel2020}; eclipse or transit mapping, or both, i.e. the use of time-resolved information during ingress and egress at primary and secondary eclipse (to access latitudinal information) \citep{Nikolov2015, Wardenier2021}


\subsection{AO-assisted reflected light detection}

While the transit spectroscopy technique is strongly impacted by stellar contamination and can be inconclusive for planets with no atmosphere (or a very thin one) or high-altitude clouds/hazes, this is not the case for the reflected light at high-contrast and high-resolution. The reflected light high-contrast high-resolution (HCHR) technique consists in combining an extreme Adaptive Optics (AO) and coronagraphic system with a high-resolution spectrograph \citep{Sparks2002,Snellen:2015,Lovis:2017}. 

The AO and coronagraph are used to spatially separate the planet from the star and then hide the light from the star to improve the planet-to-star contrast. An integral-field unit (IFU) then collects the light in the focal plane and sends it to a high-resolution spectrograph. An additional gain in contrast can be obtained through high-resolution spectroscopy. Star and planet spectra can be disentangled thanks to their different spectral features and Doppler shifts. This technique is sensitive to individual spectral lines in the planet spectrum, which can be either Doppler-shifted stellar reflected lines or intrinsic planetary lines originating from molecular absorptions in the planetary atmosphere. Many individual features are usually combined in a cross-correlation analysis to increase SNR. Detecting these spectral signatures constrains many properties of the planetary atmosphere and surface, such as albedo, chemical composition, abundances of molecular species, atmospheric pressure, temperature, cloud coverage, and 3D structure of the atmosphere. 

Essential observational quantities to be considered in this approach are the angular separation between planet and star, the planet-to-star contrast in reflected light, and the host star brightness. These control the planetary detectability and the amount of stellar contamination at the planet's position. The SNR on the planet spectrum is directly proportional to the planetary signal and inversely proportional to the square root of the contaminating flux from the star, which usually dominates the noise budget. In instrumental terms, the goal is to maximize the planet coupling while minimizing the stellar coupling at the planet location. The planet coupling depends on the overall throughput of the instrument and the Strehl ratio at the wavelength of observation. The stellar coupling mainly depends on the AO residual halo at the angular distance of the planet, the intensity of the stellar Airy pattern at that position, and the possible use of a coronagraphic solution to suppress the stellar point spread function (PSF). Although the technique has not been tested on sky yet, the RISTRETTO instrument will pioneer the technique on the VLT \citep{Lovis:2022, Blind2022}.

\subsection{Challenges ahead: Long period planets}
\label{sec:challenges_long_period}

\begin{figure}
\centering
\includegraphics[width=\linewidth]{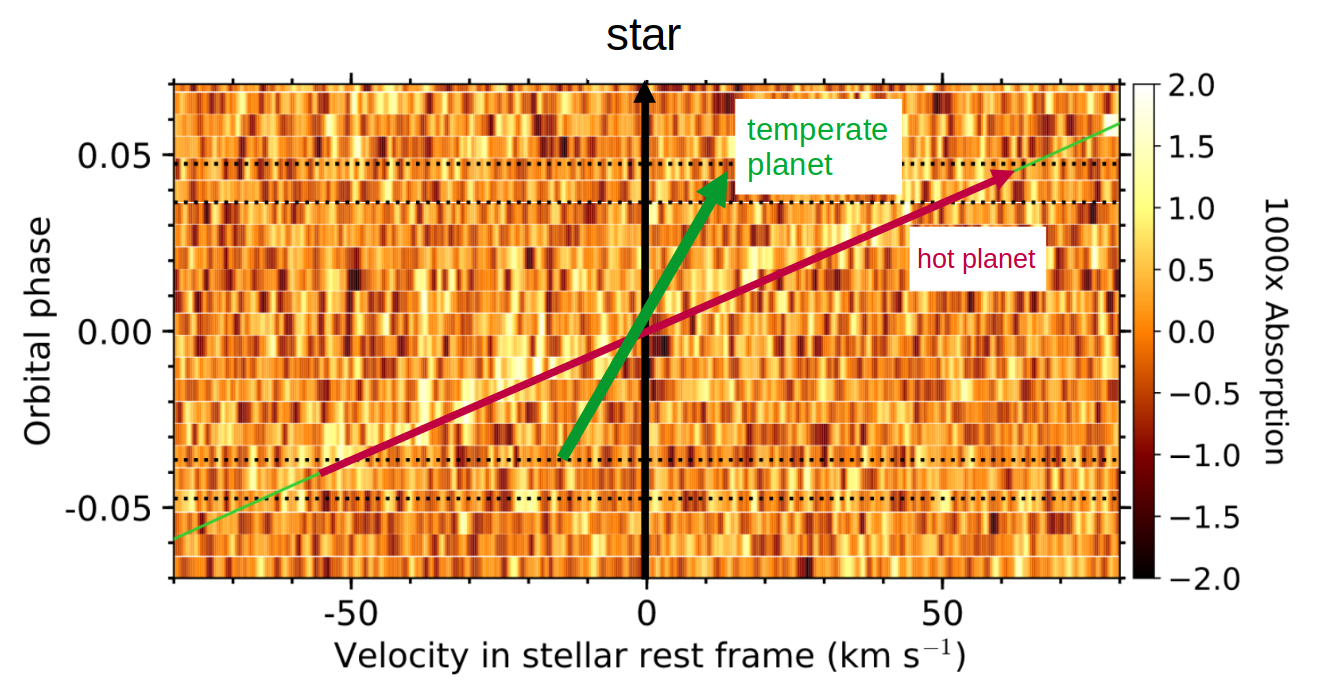}
\caption{\label{fig:example_restframes}Stellar versus planetary rest frame for a hot, i.e.\ short-orbit planet (red arrow) and a temperate planet (green arrow) in the habitable zone around the host star. For the temperate planet, the planetary and stellar (black arrow) rest frames are much closer together and changes in stellar activity may distort the planetary atmospheric signal \citep[figure adapted from][]{2020A&A...635A.205B}.}
\end{figure}

The biggest challenge in characterizing warm and temperate planets with long periods is that most of the current data reduction methods rely on the rapid change in planetary radial velocities across individual exposures to remove the stellar and telluric lines, for example, through Principal Component Analysis (PCA; \citep{Murtagh1987}). See Figure~\ref{fig:example_restframes} for a comparison example between a short and long period planet. Despite its large aperture, in the lack of a deep methodological improvement, data processing could hamper the capability of ELT to tackle this particular science case.

We assessed the sensitivity of current telluric removal algorithms, such as PCA or \textit{SYSREM} \citep{mazeh_2007_sysrem}, to slowly moving planets by measuring the fraction of the surviving signal in a simulated 5 hour observation of an emission spectrum centred at different planet's orbital phases (0.25--0.43 in steps of 0.03), for a set of orbital periods (1 to 15 days, variable steps), exposure times (100--800\,s in steps of 100\,s). We chose a target J magnitude of 14.5\,mag. SNR is calculated using a custom version of the ANDES Exposure Time Calculator (ETC) v1.1\footnote{\url{http://tirgo.arcetri.inaf.it/nicoletta/etc_andes_sn_com.html}}.

Our results show the following trends:
\begin{itemize}
    \item Phases closer to quadrature correspond to slower rates of change of RV, therefore these are the phases most affected by PCA (which is well known and expected, e.g.\ \citealt{Pino2022}).
    \item Very short period planets ($<2$\,d) are almost insensitive to the above because a 5\,h observation spans about 10\% of the orbit, including spectra far from quadrature and thus allowing the recovery of a very significant fraction ($>80\%$) of the signal.
    \item For longer-period planets, not only the radial velocity rate of change is smaller overall, but the phase range covered by a single 5\,h observation is also smaller, resulting in poorer constraints on ($v_\mathrm{sys}$, $K_\mathrm{p}$). As a result, a 15-day period planet has at most 30\% of its signal surviving.
\end{itemize}
Results are nearly independent of exposure time, except for the shorter-period planets where smearing effects for long exposure lead to a drop of 10--20\% in the recovered signal.

The above simulations imply that when estimating the detectability of a planet with current analysis techniques there might be an additional scaling factor due to the detrimental effects of PCA, to be applied on top of any scaling due to other factors such as T/P/abundance profiles or planet radius, among others. We note that these predictions apply to emission observations, that is for spectral sequence where a planet signal is present in each observed spectrum.

\begin{figure}
\centering
\includegraphics[width=\hsize]{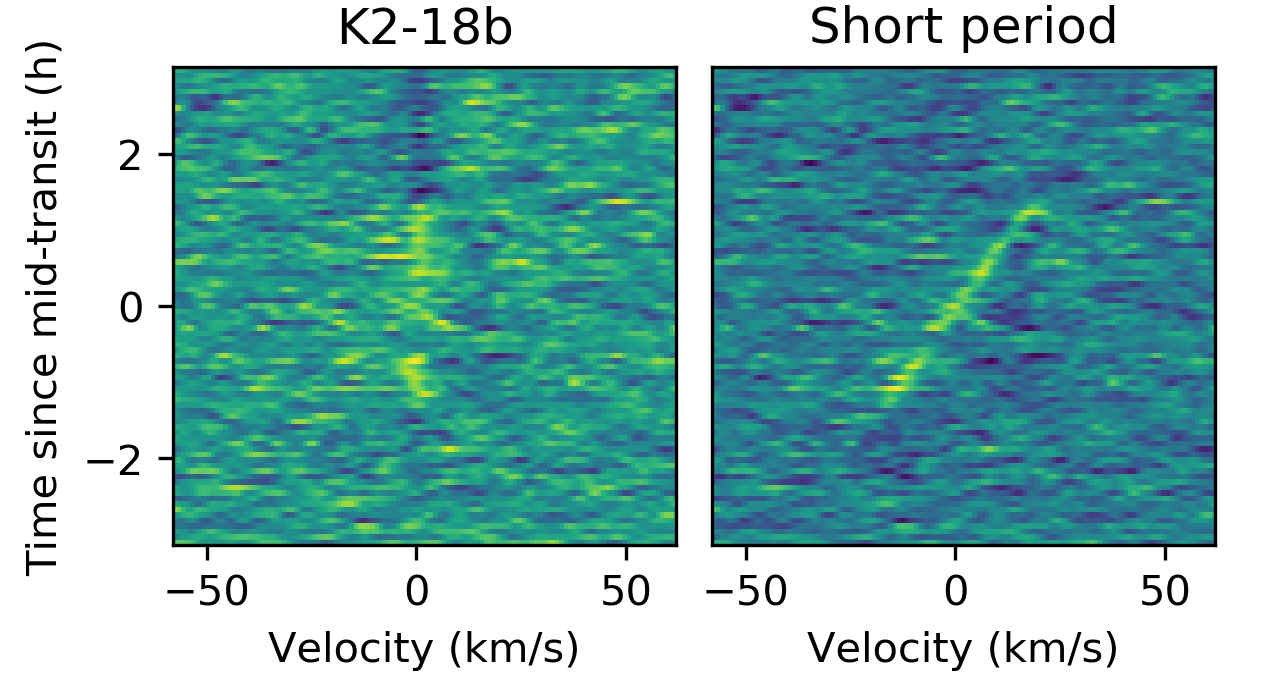}
\includegraphics[width=\hsize]{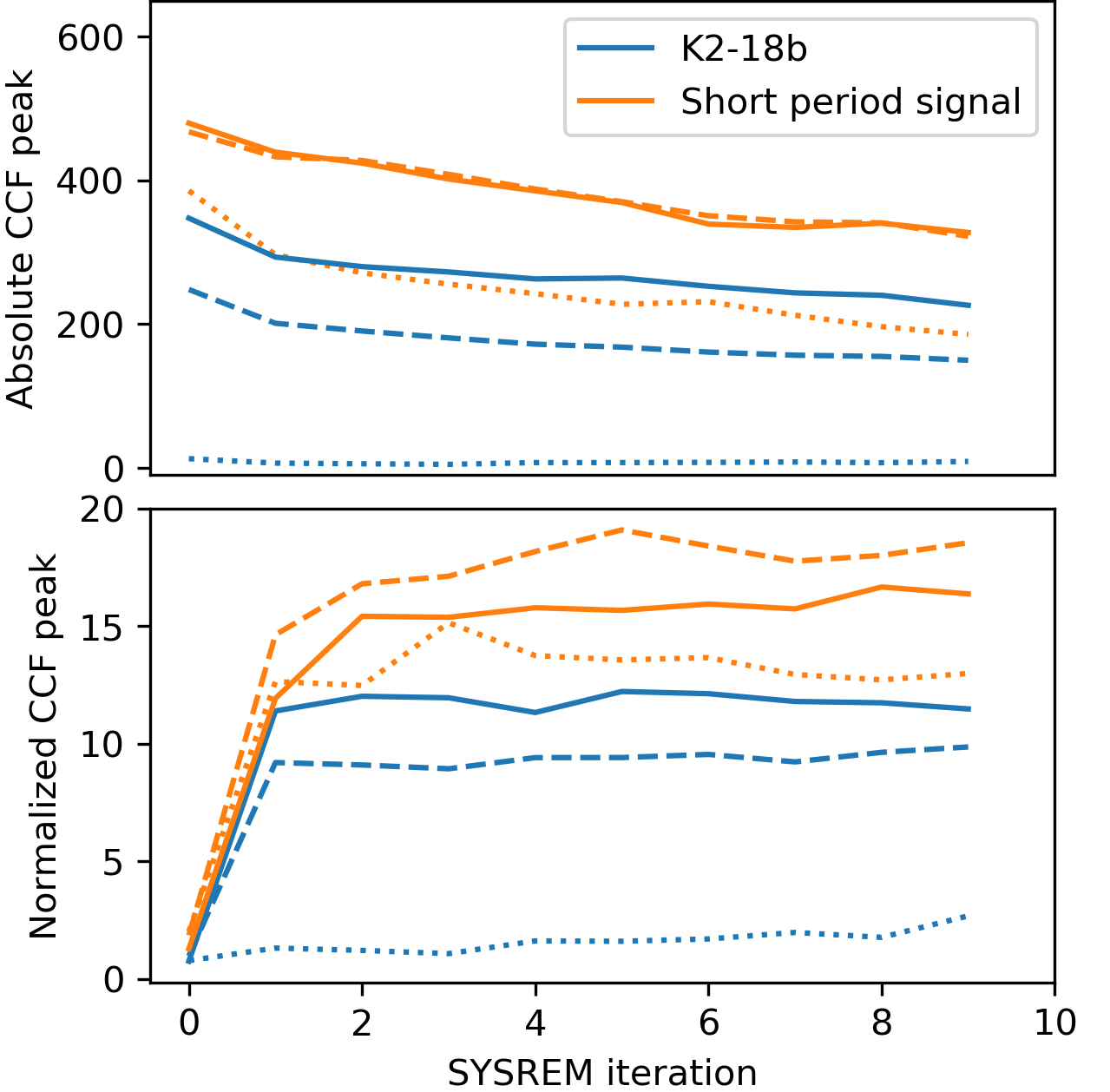}
\caption{\label{fig:sim_k218_sysrem} Top left: cross-correlation function map of the planetary signal after 10 \textit{SYSREM} iterations for K2-18b (top left) and for a hypothetical 'short period' version of the same planet moving with faster changing projected orbital velocity (top right). Middle panel: The evolution of the signal with rising \textit{SYSREM} iterations is shown. For both systems three observational scenarios are explored: 3.6\,h (solid line), 2\,h (dashed lines) and 0\,h (dotted lines) of out-of-transit baseline coverage. Bottom panel: The evolution of the remaining planet signal normalized by the noise of the \textit{SYSREM} corrected data at each iteration using the same colors and linestyles as the middle panel. 
}
\end{figure}

We also assessed the removal of signal for transmission spectroscopy of long-period planets using the example case of K2-18b. The projected velocity for this planet, that has a 32-day long orbit, changes by 1km s$^-1$ between the start and end of its 2.3\,h long transit. This causes its atmospheric signature to appear quasi-static in the data. For the assessment, we use a simulated spectroscopic time series of one transit event. We compute this simulated data using the spectrum of an M2.5 star obtained by CARMENES as a stellar template \citep{Nagel2023} and model telluric contamination on the basis of realistic 6\,h long observations. As for the emission case, we calculated the SNR expected for the J and H magnitudes of K2-18b using the ANDES ETC v1.1. We explored the effect of different duration of the observations. In each case the data covers the full transit event, but the available out of transit coverage is varied. We find that for a quasi-static signal, such as the one of K2-18b the in-transit signal is reduced by a factor of 30-55\% compared to a more favourable case in which the planet's velocity changes significantly faster (by $\approx$ 1 km s$^-1$ per exposure) during transit. We observe that for the quasi-static signal of K2-18b  the amount of signal removed depends on the amount of out-of-transit baseline available, with more out-of-transit baseline leading to less signal removal (see Figure \ref{fig:sim_k218_sysrem}). 
This effect can be well observed in the progression of both the absolute signal as well as in the signal strength relative to the residual noise in the data. For the short period case this correlation becomes more obscured when the signal is normalized by the noise level. 
This is due to larger noise levels in the regions affected by telluric lines caused by the longer observing duration. The longer out-of-transit coverage requires observations taken at high air mass, which negatively affects the noise level of the data. This effect is not observed in the quasi-static case of K2-18b as the telluric contamination remains well removed from both the planet signal and the region at which noise is evaluated.
\textit{SYSREM}/PCA not only removes signal in transit but also over-corrects the data obtained out of transit, as can be observed as a dark (anti-correlated) signature in Figure \ref{fig:sim_k218_sysrem} upper left panel. Dash et al. (under review) illustrate that such artifacts in the post-\textit{SYSREM}/PCA data i) depends on the slope of the in transit signal and ii) are completely reproducible on each tested model. Therefore, both the in-transit and the out-of-transit portion of the spectral sequence can be utilised to pinpoint the orbital solution of the exoplanet.

Here, no other approach to data analysis different to PCA, or the use of \textit{SYSREM}, has been considered. Developing additional alternative data analysis techniques for both visible and near-infrared spectral ranges combining better telluric correction and the use of better stellar templates can perhaps alleviate this difficulty.

\subsection{Challenges ahead: Stellar contamination}

The atmospheres of host stars are covered with complex, stochastic patterns associated with convective heat transport, the stellar granulation. Convection manifests in the surface layers as a particular pattern of downflowing cooler plasma and bright areas where hot plasma rises \citep{Nordlund2009}. The size of the convection-related surface structures depends on the stellar parameters, with larger granules associated to lower surface gravity stellar types \citep[e.g., ][]{2013A&A...557A...7T,2014A&A...567A.115C}. Convection is a difficult process to understand, because it is non-local, and three-dimensional, and it involves nonlinear interactions over many disparate length scales. In this context, the use of numerical three-dimensional (3D) radiative hydrodynamical (RHD) simulations of stellar convection is crucial and has become possible in the last decade large grids of simulations covering a substantial portion of the Hertzsprung-Russell diagram \citep{2009MmSAI..80..711L,2013A&A...557A..26M,2013ApJ...769...18T,2013A&A...558A..48B}. 

The direct consequence of the non-stationary stellar spectrum, in the form of either Doppler shift or distortion of the line profile during planetary transits, creates a non-negligible source of noise that can alter or even prevent detection, especially when the same spectral lines are present both in the stellar and in the planet spectrum: e.g., the diatomic molecules like CO, H$_2$O, ZrO, VO or the atomic transitions such as Fe, V, or Li, among others. Among the different stellar types, M dwarfs are key targets for high-resolution spectroscopy due to a high incidence of these stars in the solar neighbourhood \citep{2006AJ....132.2360H} and their importance as exoplanetary hosts \citep{2015ApJ...807...45D,2023A&A...670A..84K}. M dwarf stars, with masses between 0.08 and 0.4 M$_{\odot}$, are fully convective and can generate strong magnetic fields \citep{2008ApJ...676.1262B,2015ApJ...813L..31Y}. Their low effective temperatures result in a plethora of molecular line transitions throughout the optical spectrum \citep[e.g. TiO, VO, ZrO, CO, H$_2$O, MgH, OH, CaH, FeH, ][]{2018A&A...610A..19R,2021A&A...656A.162M} and atomic line transitions \citep[e.g. ][]{2018A&A...620A.180R} for which non-LTE effects on particular potassium strong lines lead to not negligible metallicity corrections \citep{2021A&A...649A.103O}.

\begin{figure}
\centering
\includegraphics[width=0.6\linewidth]{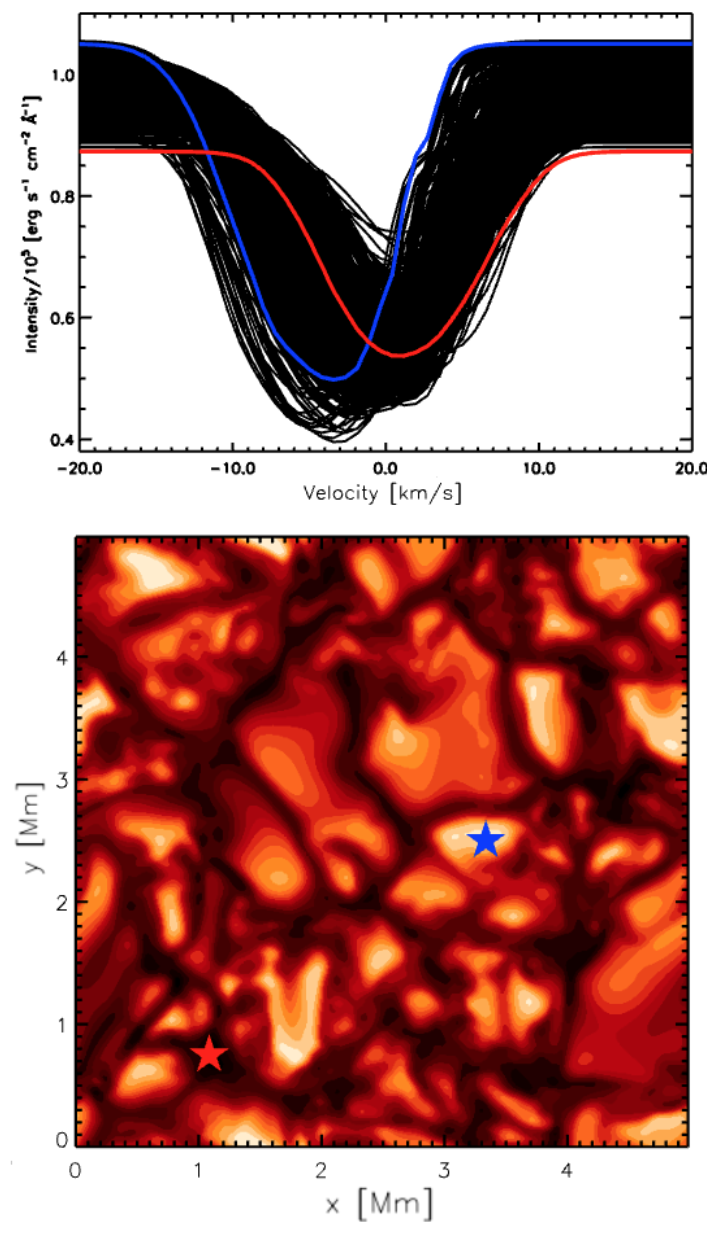}
\caption{\label{Figstarplanet} Spatially resolved profiles at disk center ($\mu$=1.0, left panel) of a CO line ($\lambda$=23015.002 \AA, log $gf$=0.221, $\chi$=-5.474 eV) across the granulation pattern (right panel) of a K-dwarf 3D radiative hydrodynamical stellar convection simulation \citep[][]{2019A&A...631A.100C}. The timescale of stellar convection variability is few minutes, much shorter than the planet transiting time. The solid red (intergranular lane) and blue (granule) lines display two particular positions (colored star symbols) extracted from the intensity map.}
\end{figure}

In the first order, the telluric bands and instrumental trends can be considered stationary or quasi-stationary in wavelength for the duration of a typical observation night and then corrected \citep{2010Natur.465.1049S,2018arXiv180604617B}, 
while at high resolution the shape and position of the stellar absorption lines are extremely variable in time (Figure~\ref{Figstarplanet}): the spatially averaged across the stellar surface convection shifts depend on the stellar parameters and range between several dozens to a few hundred m/s \citep{2013A&A...550A.103A, 2018A&A...611A..11C}.

In this context, the good and time-dependent representation of the background stellar disk with 3D RHD simulations is a natural and necessary step forward toward a better understanding of stellar properties and, in the context of exoplanet science, for a detailed and quantitative analysis of the atmospheric signatures of transiting and non-transiting planets \citep[e.g., ][]{2019A&A...631A.100C, 2019AJ....157..209F, 2023MNRAS.tmp.2691K}. 
One particular aspect concerns the 3D RHD simulations of M dwarf stellar convection. Pioneering works on M and brown dwarfs presented the challenges in modelling these objects with respect to  main sequence stars and highlighted the presence of self-excited gravity waves as an essential mixing process in their atmospheres \citep{2002A&A...395...99L,2010A&A...513A..19F}. More recently, new grids are under development \citep[e.g. ][]{2022csss.confE.161K,2022MNRAS.514.1741R}. Theoretical efforts for obtaining detailed line formation physics, using next-generation time-dependent hydrodynamics simulations, will be crucial in determining spectroscopic based parameters on these stars as well as their impact on atmospheric signatures of planets. Especially when combined with empirical solar observations that are translated into a Sun-as-a-star setting \citep[e.g.][]{2022ApJ...939...98O, 2023arXiv230903373P, 2023MNRAS.tmp.3174C, 2023ApJ...951..151C}.


\section{HZ rocky planet atmospheres: Transmission and Emission Spectroscopy}

Transmission spectroscopy is nowadays routinely used to detect molecular and atomic species in the atmosphere of hot, giant planets, using ground-based spectrographs (e.g., \citealt{Giacobbe:2021}). ANDES at the ELT does not represent a major technical advance in the application of these techniques, but it will be a major leap in capabilities. This is because in practice, the community will be moving from spectrographs on 3 to 4 m telescopes (like CARMENES, GIARPS, or SPIROU for example) to a 39m diameter telescope, as the 8-10 m class telescopes do not have stabilized instrumentation similar to ANDES. Stable spectrographs like ESPRESSO, MAROON-X or KPF, cover only the optical spectral range and thus are not suited to detect molecular species (and CRIRES+ which is limited in spectral coverage). This means that ANDES will have the capability to extend the technique down to small, rocky, temperate exoplanets transiting nearby stars. Some of the best available targets we have so far for ANDES include benchmark systems like the TRAPPIST-1 planets \citep{Gillon:2017} and LHS~1140b \citep{Dittmann:2017}, many of which are also part of JWST approved  observing programs (see e.g., \citealt{Lim:2023}).

Previous observations, in particular with JWST, have revealed several challenges on our way to use transit spectroscopy to characterize these small planets. Heterogeneities on the stellar photosphere (spots, faculae) can strongly bias the measurements \citep{Rackham:2023,Lim:2023}. Moreover, if a thick layer of clouds is present in the planet's upper atmosphere, the transit spectroscopy technique struggles to provide molecular detections \citep{Fauchez:2019}. 

A detailed feasibility study for ANDES is needed to evaluate quantitatively the number of nights required to characterize the potential atmospheres of transiting, small, nearby exoplanets with transit spectroscopy. Such study should carefully and quantitatively evaluate the two challenges mentioned above (stellar contamination, and clouds) and take into account the specific mean molecular weight of each planet to be observed.

Here we have modeled the potential contribution of ANDES by simulating the observations of a rocky and a sub-Neptune small planets, and then extend the simulations to the known population of small planetary systems. We focus on transmission spectroscopy as small rocky planets in the HZ are not good targets for emission spectroscopy.  Emission spectroscopy for larger planets is discussed in Section~\ref{sec:gasgiants}

\begin{figure}
\centering
\includegraphics[width=0.99\linewidth]{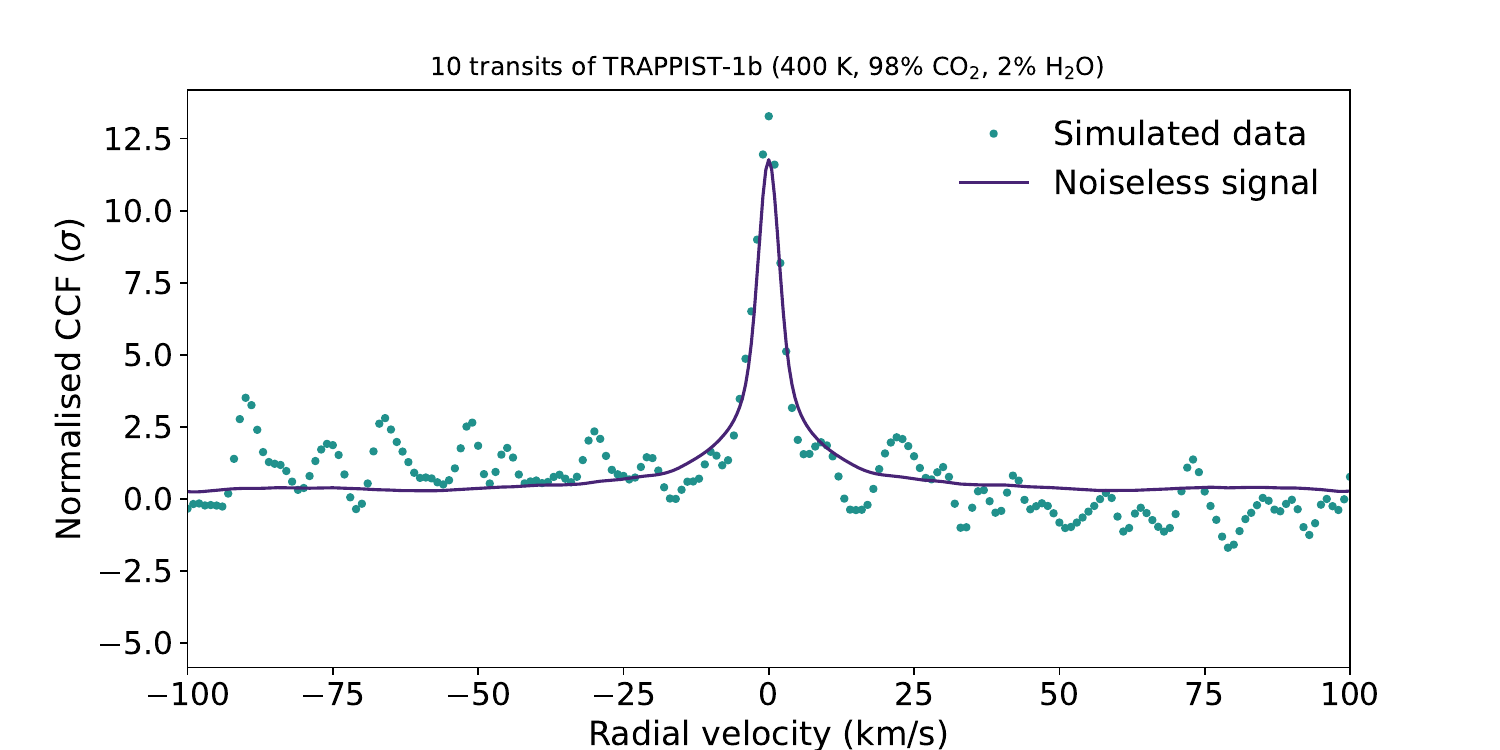}
\includegraphics[width=0.99\linewidth]{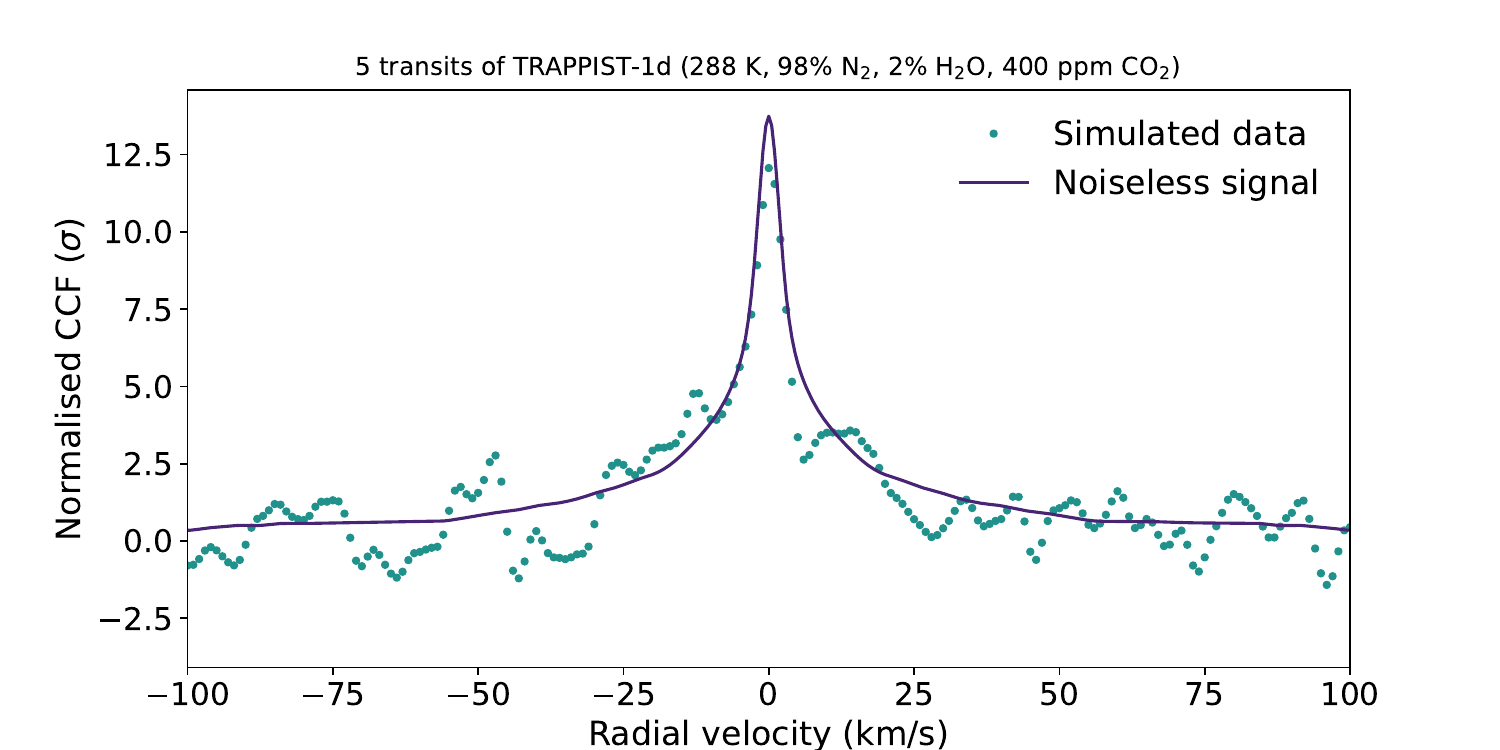}
\includegraphics[width=0.99\linewidth]{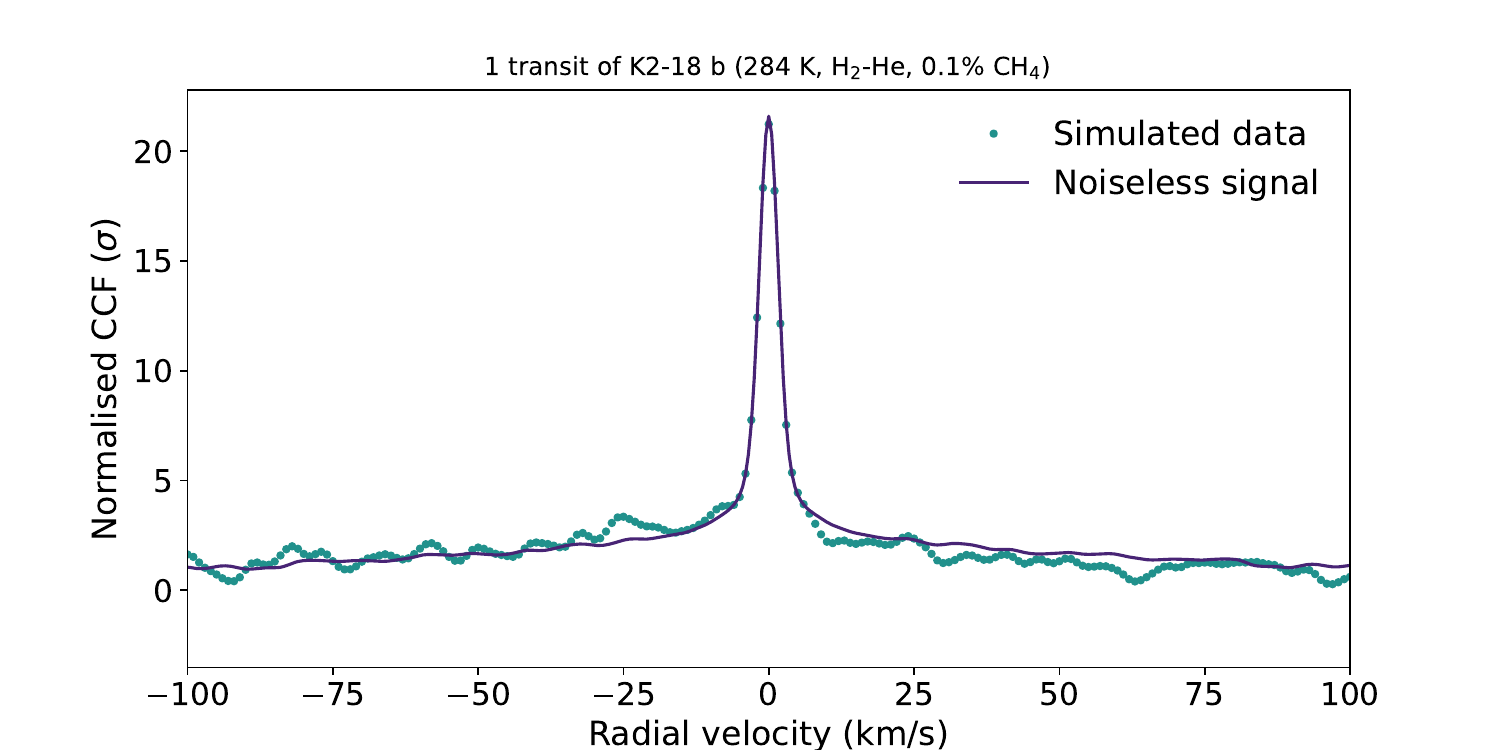}
\caption{\label{fig:ts_trappist} Simulated cross-correlation functions derived from modelled transit spectra of Trappist-1b (top), Trappist-1d (middle) and K2-18b (bottom). The y-axis indicates the significance of the detection.}
\end{figure}

\subsection{Key targets transit simulations}
\label{sec:key_targets_transit}


In order to explore the potential of ANDES, we have modeled several possible atmospheric scenarios for some key planets. In particular, we simulated the Trappist-1b, assuming a wet CO$_2$ dominated atmosphere (wet meaning 2\% H$_2$O) at the equilibrium temperature of 400\,K, and with ten transits stacked. For Trappist-1d, we simulated a wet N$_2$ dominated atmosphere with 400\,ppm of CO$_2$ (similar to Earth-like but without oxygen) and five transits stacked. Finally, we simulate K2-18b assuming a solar metallicity H-He atmosphere, with a 0.1\% of CH$_4$ as recently claimed by \cite{Madhusudhan2020}. In this case, the simulation is done for a single transit observation.

The models were built using petitRADTRANS \citep{Molliere2019}, with line absorption of CO$_2$ and H$_2$O, Rayleigh scattering of H$_2$/He, CO$_2$ or N$_2$ (depending on the species) and collision induced absorption by H$_2$-H$_2$, CO$_2$-CO$_2$ and N$_2$-N$_2$. The models also assume that the planets are tidally locked and that the atmospheres are co-rotating, leading to some rotation broadening (mostly relevant here for Trappist-1b). To create the cross-correlation templates, a continuum normalization was carried out by coarsely sampling the continuum in 50\,nm bins. No telluric contamination is considered in the modeling.

We caution that this exercise is conducted using several assumptions, namely: 

\begin{enumerate}
    \item A template that is a perfect model of the planet transmission spectrum (besides continuum normalization);
    \item The atmosphere is assumed to be clear down to 1 bar and opaque below, and with no clouds or hazes;
    \item There is no stellar spectrum modeling nor data analysis steps usually needed to remove it (e.g., PCA or SYSREM). The star is assumed to give us a flat continuum plus a wavelength-dependent SNR as the ANDES ETC dictates;
    \item We have not considered systematic noise sources. This might be important as high-resolution observations in the infrared are often not limited by photon noise but either by imperfect telluric removal or, in the case of M-dwarfs, molecular absorption by unocculted spots.
\end{enumerate}

The resulting simulated cross-correlation functions are shown in Figure~\ref{fig:ts_trappist}. It is noticeable how well one can detect the atmosphere for all three planets. In particular, the strong significance of K2-18b from a single transit is undoubtedly thanks to the broad wavelength coverage and the inclusion of the K-band (which we have assumed here) in ANDES.

\begin{figure*}
\centering
\includegraphics[width=0.49\linewidth]{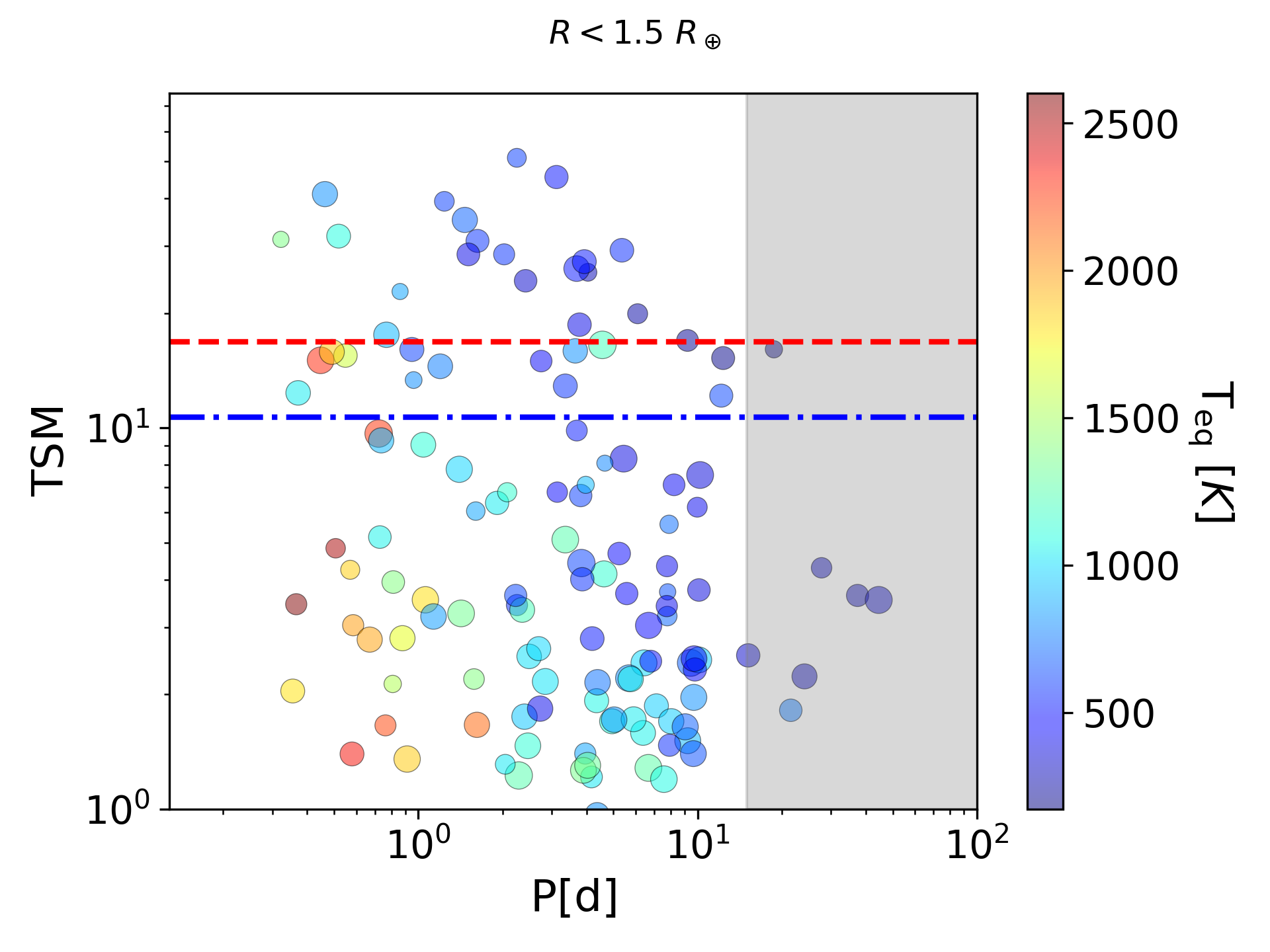}
\includegraphics[width=0.49\linewidth]{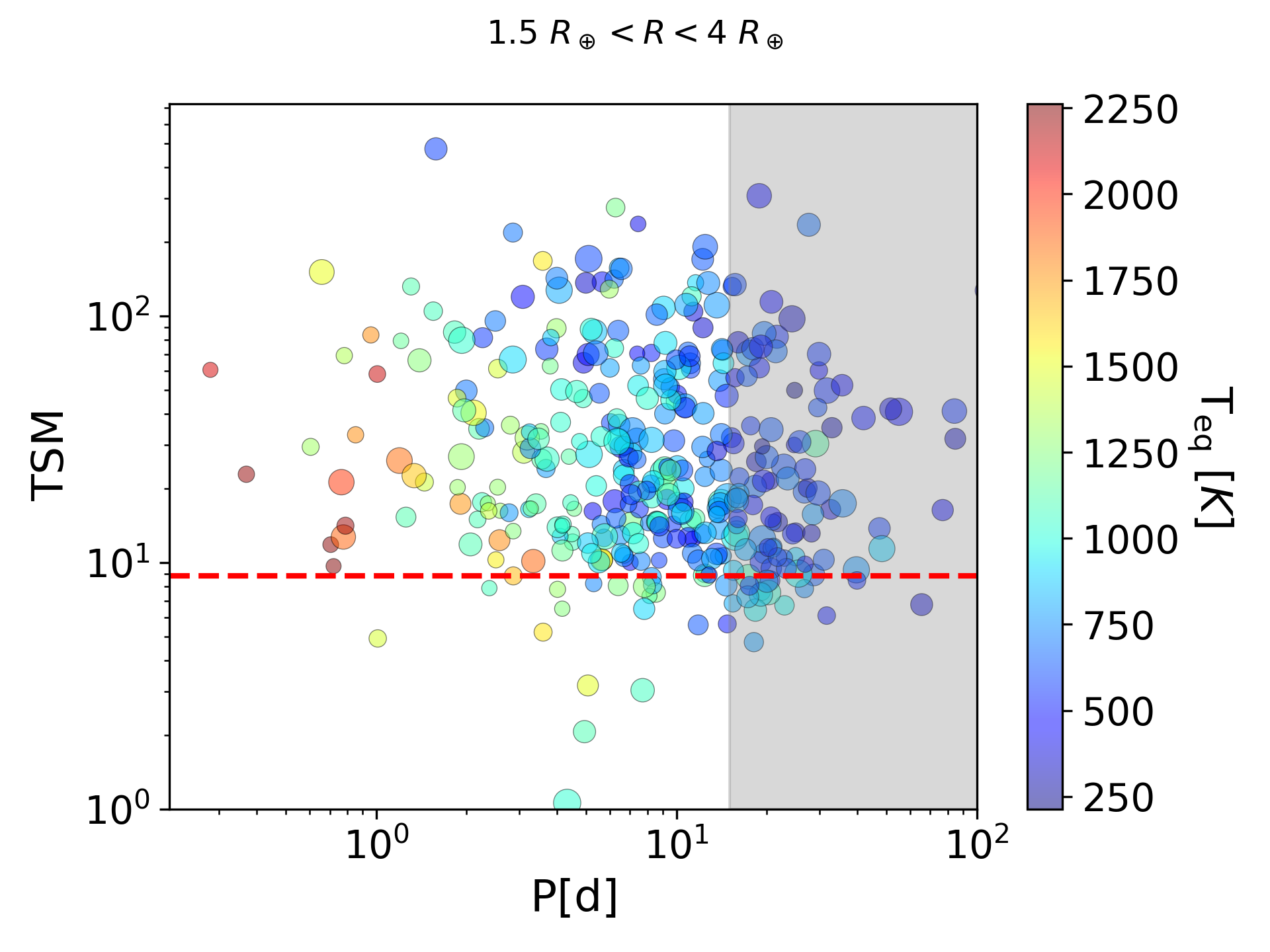}
\caption{\label{fig:trasnmissionpopulation} Atmospheric detectability with ANDES for the known sample of transiting small planets ($R < 4 R_\oplus$). 
Left panel: Earth-size planets with $R < 1.5 R_\oplus$. The atmospheres of the planets above the red (blue) broken line can be detected with ANDES at better than $5 \sigma$ in 5 transits, if they posses a $CO_2$-dominated (Earth-like) atmosphere. 
Right panel: Sub-Neptune-sized planets with $1.5 R_\oplus < R < 4 R_\oplus$. The atmospheres of the planets above the red broken line can be detected with ANDES at better than $5 \sigma$ in a single transit. 
On both panels, planets are colour-coded by equilibrium temperature, and symbol-size is proportional to planet radius. The gray shaded area represents the region at periods larger than 15 days, for which PCA- or SYSREM-based methods might erode a significant fraction of the planet signal (see Sec. \ref{sec:challenges_long_period}). }
\end{figure*}

\subsection{Small transiting planet accessible population}

Based on the simulations presented in Section~\ref{sec:key_targets_transit}, we estimate the number of currently known transiting planets with $dec < 10^\circ$ in the $ R < 4R_\oplus $ regime\footnote{Gathered from the NASA Exoplanet Archive at the date of 13th October 2023}, that ANDES could be able to characterize. For each target, we calculate the Transmission Spectroscopy Metric (TSM; \citealt{Kempton2018}), which is proportional to the expected signal-to-noise ratio based on the strength of spectral features. We then divide the sample of known planets in sub-Neptune-size ($1.5 R_\oplus < R < 4 R_\oplus$) and earth-size ($R < 1.5 R_\oplus$). We consider K2-18b as a representative member of the sub-Neptune-size category, and TRAPPIST-1b and TRAPPIST-1d as representative members of the earth-size category. Our simulations in Section~\ref{sec:key_targets_transit} show that K2-18b will be detected at $>20\sigma$ in one transit, TRAPPIST-1b ($\mathrm{CO}_2$ dominated atmosphere) at $>12\sigma$ in 10 transits and TRAPPIST-1d ($\mathrm{N_2}$ dominated atmosphere) at $>12\sigma$ in 5 transits. We thus establish the TSM-threshold to detect a sub-Neptune-size planet at $5\sigma$ in one transit as $\mathrm{TSM_{K2-18b}}/4$, an earth-size planet with a $\mathrm{CO_2}$ dominated atmosphere in 5 transits as $\mathrm{\sqrt{2}TSM_{TRAPPIST-1b}}/2.4$ (accounting for the different amount of transits), and an earth-size planet with a $\mathrm{N_2}$ dominated atmosphere in 5 transits as $\mathrm{TSM_{TRAPPIST-1d}}/2.4$. We neglect the differences in transit duration among the known planets.

Figure~\ref{fig:trasnmissionpopulation} shows the number of accessible planets for atmospheric detection via transmission spectroscopy.  For sub-Neptune-size exoplanet ($1.5 R_\oplus < R < 4 R_\oplus$), ANDES will be able to detect the planetary atmospheres in a single transit observation for about 330 out of about 370 known transiting planets with $dec < 10^\circ$ at better than $5 \sigma$, of which about 130 have $P < 7$ days and 240 have $P < 15$ days. These planets cover a wide range of equilibrium temperatures and radii within this range.

For the smallest earth-size exoplanets ($R < 1.5 R_\oplus$), the results in Figure~\ref{fig:trasnmissionpopulation} assume 5 transit observations per target, and two possible atmospheric compositions. The red line marks the detectability of a $CO_2$-dominated atmosphere, while the blue line marks the detectability of an Earth-like atmosphere.  In the first case 20 planets out of the 140 currently known are detected at $5 \sigma$, nearly all of them with $P < 7$ days. For an Earth-like atmosphere 35 planets out of the same 140 known planets are detected at $5 \sigma$, of which 30 have $P < 7$ days and nearly all have $P < 15$ days. Note that 4 and 16 earth-size planet atmospheres are detectable in the $CO_2$-dominated and Earth-like cases, respectively, with only a single transit observation.


We note that small planets ($R < 4 R_\oplus$) encompasses a diversity of possible planetary bulk compositions, from rocky earth-like composition to water worlds to puffy sub-Neptunes with H/He envelopes. In particular, the range of possible atmospheric composition of rocky earths and super-earths is not well known and subject of strong debate \citep{Liggins2022, Wordsworth2022}. This is because the composition of the atmosphere is determined by the insolation history of the system, and also by the interplay between the atmosphere and the surface or magma beneath, factoring in the solubility of volatile substances \citep{Tian2023}. While research has largely centered around the interaction between molten metals and silicate with hydrogen-rich atmospheres \citep{Kite2020, Schlichting2022}, little is known about how magma oceans in planets with large water mass fractions interact with their atmospheres. Planets that experience intense radiation could maintain high surface temperatures even if they have water-rich atmospheres \citep{Dorn2021, Vazan2022}. These magma oceans could act as large reservoirs to keep volatiles from being stripped away during the early, active life of the host stars, but this has not been empirically tested. Thus, ultimately the detectability thresholds for ANDES will be extremely target-dependent. Still the number of available targets for ANDES that our simulations show is really encouraging. Our estimations are also consistent with a similar exercise, more detailed and considering more molecular species, conducted by \cite{curriemeadows2023}. An increased well-characterized sample of small planets is needed to constrain theories for how these planets form and retain atmospheres, which is in turn crucial for understanding the habitability of their more temperate siblings.

\begin{figure*}
  \centering
  \includegraphics[width=0.98\linewidth]{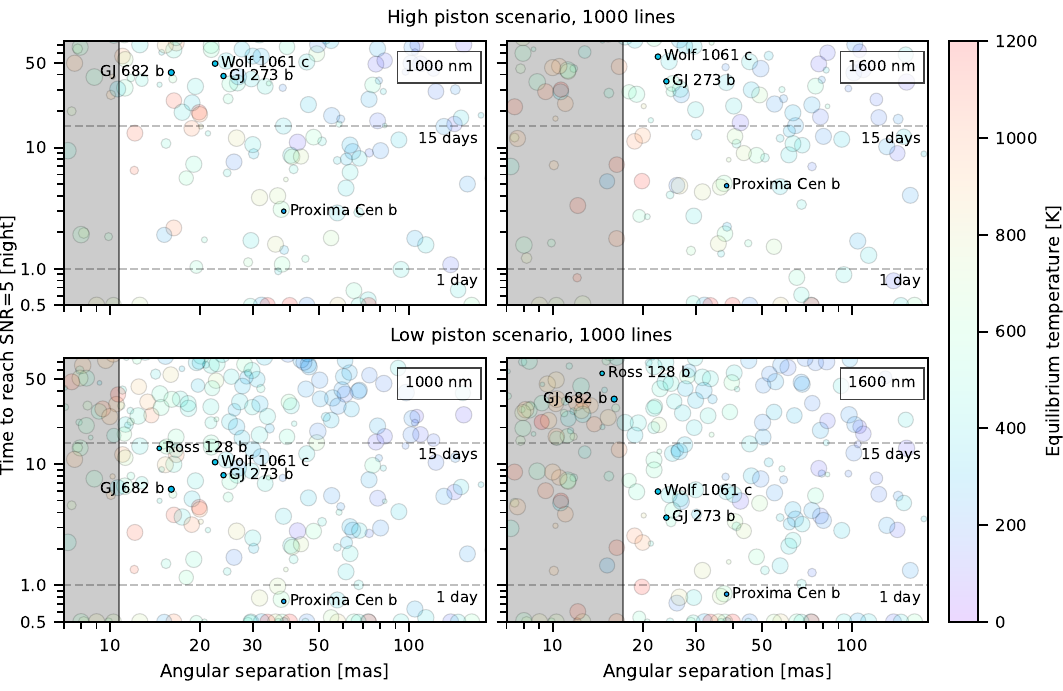}
  \caption{The currently known population of exoplanets accessible for atmospheric characterization through AO-assisted reflected light detection at different wavelengths. Left panels correspond to the Y band centered at 1 $\mu$m and right panels correspond to the H band centered at 1.6 $\mu$m. The top panels assume a high piston scenario for the AO performance, and the lower panels assume a low piston scenario (see main text for discussion). We have assumed the cross-correlation is performed over 1000 absorption lines in the planetary atmosphere. The grey shaded area correspond to angular separations smaller than $2 \lambda/D$ at the given wavelength and are considered unreachable. The five habitable zone rocky planets that constitute our current golden targets are marked and labelled. The color scale corresponds to the equilibrium temperature of the planets. The Y axis is expressed as time in nights (1 night = 8 hours) and the accumulation of planets near 0 corresponds to giant planets around bright host stars, for which SNR=5 is reached within minutes.   }
  \label{fig:samplereflected}
\end{figure*}

\section{HZ rocky planet atmospheres: Reflected light}
\label{sec:aogeneral}

The HCHR technique offers the possibility to target non-transiting, and thus more nearby, rocky planets. The HCHR reflected light technique is, by essence, sensitive to the planetary albedo, which can be strongly impacted by all reflective layers on the surface (e.g., continental or oceanic ice) and in the atmosphere (e.g., clouds) of planets. While clouds are an issue for the transit spectroscopy technique, they can be an advantage for the HCHR technique. The HCHR technique can be used first to detect planets and then to characterize their surface and atmosphere (if any).

Even surface features can, in principle, be probed in the reflected-light geometry, in contrast to the transit geometry, which is only sensitive to the upper atmospheric layers. It is therefore highly interesting to investigate temperate rocky planets with this technique, which could be used to detect the major molecular constituents of their atmosphere (e.g., H$_2$O, CO$_2$, CH$_4$, NH$_3$, O$_2$) and constrain the presence of clouds, ice caps, and liquid water on their surface. We note, however, that exoplanet reflected light detections at high resolution have yet to be robustly demonstrated to work on the sky. Some works have placed strong upper limits to this technique's capabilities \citep{Hoeijmakers2018_k9, Scandariato2021, Spring2022} using just high-resolution (without the high-spatial component) that reach down to the $\sim 1x10^{-6}$ level with optical detectors, so detection is in principle possible, but has not yet been proven.


We have simulated ANDES single conjugate adaptive optics (SCAO) observations of known exoplanets to explore the target sample that can be probed in this mode. The modeling details for the AO performance and the details of the simulations for exoplanet observations are given in Appendix~\ref{App:one}.

\begin{figure*}
  \centering
  \includegraphics[width=0.49\linewidth]{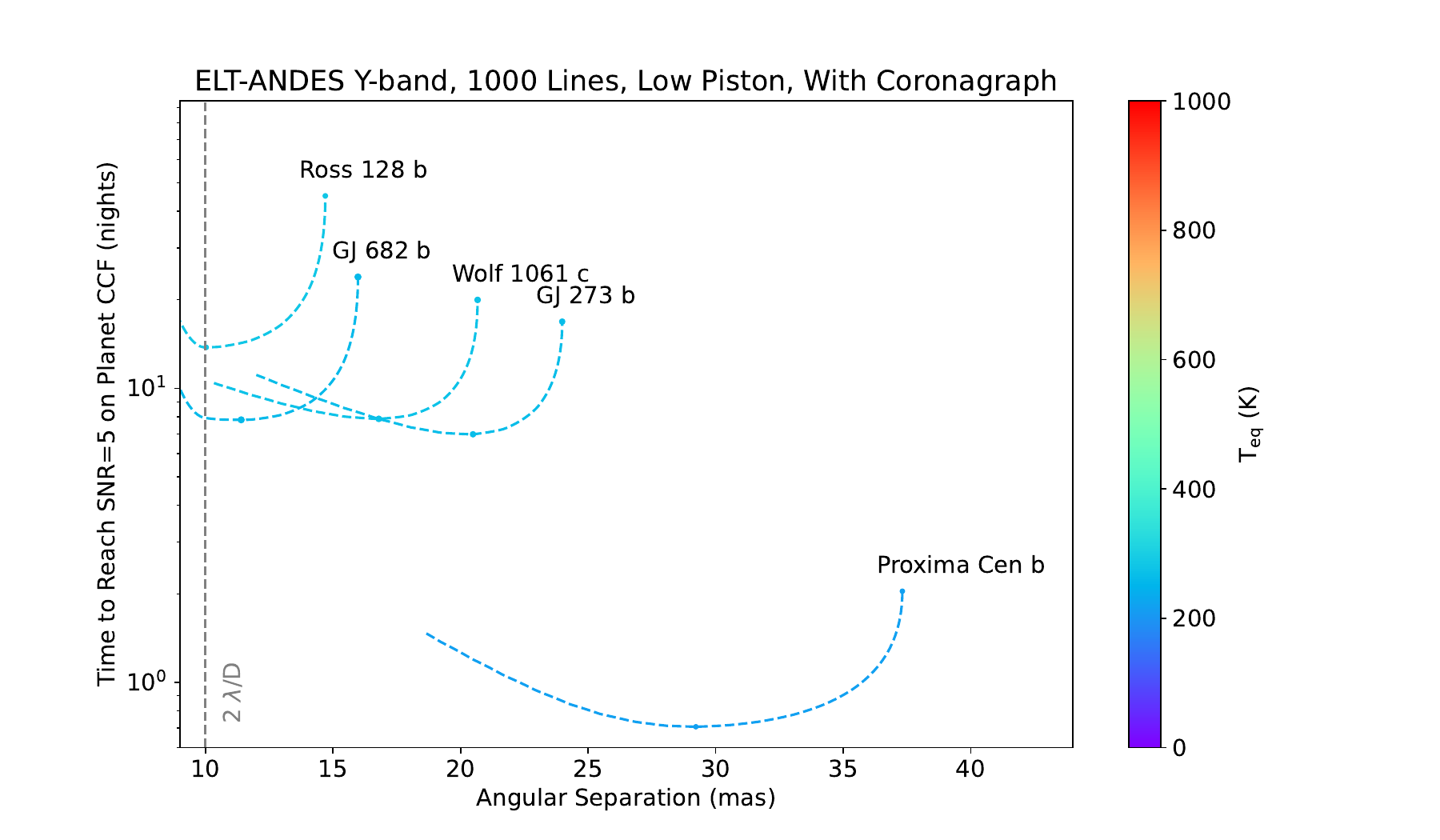} 
  \includegraphics[width=0.49\linewidth]{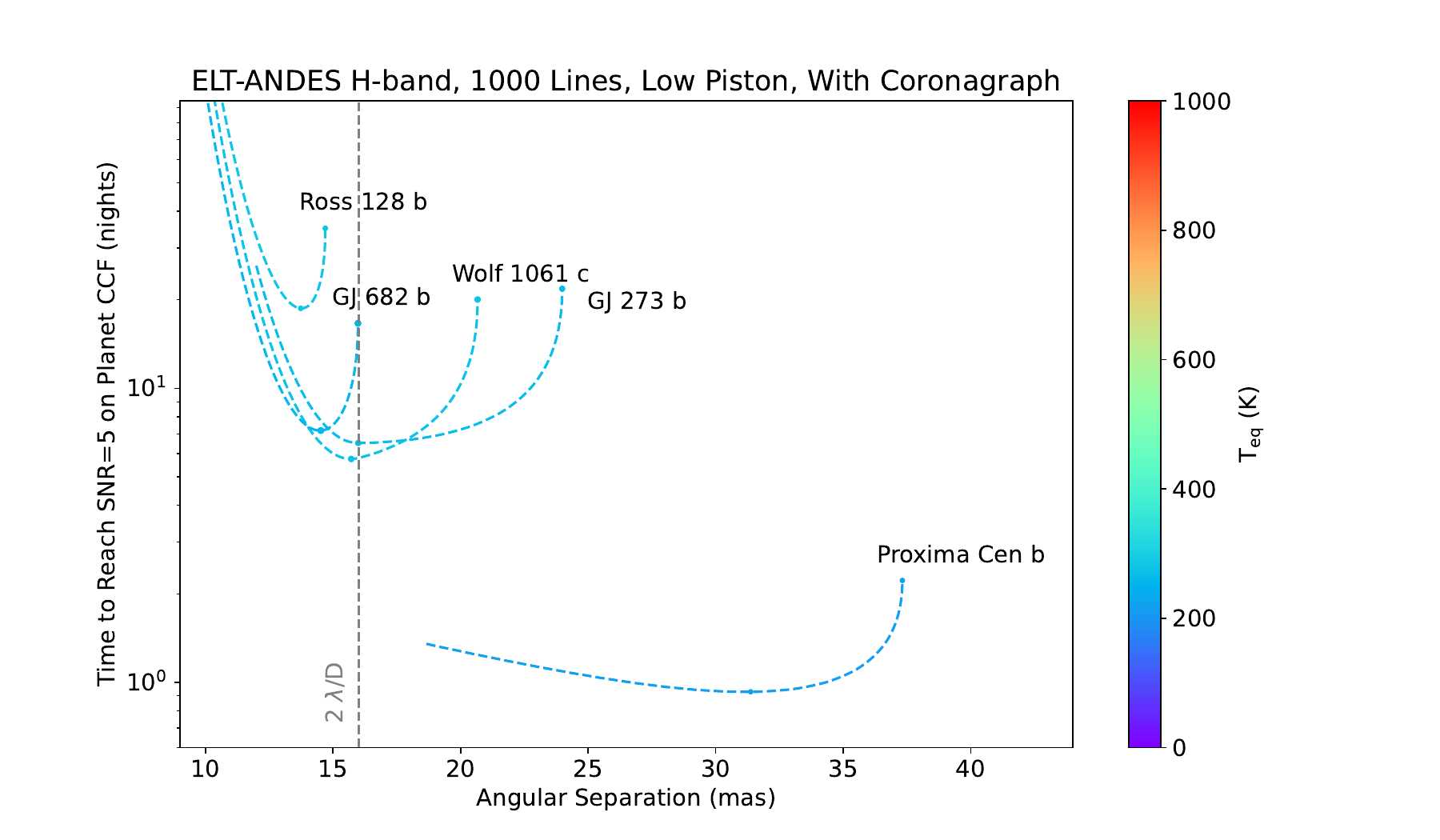}
  \caption{Observing time required to characterize the five temperate rocky planets in the ANDES golden sample as a function of the angular separation from their host star. The simulations are performed in Y band (top panel) and H band (bottom panel), assuming Strehl ratios of 0.3 and 0.6, and planetary albedos of 0.4 and 0.2, respectively. The contrast curve used assumes a low-piston scenario and a Lyot coronagraph, which smoothes out the achievable contrast vs separation compared to the non-coronagraphic case (shown in the appendix). The figure illustrates the clear gains in exposure time achieved by observing the planets closer to superior conjunction, as compared to maximum elongation.}
  \label{fig:lovis2}
\end{figure*}

\subsection{A Golden Sample for ANDES}

The results of our simulations are illustrated in Figures~\ref{fig:samplereflected} and ~\ref{fig:lovis2}, which shows the required exposure time for an SNR=5 detection as a function of angular separation. We can see that about 30 planets of all kinds can be characterized with ANDES in less than one night of observing time. Some "easy" giant planets, both hot and cold, can be probed in less than one hour. The sample includes a large diversity of objects in terms of mass and equilibrium temperature, from warm Jupiters and Neptunes to temperate rocky planets. Among these is Proxima b, which can be characterized in just one night of observations at maximum elongation and significantly less when closer to superior conjunction. One of the primary science goals of ANDES is to detect potential biomarkers in the atmospheres of temperate rocky worlds. From the results of the simulations we thus define a "golden sample" of 5 such planets which are the most favorable currently known in terms of SNR. These are: Proxima b, GJ 273 b, Wolf 1061 c, GJ 682 b, and Ross 128 b. Their main properties are listed in Table~\ref{tab:goldensample}.

\begin{table*}[h]
\caption[width=\hsize]{\label{tab:goldensample}  The Golden Sample planets for ANDES, i.e., the currently known five rocky planets orbiting in the habitable zone of their host stars most favorable for ANDES atmospheric characterization. Along with several planet properties, we tabulate the number of nights (1 night = 8 hours) necessary for detecting each planet's atmosphere.}
\centering
\begin{tabular}{@{}lcccccccccc@{}}
\hline \hline \vspace{-0.25cm} \\
Name & SpecTyp ($T_\mathrm{eff}$) & $d$ &  V & $P$ & $m\sin i$ & $R_\mathrm{p}$ & $T_\mathrm{eq}$  & sep & contrast  & nights\\
& [K] & [pc] & [mag] & [d] & [$m_\oplus$] & [$R_\oplus$] & [K] & [mas]  & [$10^{-8}$] \\
\hline \\
Proxima Cen b & M (2900 K) & 1.30 & 11.01 & 11.19 & 1.1 & 1.07 & 217 & 37.3 & 11.2 &  0.67 \vspace{0.15cm}\\
Ross 128 b & M (3163 K) & 3.37 &  11.12 & 9.87 &  1.4 & 1.15 &  283 & 14.7 &  12.5 &  13 \vspace{0.15cm}\\
GJ 273 b & M (3382 K) & 3.80 &  9.84 & 18.65 &  2.9 & 1.64 &  266 & 24.0 &  7.52 &   6.5 \vspace{0.15cm}\\
Wolf 1061 c & M (3309 K) & 4.31 &  10.10 & 17.87 &  3.4 & 1.81 &  275 & 20.7 &  9.57 &   5.8 \vspace{0.15cm}\\
GJ 682 b & M (3237 K) & 5.01 &  10.94 & 17.48 &  4.4 & 2.11 &  259 & 16.0 &  16.0 &  7.2 \vspace{0.15cm}\\
\hline
\end{tabular}
\end{table*}

In these simulations, we chose to observe in Y and H bands. According to the SCAO performance study, a Strehl ratio of 0.6 can be achieved in median seeing conditions in H band, while a value of 0.3 is achieved in Y band. The raw contrast curves include both a high and low-piston scenario. The low-piston scenario provides a contrast of typically ${\sim}1.5\cdot 10^{-3}$ at 25--45\,mas from the star. Below 25\,mas, the achievable contrast is limited by the first Airy rings of the stellar PSF. A coronagraphic solution would thus be highly beneficial to access the golden sample targets Ross 128\,b, Wolf 1061\,c and GJ 682\,b, located at 15--21\,mas from the star at maximum elongation. If this can be achieved, a basic characterization of all 5 golden sample targets could be done in about 33 nights of observing time. The population of potentially habitable planets in the immediate solar neighborhood could thus be probed for the first time by ANDES.

\subsection{Expected properties and Detectable species in or near the Habitable Zone of small planets: Proxima b}


To estimate the detectability of the main atmospheric constituents of the best target of our golden sample -- Proxima~b -- with ANDES, we adapted the methodology of \citet{Molliere:2019} to HCHR observations to compute cross-correlation functions (CCF) and then SNR. We accounted for the contamination by Earth's telluric lines, as well as the relative velocity of Proxima~b, Proxima Cen and the Earth. Calculations were performed in the Y, J and H bands, using the contrast curves in the low-piston scenario, in line with calculations presented in the previous subsections. We first assumed fixed (i.e., wavelength-independent) planet-to-star contrast ratios using typical values derived from global climate model (GCM) simulations of \citet{Turbet2016} at maximum elongation. We find that, assuming a planet-to-star contrast ratio of 10$^{-7}$ ($3 \times 10^{-8}$ and $2 \times 10^{-7}$, respectively), Proxima~b can be detected with ANDES at 5\,$\sigma$ in about 7\,h (60 and 2\,h, respectively).

\begin{figure*}[t!]
  \centering
  \includegraphics[width=1.0\linewidth]{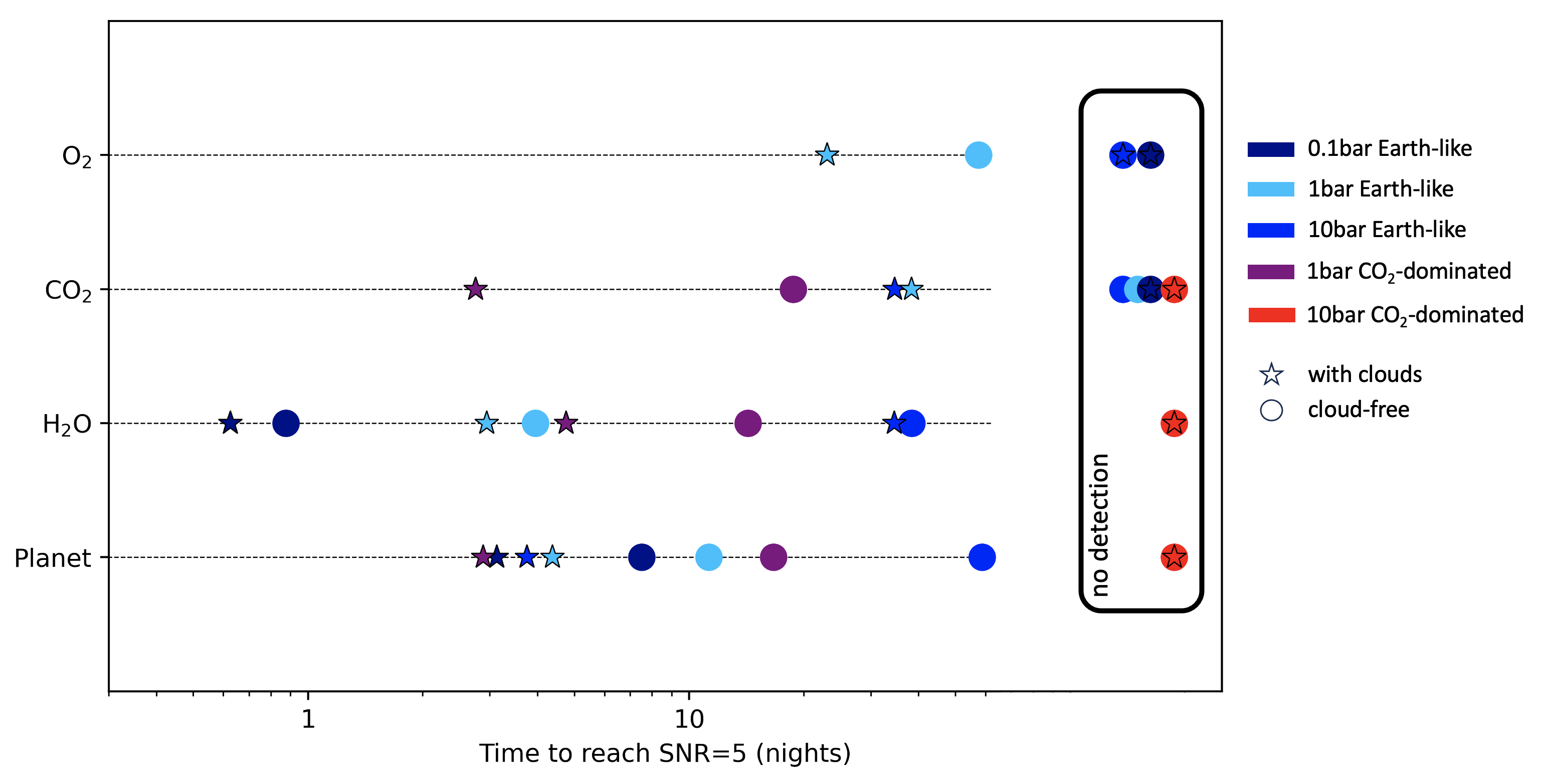}
  \caption{Detectability in reflected light (in Y, J, H bands) with ANDES of Proxima b and its atmospheres. The top three lines indicate the integration time required to reach a SNR=5 molecular detection (O$_2$, CO$_2$, H$_2$O), calculated with a CCF using Doppler-shifted molecular absorption lines. The last bottom line indicates the integration time required to reach a SNR=5, calculated with a CCF using Doppler-shifted stellar lines. Required integration times strongly vary from one atmospheric composition to the other, and are significantly lower when considering the effect of clouds. The symbols inside the right side legend indicate cases that are not detectable in less than 60 nights.}
  \label{fig:molec}
\end{figure*}

In a worst-case scenario, if Proxima~b (and the other planets of our 'golden sample') happen to have no atmosphere (which also implies no clouds) and an extremely dark surface, the planet-to-star contrast ratios can be relatively low. Typically, for Proxima~b assuming a dark surface (i.e., a Mercury-like albedo of 0.1), no atmosphere, then contrast ratios should be around 3~$\times$~10$^{-8}$ at maximum elongation (for a most probable inclination of 60$^{\circ}$). This very unfavorable scenario cannot be ruled out at the moment for this category of planets, about which we know so little today, but ANDES precisely has the capability to explore this. One might be tempted to give credence to this hypothesis, given the recent results of JWST MIRI on TRAPPIST-1b \citep{Greene:2023} stating the planet has a low albedo and possibly no atmosphere, but there are several things to bear in mind: (1) Additional JWST MIRI observations and models are underway on planet b and the other planets in the TRAPPIST-1 systems, which could invalidate this interpretation (e.g., thermal phase curve of the planet were obtained as part of JWST Cycle 2 observations) ; (2) if we had first observed Mercury – the innermost planet – in the solar system we would have seen a planet with no atmosphere and a very low albedo, whereas the other rocky planets have much higher albedos (around 0.3 for the Earth; 0.7 for Venus). \citet{Krissansen-Totton:2023} discussed this in the context of the TRAPPIST-1 planets ; (3) it is dangerous to draw too broad conclusions about rocky planets around M stars from a single system, and that is precisely where ANDES -- with its 'golden sample' -- would make it possible to probe in depth this population of planets.

Even in this worst-case scenario, the planet Proxima~b would be detected (and its albedo determined) within 7~nights of observing time. 

To simulate in a more advanced way the possibility of detecting the planet Proxima~b in the event it has an atmosphere, and to evaluate the capacity of ANDES to detect molecular species, we computed a grid of high-resolution planet-to-star contrast ratios based on the results of 3D line-by-line radiative transfer calculations of Proxima b. This grid was built using a grid of 3D Global Climate Model simulations described in \citet{Turbet2016}. Our assumptions about the planet in our simulations can strongly affect the expected contrast ratios and, thus, our ability to detect a signal in reflected light with ANDES. They show that the no atmosphere, low albedo scenario is pessimistic for several reasons listed below:

\begin{itemize}
   \item Even if the planet has a very low surface albedo, with a thick enough atmosphere, Rayleigh scattering by the atmosphere causes the geometric albedo and, thus, the planet-to-star contrast ratio to rise at visible wavelengths. Taking advantage of this fact however would require that the AO-assisted IFU observing mode of ANDES covered the I band, which is currently only a goal and presents substantial technical difficulties.  
    \item If water is present on the planet, it will form surface ice or clouds, likely both simultaneously, which will further boost the planet-to-star contrast ratio up to $1.0 \times 10^{-7}$ at visible wavelengths. Note that the ocean glint could increase this contrast ratio by up to typically 30$\%$ in the simulations we did (i.e.,\ up to $1.3 \times 10^{-7}$). Other types of clouds or hazes would also likely help increase the contrast ratio.
    \item Compared to an Earth-like 1\,bar atmosphere, increasing the surface atmospheric pressure can boost the planet-to-star contrast (due to Rayleigh scattering and more clouds forming in the model) despite less sea ice forming due to higher surface temperatures, decreasing the surface atmospheric pressure can also boost the signal due to more sea ice forming. We obtain a similar order of magnitude for the planet-to-star contrast ratio when assuming N$_2$-dominated atmosphere, O$_2$-dominated atmosphere, CO$_2$-dominated atmosphere, etc. In any case, all the habitable planets (i.e.,\ with global oceans) we have simulated have contrast ratios of the order of $1.0 \times 10^{-7}$ at maximum elongation (and possibly more, see below).
    \item So far, we have been using an inclination of 60$^{\circ}$ for the inclination of Proxima~b in our calculations. However, if the planet is more inclined and thus more massive and bigger (while still in the rocky planet regime), the reflected light signal could be boosted up to typically $2 \times 10^{-7}$. Note that the difference is minimal for inclinations $> 60^{\circ}$, but can be massive for inclinations $< 60^{\circ}$. This contrast boost also applies if the planet has no atmosphere and a very low albedo.
    \item Even more important, so far, for our calculations, we have been using the assumption that we are working at the maximum elongation of the planet (37~milliarcseconds for Proxima~b). The fact that ANDES can work at lower inner working angles can help us to strongly boost the planet's signal and reduce the stellar coupling at the planet location. For instance, targeting Proxima~b when the planet is at 30~mas (instead of 37~mas) could reduce the required integration time by a factor up to 5 (see Fig.~\ref{fig:lovis2}). This gain also applies if the planet has no atmosphere and a very low albedo. This aspect is a main motivation for ANDES to work as close as possible to the diffraction limit, enabling us to probe the planets of the “golden sample” at phase angles that are more favorable for detecting reflected light (relative to the maximum elongation geometry).
\end{itemize}

All the effects mentioned above can be combined, further amplifying the planet-to-star contrast ratio.

In addition, we evaluated the capability of ANDES to detect and characterize an atmosphere around Proxima~b (if any) using the methodology described above, but using a CCF with atmospheric molecular templates instead of stellar lines reflected by the planet. Using GCM simulations with a global ocean and an Earth-like atmosphere with the self-consistent formation of water vapor and clouds \citep{Turbet2016}, we evaluated it would take 20~hours of observation (300 and 180~hours, respectively) for ANDES to detect H$_2$O (CO$_2$ and O$_2$, respectively). Figure~\ref{fig:molec} synthesizes the number of nights required to detect various molecules, depending on the atmosphere considered and whether clouds are included or not. Simulations show a significant spread in the required number of nights to characterize Proxima~b, but molecular detections are feasible for a moderate amount of observing time in many scenarios. Surprisingly, for some atmospheric scenarios we note that water absorption lines (or, to put it more accurately, water troughs) are easier to detect than the stellar lines reflected on the planet. If Proxima~b happens to have oceans and a thin atmosphere, the planet could be detected at a SNR=5 through water lines in about 1~hour of observing time if targeted at the optimal angular separation of 30~mas. Tenuous atmospheres are easier to detect in Y, J and H bands for two reasons: (i) many more molecular lines are resolved (in thick atmospheres, molecular lines saturate) which facilitate molecular detection by CCF; and (ii) the overall absorption by the atmosphere is much weaker, which tends to increase the surface albedo, increasing the contrast ratio and the planet detection by CCF.

Last but not least, in addition to atmospheric molecular detections, very interesting science could also be pursued by:

\begin{itemize}
    \item Measuring the broad-band spectral variations of the planet-to-star contrast ratio. For instance, some cases have a sharp decrease of the contrast ratio above 1 micron, which is induced by the presence of sea ice (that has an albedo that strongly decreases in the infrared). Another interesting and critical aspect that we see in our simulations is that some molecular lines can disappear from the planet-to-star contrast ratio spectra when the atmospheric column of the gas and atmospheric pressure are so large that the saturated lines spill over into one another. In this case, detection of the molecule is still possible, but this requires measuring albedo variations in and out of the molecular bands (and no more a cross-correlation with individual molecular lines).
    \item Measuring the contrast ratio variation for several phase angles, which could be used to probe, for instance, the properties of clouds, surface, or both.
\end{itemize}

\begin{figure}
\centering
\includegraphics[width=0.99\linewidth]{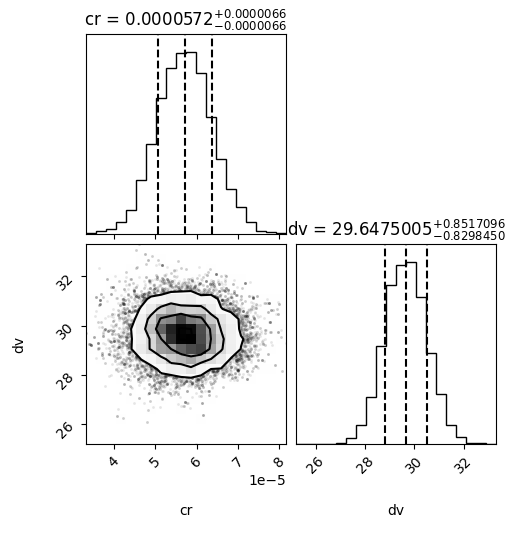}
\caption{Atmospheric detection of Proxima b in reflection light through a log-likelihood MCMC method assuming a clear water-rich Earth-like atmosphere and 20 hours of exposure time. The posterior distributions are those of the local planet contrast ($cr=5.7 \pm 0.7 \times10^{-5}$) and the orbital velocity ($dv=29.6 \pm 0.8$) both detected with a  SNR 8.6 and 35, respectively.  \label{fig:beaudoin}}
\end{figure}

The above results are confirmed through more detailed end-to-end simulations  (Beaudoin et al. in prep) of ANDES IFU assuming low-piston AO performance. All (10 mas) spaxel signals, dominated by the host star’s spectrum modulated by telluric absorption over the $YJH$ spectral range, are used to construct a master reference spectrum used to extract the planetary signal (local planet contrast and orbital velocity) though a log-likelihood MCMC method. Figure~\ref{fig:beaudoin} shows the posterior distributions for a 20-hr simulation of Proxima b at maximal elongation (37 mas) assuming an Earth-like (water-rich) clear atmosphere, a planet/star contrast of $1.3\times10^{-7}$ and an orbital velocity of with $\sim$30 km/s. At this separation, the local planet contrast ratio ($cr$) is  $5.7\times10^{-5}$. As shown in Figure~\ref{fig:beaudoin} $cr$ is recovered with a SNR of 8.6 equivalent to a 5-$\sigma$ detection in 6 hours i.e less than one night.  The same simulation performed with the O$_2$ component only of the atmosphere would require $320$ hours  ($\sim$45 nights) for a 5-$\sigma$ detection.


\section{Atmospheric Characterization from inflated Jupiters to super-Earths}

\label{sec:gasgiants}
\subsection{Atmospheric composition and dynamics of gas giants}

Due to their diversity, exoplanets constitute an ideal laboratory to study atmospheric dynamics in regimes not accessible in the Solar System. Here, we discuss the prospects of tracing atmospheric dynamics with ANDES. While not included in the baseline design, extending ANDES to the K-band is a goal that is particularly powerful for studying dynamics. K-band is discussed in more detail in Section~\ref{sec:Kband}, but some of the calculations presented below already use it.

Models predict that various circulation patterns should arise as we move in the irradiation - internal heating diagram (Figure~\ref{figure}). Atmospheric dynamics theories have been successfully captured some of the trends observed across this parameter space \citep{Showman2020}. However, many open questions remain, several of which can be tackled thanks to ELT ANDES, for example:

\begin{center}
\begin{figure*}
  \resizebox{\hsize}{!}
    {
    \includegraphics[width=\textwidth]{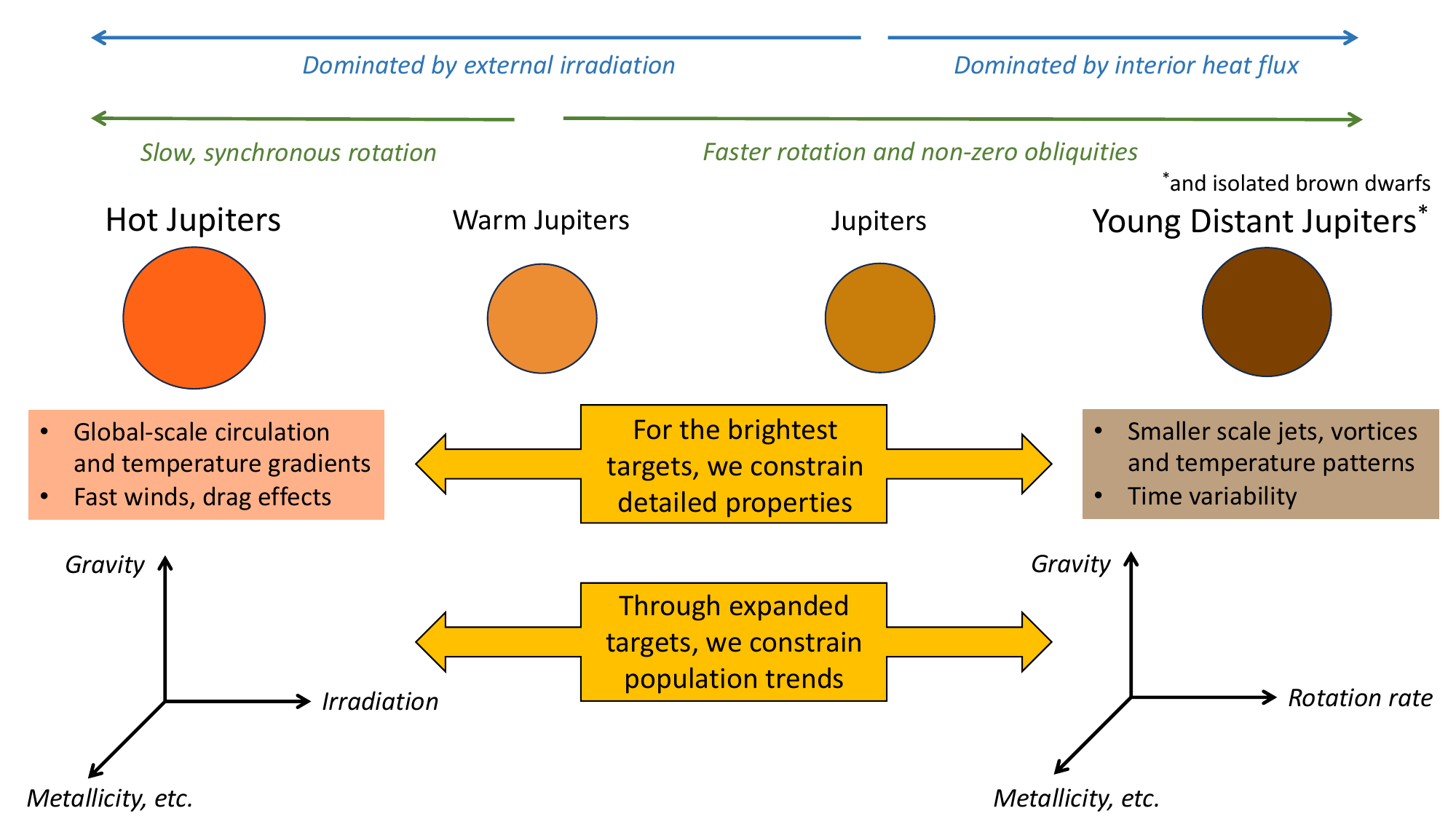}
    }
    \caption{The study of exoplanets enables us to expand our understanding of atmospheric dynamics across a much wider range of physical conditions than realized in the Solar System. For the atmospheric dynamics of gaseous planets, Jupiter sits at a balancing point where the irradiation from the Sun is comparable to the heat flux emerging from its interior. In contrast, Hot Jupiters and Young Distant Jupiters (as well as isolated brown dwarfs) occupy extremes of being dominated by external or internal heating, respectively. Thanks to the high sensitivity of ANDES to line shapes, ultimately shaped by atmospheric dynamics, we may hope to constrain complex physical details of gas giants across ages and irradiation patterns, revealing their atmospheric circulation: rotation rates, wind speeds, and spatial patterns, and the temperature structures whose sizes and gradients are shaped directly by the atmospheric dynamics. In addition to detailed characterization of the brightest systems, with ANDES, we will be able to study trends across a larger number of planets, placing valuable empirical constraints on population-level trends. \label{figure}}
\end{figure*}
\end{center}

\begin{enumerate}
    \item \textit{(Very-)Hot gas giants:} Q1: What is the altitude- and longitude-resolved geometry of winds, and what drives it across the parameter space (surface gravity, rotation rate, stellar flux, metallicity)? How do composition, clouds/hazes, and thermal structure influence the wind structure, and how do winds impact chemical and thermal gradients? \citep[e.g.,][]{Lee2016, Drummond2020} Q2: What is the role of atmospheric drag, and which is/are its physical origin/s across the parameter space? \citep{Li2010,Perna2010,Koll2018}
    Q3: What is the magnitude and origin of any variability in atmospheric winds strength and geometry? \citep{Rogers2017,Komacek2020,Pai2022}
    \item \textit{Warm gas giants (up to 1000\,K):} Q1: What is the rotational axis (period and obliquity) of warm gas giants, and to what extent does it impact atmospheric dynamics? \citep{Ohno2019,Rauscher2023} Q2: What is the magnitude of stratospheric winds in less irradiated planets \citep{Benmahi2021, Cavalie2021}?
    \item \textit{Young distant Jupiters and isolated brown dwarfs:} Q1: What is the underlying mechanism driving atmospheric circulation on these objects, forcing from the convective region below or a form of instability between clouds and the underlying atmospheric structure? \citep{Freytag2010,Showman2019} Q2: Can we link observed properties such as chemical disequilibrium, evidence of patchy clouds, and variability caused by changing atmospheric features to a global picture of atmospheric circulation for these objects? \citep[e.g.,][]{Karalidi2016,Miles2020,Vos2020}
\end{enumerate}

For young Jupiters and brown dwarfs, vertical circulation has been advocated to explain the presence of specific spectral features (e.g.,\ dust, disequilibrium species). At the same time, turbulence likely plays a role in determining their atmospheres' patchiness, which likely causes the observed rotational modulation. In brown dwarfs, this has been further reinforced through Doppler imaging enabled by K-band measurements \citep{Crossfield2014}, and wind speeds have been inferred by comparing the radio and infra-red variability \citep{Allers2020}. These techniques are currently only applicable to a few objects with masses outside the planetary regime. While mapping may be possible to a degree in L-band on the ELTs \citep{Plummer2023}, ANDES will allow us to Doppler-map many tens of variable brown dwarfs and the brightest directly imaged planets if the K-band (clearly superior for mapping when compared to L-band, and a goal for ANDES) is added to the instrument, see Section~\ref{sec:Kband}.

For transiting exoplanets, information on dynamics is encoded in spatial variations in longitude and latitude across the atmosphere. Space-borne phase-curves \citep{Harrington2006, Knutson2007, Cowan2008} and eclipse mapping \citep{Williams2006, Rauscher2007} - especially in combination with the large aperture and spectroscopic capability of JWST - have led the way into observational constraints of atmospheric dynamics in transiting exoplanets \citep{Parmentier2018}. The challenge is that dynamical information has to be inferred indirectly, for example through shifts in the hot-spot location, or to account for the energy balance between day and night-side hemispheres.

High spectral resolution observations offer a complementary avenue to probe dynamics of transiting planets \citep{MillerRicci2012, Showman2013}, through the measurement of precise line-shapes and longitudinally-resolved atmospheric properties and Doppler-shifts \citep{Flowers2019, Seidel2020, Ehrenreich2020, Beltz2021, Pino2022, Brogi2023}. In addition, they can be used to constrain the rotation rate of young Jupiters and brown dwarfs \citep{Snellen2014,Bryan2020}.


Here, we assess the capabilities of ANDES as an atmospheric dynamics survey machine but also as an instrument to unlock unprecedented details about the atmospheric dynamics of the best individual objects. To this end, we designed two sets of simulations that, despite not representing all possible observing strategies, showcase the potential of ANDES.

We first assess the theoretical precision that ANDES can reach on the systemic velocity ($v_\mathrm{sys}$) and the orbital velocity ($K_\mathrm{p}$) of a transiting planet based on its spectrum. To this end, we consider a Jupiter-size planet around a K1 star (analogous to the HD 189733b system), including $\mathrm{CO}$ and $\mathrm{H_2O}$ at roughly solar abundances and in chemical equilibrium. This study's simulated wavelength range assumed a wavelength range from Y-, J-, H-, all the way to K-band, the last being required to leverage the strong CO lines best (see Section~\ref{sec:Kband}). We simulate a time-series observation of the primary transit of this system using a custom version of the ANDES ETC v1.1 (Sanna et al., 2023\footnote{https://drive.google.com/file/d/18dM5OzTxdvshHY86XPMlfDsbm0hN6juZ/view}). We include telluric lines at fixed precipitable water vapor, scaling them with airmass, and process observations through a custom pipeline based on Principal Component Analysis (PCA). 

For targets in the $J=12-13$ magnitude range (bright end), detections are so strong that theoretically one could constrain $v_\mathrm{sys}$ within 10\,$\mathrm{m~s^{-1}}$ (i.e.\ below the instrumental required stability) and $K_\mathrm{p}$ within a few 100\,$\mathrm{m~s^{-1}}$ in one transit observation. Current peak VLT/CRIRES-like performances (sub-$\mathrm{km~s^{-1}}$ in $v_\mathrm{sys}$ and a few $\mathrm{km~s^{-1}}$ in $K_\mathrm{p}$), typically obtained at $J\sim9$ mag, are still attainable down to $J=15$ mag.

We then assess to what extent ANDES can identify shifts among different species simultaneously present in the atmosphere. We simulate an atmosphere containing CO (again requiring K-band, an ANDES goal) and $\mathrm{H_2O}$ with a range of relative velocities. We perform a likelihood-test ratio to determine whether the simulated data justify the additional complexity of the model requiring relative shift between the species. We find that the higher complexity model is confidently preferred for a relative shift of 1.0\,km\,$\mathrm{s}^{-1}$, at 3.6$\sigma$ and 7.5$\sigma$ at $J=13.5$ and 13.0, respectively. At the faint $J=15.5$ end, ELT ANDES will be sensitive to $\sim5~\mathrm{km~s^{-1}}$ shifts, comparable to current peak performances of CRIRES+ at $J = 9$. We note that the actual sensitivity might depend on the species present in the atmosphere (in this case, CO and H$_2$O).

\begin{figure}[t!]
  \centering
  \includegraphics[width=1.0\linewidth]{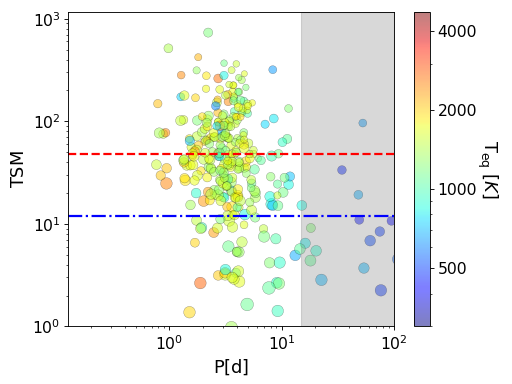}
  \caption{Detectability of atmospheric dynamics with ANDES for the known sample of transiting gas giants. The detailed atmospheric dynamics characterization sample is located above the horizontal red dashed line, which corresponds to the TSM of HD189733b scaled to a J magnitude of 12. For all planets above the blue dash-dotted line, a first determination of atmospheric dynamics may still be possible. Planets are colour-coded by equilibrium temperature, and symbol-size is proportional to surface gravity. The gray shaded area represents the region at periods larger than 15 days, for which PCA- or SYSREM-based methods might erode a significant fraction of the planet signal (see Sec. \ref{sec:challenges_long_period}).}
  \label{fig:atmo_dynamics_survey}
\end{figure}

These simulations are somewhat idealized. However, it is clear that the precision achieved is exquisite, and it might allow us to perform a first census of atmospheric dynamics for a statistically significant sample of gas giants. We evaluated the potential size of this sample as follows. First, we selected from the NASA Exoplanet Archive\footnote{Updated to the 13th of October 2023} all planets known at declination smaller than $<10^\circ$ that have mass larger than 0.4 Jupiter masses, and that transit their host star. Finally, for each target we compute the TSM and compared it to the TSM of HD189733b scaled to a J magnitude of 12, to determine the sample of planets for which detailed atmospheric dynamics characterization is possible ($\mathrm{km~s^{-1}}$-level differences among velocities of different species detected, precisions of down to $\sim 10-100~\mathrm{m~s^{-1}}$ on $v_\mathrm{sys}$ and $K_\mathrm{p}$), and to a J magnitude of 15, to determine the sample for which a first assessment of atmospheric dynamics might be possible ($\sim 5\mathrm{km~s^{-1}}$-level differences among velocities of different species detected, sub-$\mathrm{km~s^{-1}}$ precision on $v_\mathrm{sys}$ and $K_\mathrm{p}$). Out of about 329 known planets selected according to our criteria, about 140 have TSMs that qualify them for the detailed atmospheric dynamics characterization sample (130 with orbital period shorter than 7 days), and 260 have TSMs that might in principle allow a first assessment of atmospheric dynamics (240 with orbital period shorter than 7 days). Ultimately, this sample size is comparable to the ARIEL tier 1 sample size \citep{Tinetti2018}, opening exciting opportunities to synergistically measure atmospheric properties from space and wind speeds and geometries from the ground. We illustrate the sample in Fig. \ref{fig:atmo_dynamics_survey}. We note that the numbers quoted here ($\sim~$100 planets for detailed characterization) are comparable to those presented for the K-band (many tens of planets) in Section~\ref{sec:kband_close_in_dynamics}, and that the study presented here also included the K-band. The numbers between the two sections are not directly comparable because the characterizability was assessed in different ways (shifts between different species vs. line shape sensitivity to dynamics). That being said, both point to the K-band likely playing an important goal for the characterization of atmospheric dynamics. A more detailed analysis of how the studies described above depend on the inclusion of K-band would be beneficial.

\begin{figure*}[t!]
  \centering
  \includegraphics[width=0.51\linewidth]{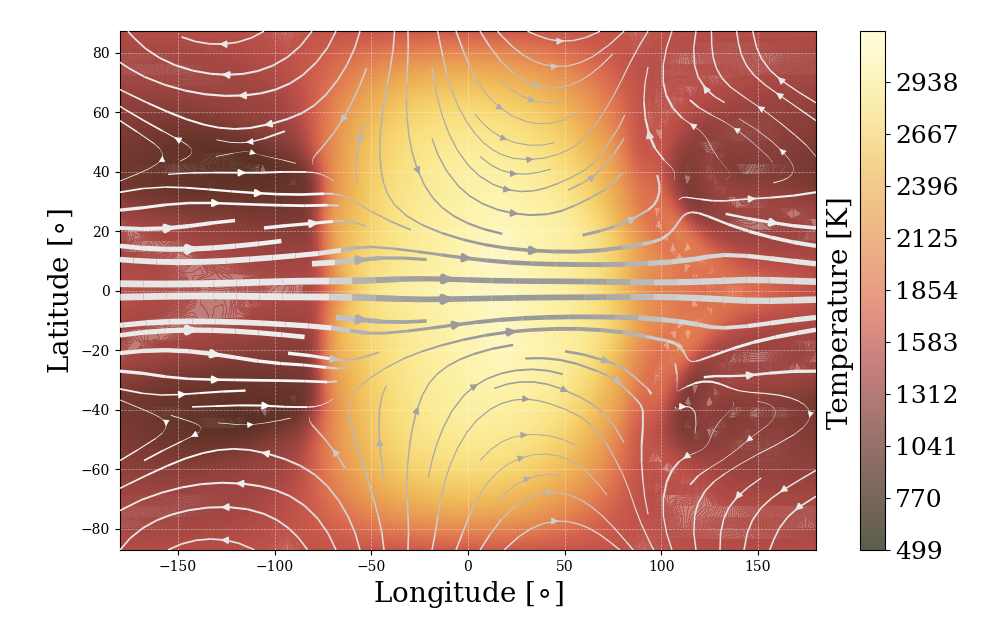}
  \includegraphics[width=0.39\linewidth]{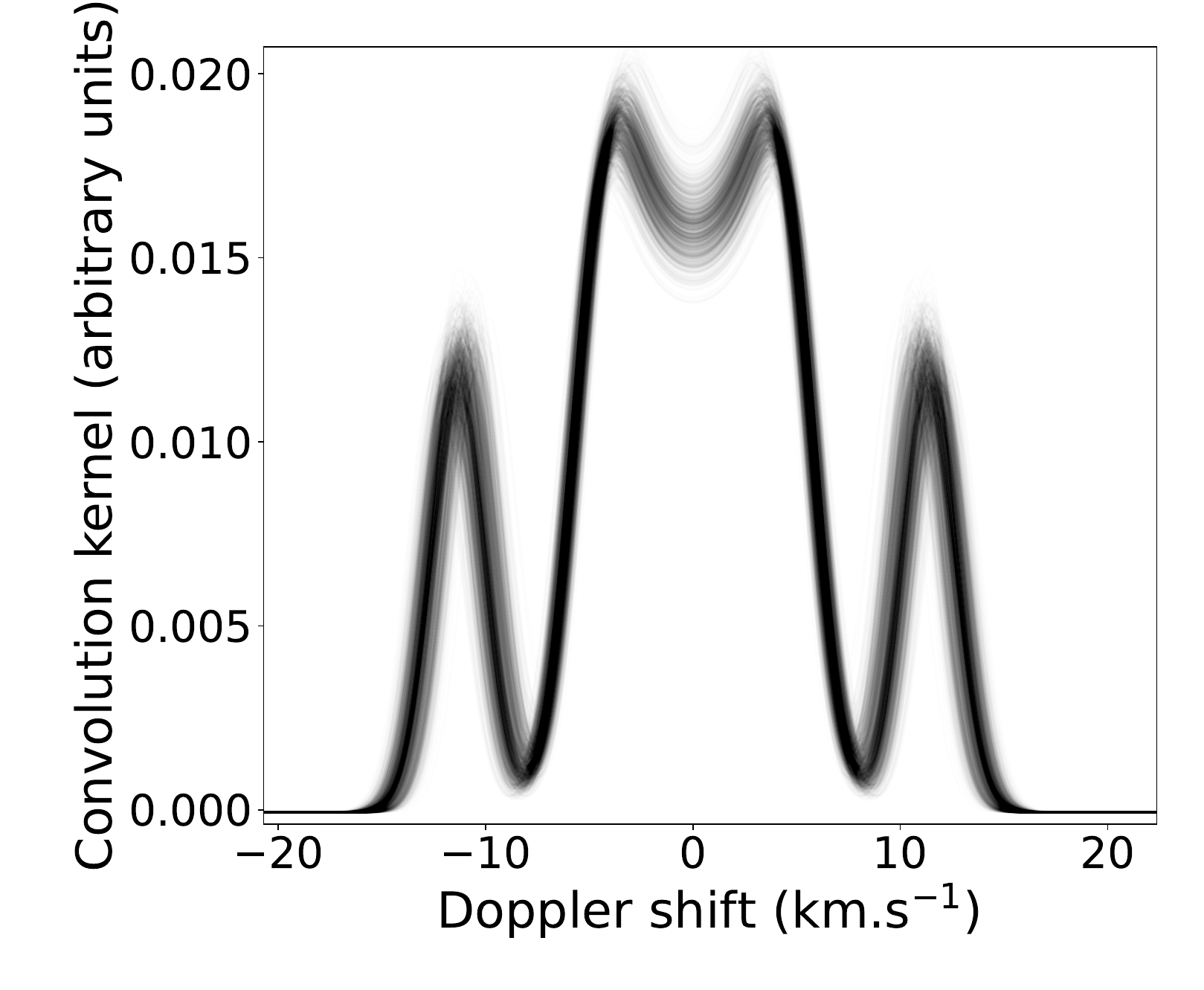}
  \caption{Left: Temperature (colors) and winds (arrows) at the 0.01 bar level from a 3D model of WASP-76 b by \cite{Beltz2022a} used as an input model for our synthetic ANDES spectra. The maximum wind speed is 9.1 km.s$^{-1}$.  Right: Recovered convolution kernel as a function of Doppler shift from the posterior distribution of our WASP-76 b synthetic ANDES data analysis through nested sampling.}
  \label{fig:W76_Beltz}
\end{figure*}

Focusing on what can be accomplished on the most favorable targets using ANDES' baseline design, we have simulated more refined ANDES J-band observations (1.15--1.35\,$\mu \mathrm{m}$) of WASP-76b (as an example of a very bright target), based on the three-dimensional water absorption by \citet{Beltz2022a,Beltz2023} including Doppler shifts from atmospheric dynamics and rotation, as shown in Fig.\ref{fig:W76_Beltz}. We scaled the noise from an actual WASP-76b SPIRou transit observed in 2020, accounting for a factor of 100 due to the larger collecting surface of ELT compared to CFHT, and included SPIRou-like blaze functions. As a control experiment, we simulated a similar observation scaled to the size of CFHT. Synthetic telluric absorption was calculated using the ESO sky model calculator, accounting for the airmass at each exposure. The synthetic observation sequences were then analysed following a custom PCA-based method \citep{Brogi2017, Klein2023}.



Since the model that we used to synthesize these observations is inherently three-dimensional, the line shapes should significantly deviate from Voigt profiles. To test this in a statistical framework, we have employed a nested sampling algorithm to fit for the temperature, the water content, the rotation speed of the planet, and in addition, the speed and latitudinal width of equatorial superrotation following the formalism of \cite{Klein2023}. With SPIRou synthetic and real data, the code could not resolve the rotation-wind degeneracy, and although temperature and water content were retrieved, the dynamical information was impossible to obtain. ANDES synthetic data, on the other hand, constrain both the rotation rate and super-rotating winds with a precision of 300\,$\mathrm{m~s^{-1}}$. Moreover, the latitudinal width of superrotation (25 degrees) was recovered with a 1.7-degree error bar. The recovered line shape from our posterior distribution is shown in Figure\ref{fig:W76_Beltz}.

This test points to ANDES's ability to resolve line shapes in the best transiting planets to a precision sufficient to discriminate between different atmospheric dynamics patterns, even with the very simplified pseudo-1D analysis framework we used. An extension beyond the H-band in wavelength coverage should result in better precision and allow the study of additional molecules.




\subsection{Unlocking the atmospheres of young giant planets}

Since 2014, the use of SPHERE and GPI enabled to acquire for the first time low resolution (R$_\lambda \sim$ 50) emission spectra of tens of young giant planets exhibiting broad features due to unresolved molecular (H$_2$O, CH$_4$, VO, FeH, CO) absorptions  \citep{2018SPIE10703E..05C}.
These spectra can be compared to predictions of atmospheric models to infer first-order information on the bulk properties of the atmosphere, mainly the effective temperature, the surface gravity (pressure), and the properties of clouds. The direct imaging community's current atmospheric models are mainly 1D models involving at least thermodynamics, radiative and convective energy transport, and gas-phase chemistry \citep[see for review][]{2019ARA&A..57..617M}. They solve the pressure-temperature structure in radiative-convective equilibrium and determine the chemical species that are supposed to form. Sophisticated models of clouds of different compositions (silicates, sulfites) have been considered for over a decade. 

The analysis of these low-resolution spectra of young giant exoplanets showed that as for young brown dwarfs, young exoplanets of a given spectral type with lower surface gravity can be up to 200--500\,K cooler than their older counterparts. This discontinuity is particularly noticeable at the M-L transition and evidenced by a lack of methane at the L/T transition, as seen for HR\,8799\,b \citep{2018A&A...616A.146P}. These results confirm the early photometric characterization that showed the peculiar properties of young L/T type planets like 2M1207\,b \citep{2004A&A...425L..29C}, HR\,8799\,bcde \citep{2010Natur.468.1080M}, HD\,95086\,b \citep{2013ApJ...772L..15R}, or HD\,206893\,b \citep{Hincley2023} owing to the low-surface gravity conditions in their atmospheres, leading to enhanced production of clouds probably composed of sub-$\mu$m dust grains made of iron and silicate. With cooler temperatures than typically 1000\,K, the clouds in their atmospheres start to fragment and sink below the photosphere, as recent variability studies show high-amplitude photometric variations for several of these young, planetary-mass companions \citep[e.g.][]{2019AJ....157..128Z,2022ApJ...924...68V}. The presence of circumplanetary disks, as directly evidenced for PDS 70\,c by ALMA sub-mm observations \citep{2021ApJ...916L...2B}, might also affect the spectral energy distribution of the youngest planets mixing contribution in infrared from the photosphere and the (spatially unresolved) circumplanetary disk \citep{2023NatAs...7.1208W}. 

With the increase in spectral resolution ($R_\lambda > 1000$), the line blending of molecular absorptions starts to be resolved, yielding unprecedented accuracy constraints on surface gravity (pressure), chemical abundances, and the complex molecular chemistry of exoplanetary atmospheres. Atmospheric model degeneracy can be better suppressed and missing opacities identified. Note that the combination of spectral diversity with higher spectra resolution and high-contrast imaging is particularly interesting in boosting the detection performance of young exoplanets using the molecular mapping technique \citep{2018A&A...617A.144H}. As for characterization, high-contrast spectroscopic observations at medium resolution with AO-fed spectrographs, e.g. by OSIRIS at Keck and SINFONI at VLT, led to the determination of molecular abundances such as water, carbon-monoxide or methane for several young planets like HR\,8799 b and e \citep{2015ApJ...804...61B}, $\beta$~Pic~b \citep{2018A&A...617A.144H}, HIP\,65426\,b \citep{2021A&A...648A..59P}. More recently, even the 12CO/13CO isotopologue ratio, connected to the carbon-ice fractionation process, was measured for the first time in the atmosphere of the young exoplanet TYC-998-760-1~b suggesting a formation beyond the CO snowline \citep{2021A&A...656A..76Z}. We note that all studies presenting findings based on the presence of CO (and its isotopologues) required observations in the K-band, which is a goal for ANDES and discussed in Section~\ref{sec:Kband}. Accessing the planet's atmospheric composition (C/O, D/H, N/O, N/C, and isotopologue ratios) and metallicity is a path forward in the exploration of the formation mechanisms, the birth location, and the potential migration/dynamical history of the exoplanets \citep[e.g.][]{2011ApJ...743L..16O,2016MNRAS.461.3274C,2021A&A...654A..71S}. However, the observed atmospheric composition results from various physical complex processes . They are connected to the disk composition, physical and thermal structure, chemistry, and evolution that will define the building blocks composition of the giant planet, the planetary formation processes that will affect the solid-gas accretion phase, the vertical mixing/diffusion between the planet’s bulk and atmosphere, or simply the dynamical evolution over time, which overall will make any direct interpretation challenging at this stage, but essential to ultimately connect the planet’s physical properties and atmospheres with their origins \citep{mollieremolyarova2022}.

With the advent of near-infrared spectrographs offering higher spectral resolution ($R_\lambda \geq 30\,000)$ and combined with high-contrast, as this is the case for HiRISE/CRIRES+ at VLT or Kpic at Keck \citep{2017SPIE10400E..29M,2022SPIE12185E..0SV}, the radial and rotational velocities of the young giant planets can be directly determined to explore the global 3D orbital properties, but also measure the rotational period and the planet's obliquities in connection with their history of formation \citep{2014Natur.509...63S,2021AJ....162..148W,2023A&A...670A..90P}. Moons similar to those around Jupiter are also expected
to form in circumplanetary disks as a by-product of planet formation \citep{2020ApJ...894..143B} and the potential detection of these elusive objects from radial
velocity monitoring of self-luminous directly imaged planets will be feasible with ANDES at ELT \citep{2023AJ....165..113R}. The novelty of ANDES in this field is clearly the gain in angular resolution, contrast combined with high spectral resolution compared with current or near-term facilities.

In summary, given the current estimations of performance and limitation, the niche for ANDES in SCAO-IFU mode is the characterization of young giant planets detected in the planet-forming region zone (1--10\,au) not characterized by 10-m class Telescopes. Access to a spectral resolution of 100\,000 in near-infrared will give access to the molecular abundances, the isotopologues, the radial and rotational velocities, and the exploration of the global atmosphere circulation of the planet (winds, jets). For very bright exoplanet and brown dwarf companions, the possibility of acquiring Doppler imaging of the exoplanet/brown dwarf companion can be envisioned as detailed Section~\ref{sec:Kband}. The targets will be the dozens of known/new young exoplanets and brown dwarfs that have been/will be imaged or detected before the first light of the ELT from the ground or in space (Gaia, JWST), and also potentially discovered by MICADO, HARMONI, and METIS at ELT 
Therefore, ANDES will be a game changer in addressing the following science cases for young Saturns and Jupiters: i/ their direct detection and characterization down to the snow line with potential hints from Gaia or radial velocity studies, ii/ the study of the mass-luminosity \& initial entropy relation in relation with their formation and evolution processes, iii/ the exploration of the chemical and cloud composition mapping and variability in connection with the global 3D atmospheric circulation, iv/ the physics of accretion for young protoplanets in connection with their circumplanetary disks, v/ the 3D orbital and dynamical characterization including the measurement of obliquities of young giant planets, and ultimately 
vi/ the potential first discoveries of young exomoons.





\subsection{Understanding the nature of sub-Neptunes and super-Earths}
\label{sec:waterworlds}

The broad diversity of exoplanets so far discovered is starting to reveal demographic trends that provide the first clues to understand the underlying planet formation and evolution processes. Focusing on small planets  ($R<4\,R_\oplus$), Kepler revealed that about 25\% of Sun-like stars in the Milky Way host planets with no counterpart in the solar system: super-Earths ($R=1$--$2\,R_\oplus$) and sub-Neptunes ($R=2$--$4\,R_\oplus$), with orbital periods shorter than 100 days \citep{Batalha2013, Marcy2014}.  This small planet population presents a bi-modal radius distribution \citep{Fulton17, Berger2020} known as the "radius valley".

Canonically, the dearth of planets between 1.4 and 2.0$\,R_\oplus$ has been explained as a transition region between rocky planets that held on to a primordial H2/He envelope (sub-Neptunes) and stripped cores (super-Earths).
Atmospheric loss models reproduce the correct position of the valley \citep{Owen2017, Ginzburg2018}.
However, water-rich planets are expected to populate the regions close to the star due to type I migration and do not fit into this simple paradigm  \citep{Venturini2020}. This indicates that  our knowledge of the true nature of those planets is still limited.

Using a refined sample of small planets, \citet{Luque2022} have recently proposed that for M dwarf hosts, the radius valley is a consequence of interior composition rather than an indicator of atmospheric mass loss. These planets seem to be distributed into three main populations: 1) those that follow a bulk density comparable to Earth, 2) those with cores consisting of rock and water-dominated ices in 1:1 proportion by mass, and 3) planets with a significant envelope made of H/He. Formation models including type I migration support the change in paradigm that these observations propose: the valley separates dry from water worlds rather than rocky planets with or without H/He envelopes. Proving that these results hold true for FGK stars would transform our interpretation of the Kepler-planet demographics. To this end we need more precise masses for a handful of high-value targets. However mass and radius alone will still not be enough \citep{Rogers2023}, and ultimately the characterization of the atmospheres will be needed to break the degeneracies in the internal composition modeling.

The range of possible atmospheric composition of rocky earths, super-earths and sub-Neptunes is not well known and subject of strong debate \citep{Liggins2022, Wordsworth2022}. This is because the composition of the atmosphere is determined by the insolation history of the system, and also by the interplay between the atmosphere and the surface or magma beneath, factoring in the solubility of volatile substances \citep{Tian2023}. While research has largely centered around the interaction between molten metals and silicate with hydrogen-rich atmospheres \citep{Kite2020, Schlichting2022}, little is known about how magma oceans in planets with large water mass fractions interact with their atmospheres. Planets that experience intense radiation could maintain high surface temperatures even if they have water-rich atmospheres \citep{Dorn2021, Vazan2022}. These magma oceans could act as large reservoirs to keep volatiles from being stripped away during the early, active life of the host stars, but this has not been tested observationally. Mini-Neptunes atmospheres are somewhat easier to characterize due the relatively large extent of their atmospheres induced by the remaining primordial gas (H/He). For example water absorption has been detected in K2-18b \citep{Tsiaras2019, Benneke2019} and HD 106315c \citep{Kreidberg2022} atmospheres.

The best way to unlock the atmospheres of these inaccessible small planets is via infrared phase curve studies. Phase curves have been measured for several dozen hot Jupiters \citep{Knutson2007, Knutson2009, Stevenson2014, May2022}, but seldom been attempted for smaller planets, with some exceptions: 55~Cnc~e,  LHS~3844~b, and K2-141b. Observations of 55~Cancri~e ($1.9\,R_\oplus$) revealed a phase curve whose peak brightness is offset from the substellar point — possibly indicative of atmospheric circulation \citep{Demory2016}, however its observed variability is still poorly understood \citep{Mercier2022}. K2-141b is a molten lava world on an extremely short orbit showing hints of a vaporized atmosphere \citep{Zieba2022}. Although no results have been published for small rocky planets yet, the capabilities of JWST are pivotal for small planets phase curves studies. Still the potential for JWST to unlock the composition of such atmospheres may be quite limited \citep{Lustig2023, Moran2023}. ANDES however, will be able to probe these planets in search for metallic atmospheres and the detection of water and other atmospheric species that can reveal the existence or not of water worlds, and the general nature of sub-Neptune to super-Earth planets. A large well-characterized sample (see Figure \ref{sec:challenges_long_period} for ANDES potential targets) is needed to constrain theories for how these planets form and retain atmospheres, which is in turn crucial for understanding the habitability of their more temperate siblings.

\section{The study of protoplanetary disks}
\label{sec:PPDs}
The majority of the known planets are believed to have formed within $\sim$15-20 au from their hosting star, followed by radial migration inwards due to planet-disk interactions during the disk gas-rich phase (e.g., \ \cite{Pollack1996, Morbidelli2016}). Consequently, the knowledge of the composition and spatial distribution of atomic and molecular gas in the inner $<20$\,au of young ($<10$\,Myr) circumstellar disks is an essential step towards understanding planetary system's formation and evolution. Gas in this inner disk region is dissipated inside-out, as material is accreted onto the central star and removed through winds. At present, we are not able to spatially resolve the relevant regions ($<$ 100 mas at 150 pc). Therefore, gas excitation and spatial distribution information is obtained only through high-resolution spectroscopy of species probing different conditions.

The main scientific objective of ANDES in the field of protoplanetary disks (PPDs) will be to settle the properties of the gas in the inner star-disk region, where different competing mechanisms of disk gas dispersal are at play, namely magnetospheric accretion, jets, photo-evaporated and magnetically driven disk winds. High spectral ($R\sim 100\,000$) and spatial ($\sim 10$\,mas) resolution are needed to disentangle and study the different phenomena in detail. The derivation of physical properties and composition of the gas in this region will constrain the mechanisms through which the forming star acquires mass and the angular momentum is removed from the system, and also on the initial condition for planet formation. 

The main targets will be circumstellar disks of young stellar objects at different ages to trace the disk evolution, including from the one side protostellar objects detectable only in the near-IR (age $\sim 10^5$\,yr) and on the other side more evolved disks (age up to $\sim 10^7$\,yr), i.e.\ disks where the inner region has been already significantly cleaned from dust. Stars with different masses and metallicities will be observed. Here, we describe in more detail the specific scientific goals in the field of PPDs to be reached with the broad spectral range of ANDES.

\subsection{Gas properties in the inner disk regions}

\begin{center}
\begin{figure*}
\includegraphics[angle=0, width=8.5cm]{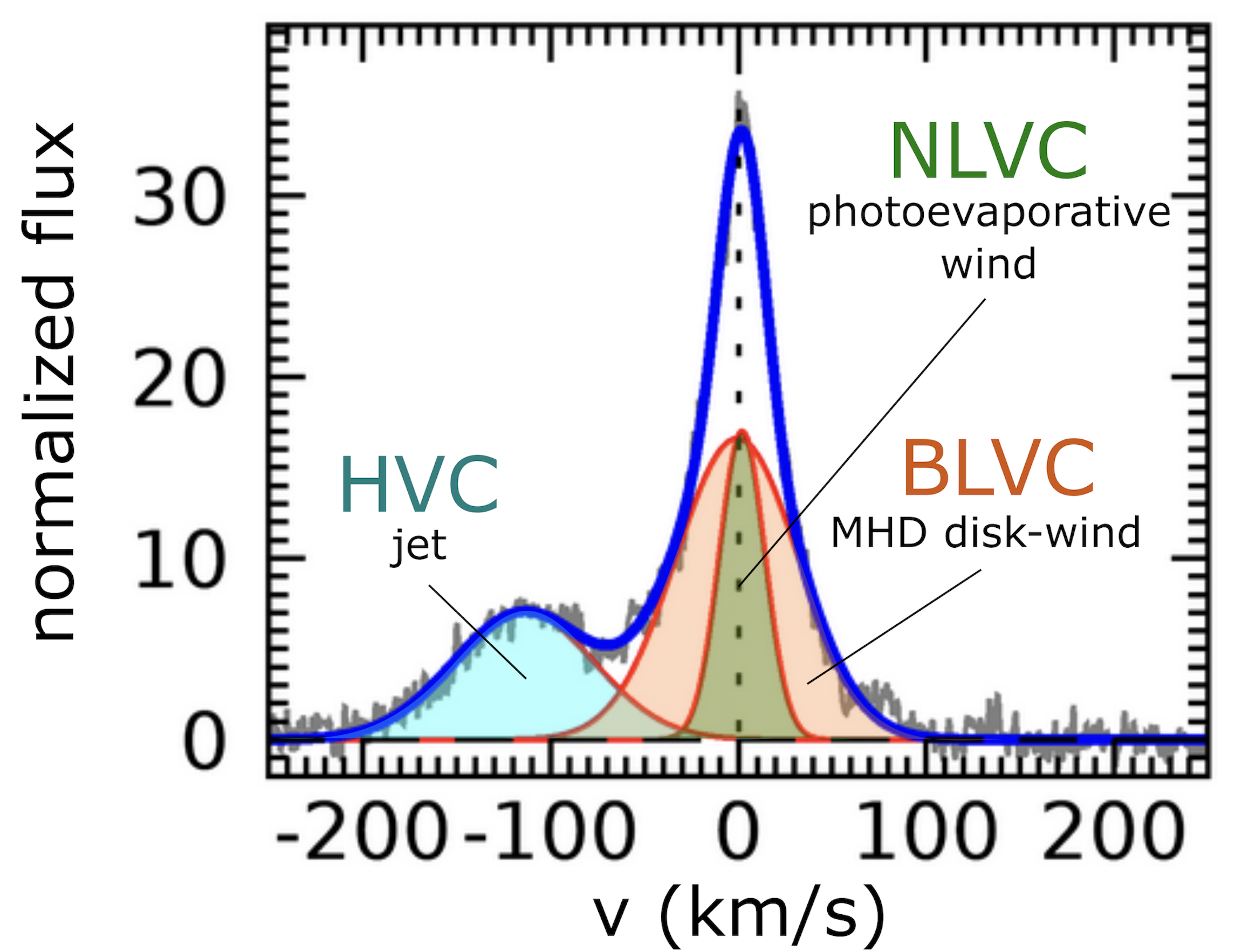}
\includegraphics[angle=0, width=10cm]{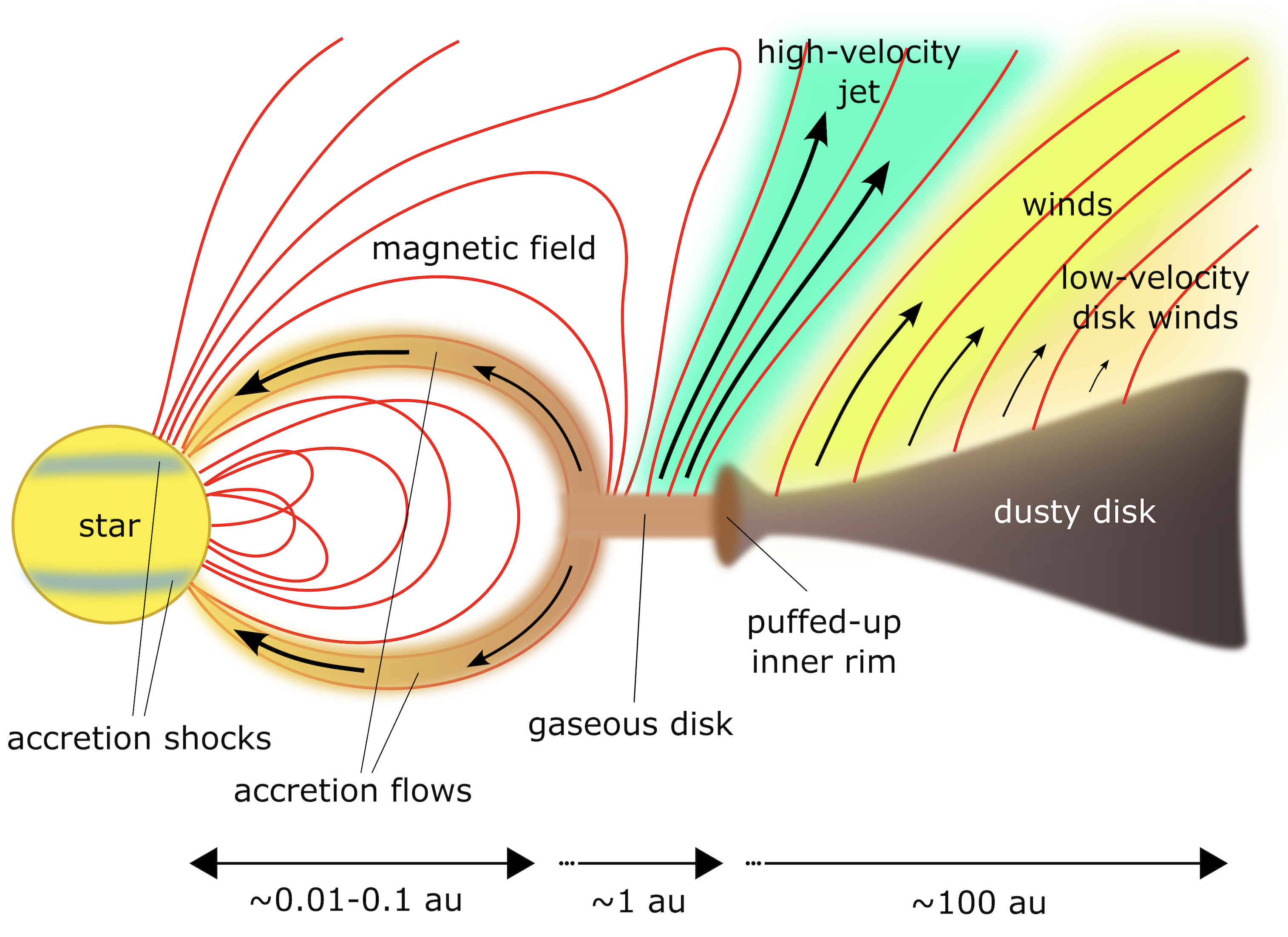}
\caption{{\it Left panel}: [O\,{\sc i}]@630 nm line profile observed with GIARPS@TNG in an accreting T Tauri star. Only the high resolution can separate the kinematical components at low velocity (the NLVC and BLVC) and allow the measurement of the low velocity peak velocities (of the order of 3-5 km/s), thus distinguishing emission from a wind from that of the gas bound in the disk. {\it Right panel}: Sketch of the star-disk interaction region ring in an accreting T Tauri star, according to the magnetospheric accretion model. The many spatially unresolved active regions can be investigated using several absorption and emission lines as diagnostics.}
  \label{fig:profile_disk}
\end{figure*}
\end{center}

Disk-winds, both photo-evaporated and magnetically driven, and bound gas in Keplerian rotation in the inner disk region can be efficiently traced by forbidden lines of atomic and weakly ionized species at low radial velocity ($\lesssim$ 10\,km\,s$^{-1}$, the so-called low velocity component, LVC; e.g., \cite{Rigliaco2013, Nisini2018, Banzatti2019}; Fig.\,\ref{fig:profile_disk}). Different lines from disk-winds are typically blue-shifted. The shift in the peaks is only of 2--3\,km\,s$^{-1}$, which settles the requirement of $R\gtrsim$100\,000. High-resolution data obtained with today's spectrographs show that superimposed with the disk-wind emission component, there is an additional component, peaking at systemic velocity, that could be associated with bound gas in the disk (Nisini et al. 2023, submitted; and references therein). Again, a resolution of $\sim$100\,000 is necessary to deblend the two different components efficiently. With the ANDES sensitivity, it will be possible to separate the two components in a variety of forbidden lines tracing different excitation conditions and thus derive the physical parameters, such as density, temperature, and ionization fraction, that are of paramount importance to disentangle the different wind models and thus quantify their impact on the disk evolution and gas dispersal. The Integral Field Spectroscopy (covering, e.g., the [O\,{\sc i}] at 557,630\,nm and [S\,{\sc ii}] at 406,673\,nm lines) will allow the use spectro-astrometric methods to resolve the spatial regions involved. These regions are at significantly smaller scales than the instrumental resolution (few au), as demonstrated by, e.g., \cite{Goto2012} and \cite{Whelan2021}.

Besides forbidden lines, molecules are also expected to be abundant in the gas phase at larger distances with respect to atomic emission, or originating in deeper layers of the optically thin gaseous disk. They will be sufficiently excited to produce rovibrational features in the infrared, providing important information on the thermochemical structure of the disk, which in turn might have some influence on the final composition of the atmospheres of inner giant planets. The main molecular features that can be observed in the near-IR are the overtone CO bands at $\sim$2.3\,$\mu$m, the H$_2$ ro-vibrational lines, and a variety of weak H$_2$O and OH emission lines in the K-band, which is an ANDES goal. Again, other atomic features, like Na\,{\sc i}@2.206,2.208\,$\mu$m and Ca\,{\sc i}@2.3--2.4\,$\mu$m, can be observed in K-band, which trace the very inner and hot regions of PPD disks. Atomic and molecular diagnostics are characterized by different emitting temperatures, from ~1500 K (e.g., H$_2$O) up to temperatures higher than 3000 K (e.g., CO) and originate at different disk radii. These lines are observed in emission in strong accretors, such as class I protostars or highly active T Tauri stars, and as such, can give important clues on the influence of irradiation from accretion UV photons in planet-forming regions.

Moreover, as the near-IR traces the hotter molecular line emission regions, these kinds of observations will be a natural complement of observations of the fundamental transitions of the same molecules performed in the mid-IR by JWST and METIS.

\subsection{Angular momentum extraction by jets and molecular winds}

A long-standing problem in the study of PPD evolution is how angular momentum is transported in order to allow matter to be accreted by the central star. There are mainly two mechanisms on how accretion is driven inside a PPD: 1) In a viscous disk, turbulent viscosity (often referred to as $\alpha$-viscosity) leads to redistribution of angular moment towards the outer disk, which in turn allows gas to accrete onto the central star (see, e.g., \cite{Hartmann2016}, and references therein); 2) MHD-wind driven accretion in which a large scale magnetic field threads the disk: the gas follows these magnetic field lines and removes angular momentum, which in turn allows for accretion (see, e.g., \cite{Pascucci2022}, and references therein). Both these evolution paradigms are, in principle, able to reproduce observed accretion rates and other constraints but produce very different dynamics and, more importantly, very different evolutionary paths for the PPDs evolution \citep{Tabone2022, Manara2023}. From an observational point of view, it is not clear whether the level of turbulence in PPDs is sufficiently high to drive accretion \cite[see recent review by][]{Rosotti2023}. Nevertheless, it also needs to be clarified if, how, and to what extent MHD-winds can efficiently extract angular momentum from the disk and thus drive accretion.

The only way to verify the wind-driven accretion scenario is to measure the amount of mass and angular momentum taken away by the outflows, and this can be efficiently done only with an instrument like ANDES, which combines high angular and spectral resolution with high sensitivity.

In particular, spectra of T Tauri stars show, in addition to the LVC, forbidden line emission at high velocity, i.e., larger than 40\,km\,s$^{-1}$, which is associated with the high speed and collimated jets that are often observed at large distance from the star (e.g., \cite{Hartigan1995}; Fig.\,\ref{fig:profile_disk}). Collimated jets are the mechanism through which the accreting system gets rid of the angular momentum in excess, thus evolving at rotational velocities below the break-out velocity. The relevant information to understand the mechanism of jet formation is stored within a region with a spatial scale of around 10\,au from the star, needing an IFU facility with a $\sim$100\,mas FoV. ANDES will allow us to distinguish between the various jet launching scenarios (e.g.,\ from the stellar surface, the magnetosphere/disk interface, or the above-mentioned disk-winds; \cite{Ferreira2006}) that predict different spatial scales for the emission. 

In addition, with the IFU, it will be possible to constrain the rotation of the star-disk system and the consequent removal of angular momentum by the jet (e.g., \cite{Bacciotti2002}, \cite{Chrysostomou2008}). Some evidence that collimated jets are indeed able to remove angular momentum from the inner ($<$1\,au) regions of PPD have been given by \cite{Bacciotti2002} with HST/STIS observations of optical jets and by \cite{Lee2017} with ALMA SiO observations of an embedded protostar.
The velocity gradient between the two sides of the rotating jet, which needs to be resolved spatially, is expected to be of the order of a few km\,s$^{-1}$, which is, therefore, the velocity accuracy needed to perform these observations. ANDES will be the perfect instrument to study this aspect through IFU observations of the [O\,{\sc i}]@630\,nm line in the optical and the [Fe\,{\sc ii}]@1.64\,$\mu$m line in the H-band. 

Moreover, possible observations in the K-band (an ANDES goal) will also be important to probe the molecular component of the flow. In particular, high-resolution spectral imaging of the bright H$_2$ line at 2.12\,$\mu$m, a tracer of molecular winds, will provide information on the role of these molecules in the global evolution of PPDs (e.g., \cite{Beck2009, Gangi2020, Gangi2023}). For the first time in the IR regime, it will be possible to measure the mass loss through molecular winds and the angular momentum loss, via the measurement of wind rotation signatures.  Rotation of MHD winds created in the 1-10\,au disk regions, such as those traced by the H$_2$ NIR lines, will be fundamental to possibly validate the recent models of wind-driven PPD evolution in contrast with the so far considered paradigm of viscous disk evolution.

\begin{figure}
\begin{center}
\includegraphics[width=8cm]{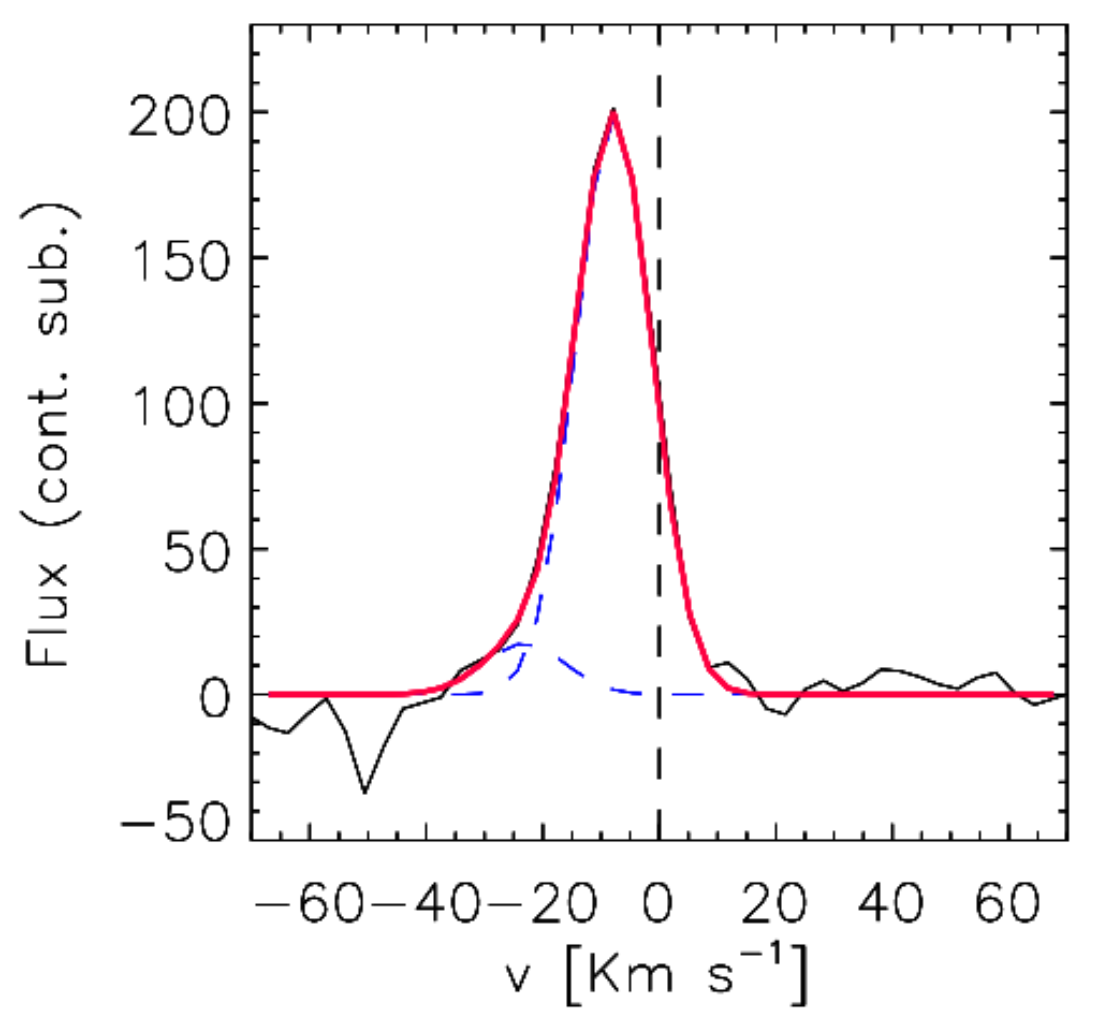}
\end{center}
\caption{
Continuum-subtracted GIARPS H$_2$ line profile of an accreting T Tauri star observed on-source. The fit to the profile obtained by adding multiple Gaussian components (dashed blue lines) is plotted in red. Fluxes units are 10$^{-15}$\,erg\,s$^{-1}$\,cm$^{-2}$\,\AA$^{-1}$. The target displays a blue-shifted line peak with velocity ranging between $-17$ and $-5$\,km\,s$^{-1}$ indicating an origin in winds.}
  \label{fig:comp}
\end{figure}

An example of a molecular wind as observed by SINFONI@VLT (at an angular resolution of 0.12") is given in Fig.\,1 by \cite{Agra-Amboage2014}, while a spectrum of the H$_2$ line at 2.12\,$\mu$m obtained with GIARPS (R$\sim$45\,000) at the {\it Telescopio Nazionale Galileo} is plotted in Figure\,\ref{fig:comp}. The dynamical properties of the wind imaged by SINFONI cannot be at present disentangled due to the need for a spectro-imager with a spectral resolution high enough to discriminate its kinematic behaviour morphologically. The shown spectrum demonstrates that a resolution of the order of 100\,000 is needed to resolve the line and determine any velocity displacement at a level of some km/s due to the outflow rotation. ANDES performances (angular resolution of 15\,mas and spectral resolution of $R\sim$100\,000) will allow us to obtain in one shot both high spatial and spectral information at a much higher resolution than now possible to exploit the morphology and kinematics of the wind launching region. By combining ANDES observations in the H-band, it will be possible for the first time to directly compare the molecular and atomic wind components, providing a global picture of the wind-driven scenario in young stars  (see Sect.\,\ref{sec:PPDs_Kband}).

\subsection{Accretion disks of embedded sources in different environmental conditions}

Recent surveys performed with the X-shooter@VLT and GIARPS instruments on samples of T Tauri stars in nearby clouds have provided a detailed characterization of the accretion and stellar properties of populations of young low-mass stars (e.g.,\ \cite{Alcala2014, Alcala2017, Alcala2021, Manara2021, Gangi2023}). These studies are now starting to probe the accretion evolution and how it influences the evolution of the circumstellar disk. ANDES can extend these kinds of studies in different directions: 1) probing accretion in younger and embedded sources, where present instrumentation is not sensitive enough to get a sufficiently high $S/N$ ratio; 2) exploring the accretion phenomenon down to the brown-dwarf regime; 3) study if and how accretion depends on different environmental conditions, such as metallicity, density, and photoevaporation. 

From the one side, characterizing the stellar and disk properties of young embedded accreting protostars (class 0/I) is extremely challenging due to their large extinction and the strong IR excess coming from their circumstellar envelope. Currently, weak photospheric features and emission lines from the inner disk have been detected with K-band low/medium-resolution spectra of small samples of very bright class 0/I sources with moderate IR excess (e.g., \cite{Nisini2005, Greene2008}). However, no measurement was performed on fainter sources where high sensitivity spectra at high resolution in the IR are needed to resolve photospheric lines (e.g., at 2.1--2.3\,$\mu$m) useful to derive spectral type and veiling. Again, the K-band contains several emission lines, which are useful extinction indicators \citep{ConnelleyGreene2014}.  This information, together with the observations of other accreting diagnostics detectable in the K-band, like the Br$\gamma$ (at around 2.17\,$\mu$m), will enable us to fully characterize for the first time large samples of protostars (see Sect.\,\ref{sec:PPDs_Kband}). 

On the other side, studies in several young clusters of the Large and Small Magellanic Clouds and the outer Galaxy (\cite{DeMarchi2011,DeMarchi2017} \cite{Biazzo2019}, \cite{Vlasblom2023}, \cite{Tsilia2023}) suggest that metal-poor stars accrete at higher rates compared to solar-metallicity stars in nearby Galactic star-forming regions; moreover, they also suggest that the cluster stellar density of the forming environment can affect the mass accretion rate and thus the star formation process itself. Also, it has been argued on theoretical grounds that the efficiency of dispersal of circumstellar disks depends on stellar metallicity, i.e.,\ the formation of planetesimals around stars may be faster in a high metallicity environment (\cite{Ercolano2010}). Therefore, simultaneous measurements of accretion rates and metallicity in young stellar objects in different environments are crucial. Accretion luminosity can be measured from permitted lines, such as H$\beta$, Ca\,{\sc ii}, Pa$\gamma$, or from the Balmer jump between 340 and 370\,nm. This measurement does not need high spectral resolution. However, to measure the mass accretion rate from the accretion luminosity and to compare accretion properties with properties of the central star, such as mass, age, luminosity, and metallicity, it is required to observe the narrow photospheric features sensitive to spectral type, veiling, and metallicity. In the case of embedded and highly veiled young stellar objects, a high spectral resolution is needed to extract the weak photospheric features from the strong continuum. For these latter sources, observations in the IR, band H and K in particular, are particularly useful because of the presence of several useful diagnostics (see, e.g., \cite{Alcala2017}, De Marchi et al. 2023, submitted). The targets of interest will be weak in the optical and located in regions where it will be difficult to find bright close-by sources for driving the AO. They, therefore, require an AO system sensitive to stars weaker than about 20\,mag in R. 

Like in the other science cases, these kinds of observations will be a natural complement of observations performed in the mid-IR by JWST and METIS.

\subsection{Observing planets as they are being formed}
Theoretical models of planet formation are being built using only indirect observational constraints (i.e.,\ the initial conditions of the PPD and the observed planetary systems around stars as an outcome). The formation process itself has mostly remained observationally unconstrained, and thus, its study has mostly been restricted to the realm of theory. With the high spatial and spectral resolution of ANDES, this will change, and the characterization of the formation process will be feasible.

Here, we present two observational goals that will enable us to estimate constraints on the planet formation process itself and advance our understanding of planet formation.

\subsubsection{Giant planet formation: Accretion mechanisms and planetary magnetic fields}
\begin{figure}
    \centering
    \includegraphics[width=8cm]{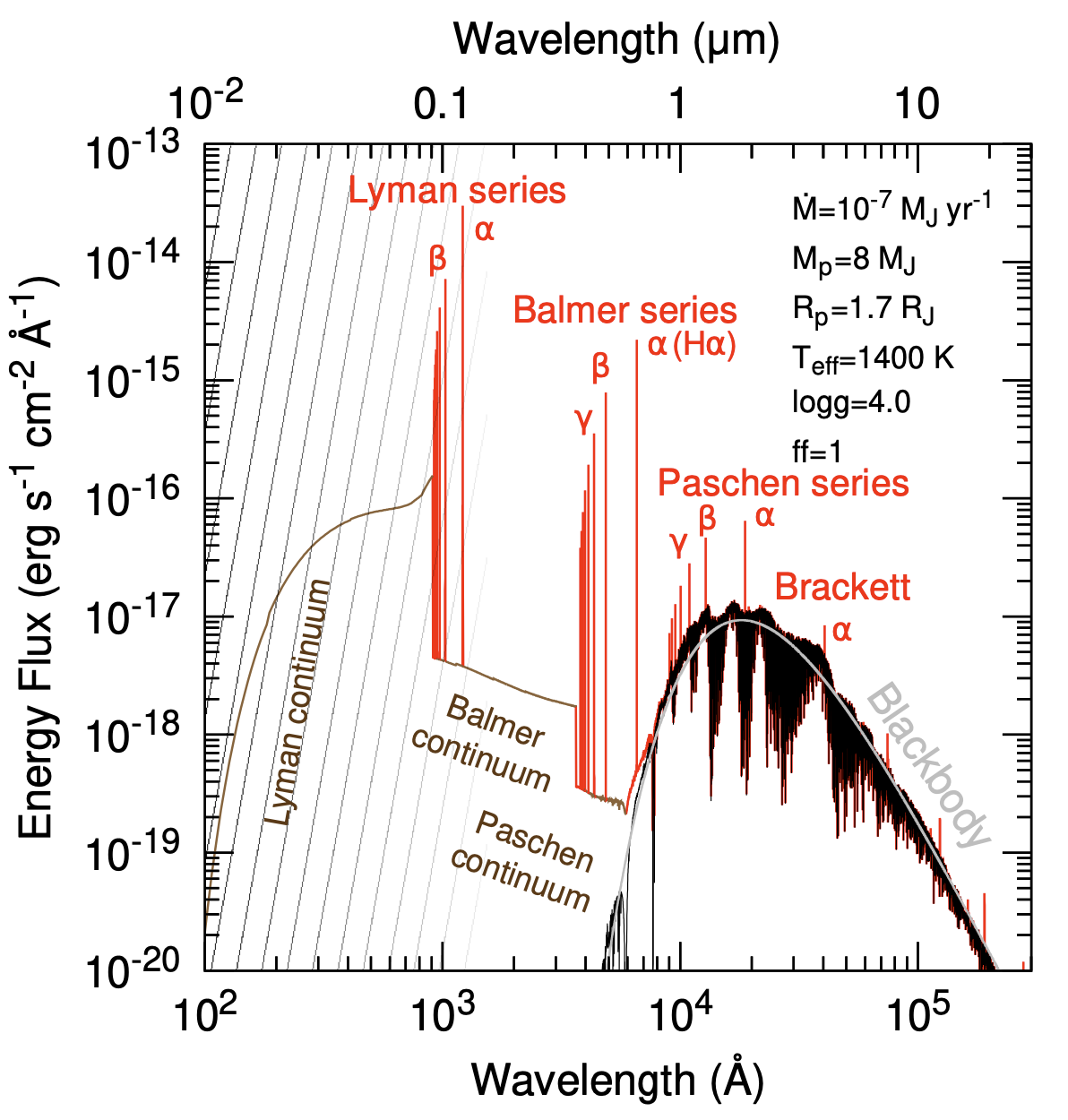}
    \caption{Image taken from \cite{Aoyama2020}. It shows a synthetic spectral energy distribution (SED) of an accreting giant planet from theoretical modelling. The blackbody of the planet's effective temperature (gray) is overplotted with photosphere emission (black). The bright emission lines of hydrogen, originating from accretion shocks, are shown in red. The SED corresponds to a giant planet of $8\, \mathrm{M}_\mathrm{Jupiter}$ accreting at $10^{-7}\, \mathrm{M}_\mathrm{Jupiter}\mathrm{yr}^{-1}$ under the hot-start assumption. Image reproduced with permission of the authors.}
    \label{fig:aoyama2020}
\end{figure}

The formation of giant planets involves a phase of strong gas accretion, which leads to much higher luminosity than at later stages due to the large amount of gravitational potential energy released during this process. Planet formation models under the paradigm of core accretion predict that forming protoplanets undergo two main formation phases \cite[e.g.][]{Lissauer2009}. First, during the \textit{attached phase}, the protoplanet is still fully embedded in the disk and has a low mass and a large radius (approximately equal to the Hill Sphere). At this stage, the protoplanet has a low effective temperature of around $200\,\mathrm{K}$ and low luminosity of $\sim 10^{-6}\mathrm{L}_\odot$. As the planet grows, it eventually enters gas runaway accretion at $\sim 100 \mathrm{M}_\oplus$ and starts to contract and detach from the nebula. A circumplanetary disc (CPD) will form and the gas falls at high velocity from either the Hills sphere or from the CPD onto the protoplanet, where it shocks which again leads to high luminosities of $10^{-4}$  to $10^{-1}\,\mathrm{L}_\odot$. This phase lasts for several $10^5$ up to a few $10^6$ years. During this so-called \textit{detached phase}, accretion tracers such as H-alpha (656.3), Paschen beta (1282), and the Bracket gamma (2166) in the K-band should be observable. In Figure \ref{fig:aoyama2020}, an example of the spectral energy distribution of an accreting giant planet is shown as predicted by theoretical models from \cite{Aoyama2020}. The H-alpha emission of accreting protoplanets has been confidently observed in one system (PDS 70b \cite{Haffert2019}) with other studies showing null \cite{Huelamo2022} or candidate (e.g.\ AB Aur \cite{Zhou2022}) detections. Quantifying the H-alpha emission, particularly by disentangling the planetary and stellar components, could provide constraints on the thermal and dynamical structure of the accretion flow \cite{Marleau2022}. Further, having spectra of accreting protoplanets could help answer the fundamental question of whether the start is \textit{hot} or \textit{cold} (i.e.,\ whether the accretion shock luminosity is incorporated into the protoplanet or radiated away efficiently in a supercritical shock  \cite[see][]{Marley2007}). This knowledge about the starting conditions has strong implications for the luminosity-mass relation of young protoplanets as it strongly depends on it \cite[e.g.][]{Spiegel2012}.

Not only giant planet formation can lead to high luminosity, but also the collisional afterglow of low-mass to intermediate-mass planets is expected to be detectable with next-generation instruments \cite[][]{Miller-Ricci2009,Lupu2014,Bonati2019}.

During and shortly after the formation process of giant planets, the magnetic field of the protoplanet is expected to be much stronger due to the high luminosity. Theoretical models and scaling relations predict high magnetic fields of $\sim 100\,-\,1000$ Gauss for these objects \cite[see][]{Christensen2009}. Jupiter has a magnetic field strength of $\sim 4$ Gauss for comparison. The magnetic field strengths and temperatures are expected to be similar to M dwarfs, where the measurement of the magnetic field strength has been achieved already, using the prominent FeH absorption band at 1\,\textmu m, by \cite[e.g.][]{Shulyak2010}, using a high spectral resolution of $R\sim100\,000$. The strength of the magnetic field would deliver the key to understanding how giant planets accrete gas. Strong magnetic fields would lead to magnetospheric accretion like stars, whereas lower magnetic fields would lead to accretion through a boundary layer. The type of accretion is crucial for the hot vs.\ cold start issue. Global hydromagnetic simulations by \cite{Gressel2013} showed that young giant planets could launch outflows and jets like stars. Here again, the strength of the magnetic field is a fundamentally important quantity. 

\begin{figure*}[t!]
  \centering
  \includegraphics[width=1.\linewidth]{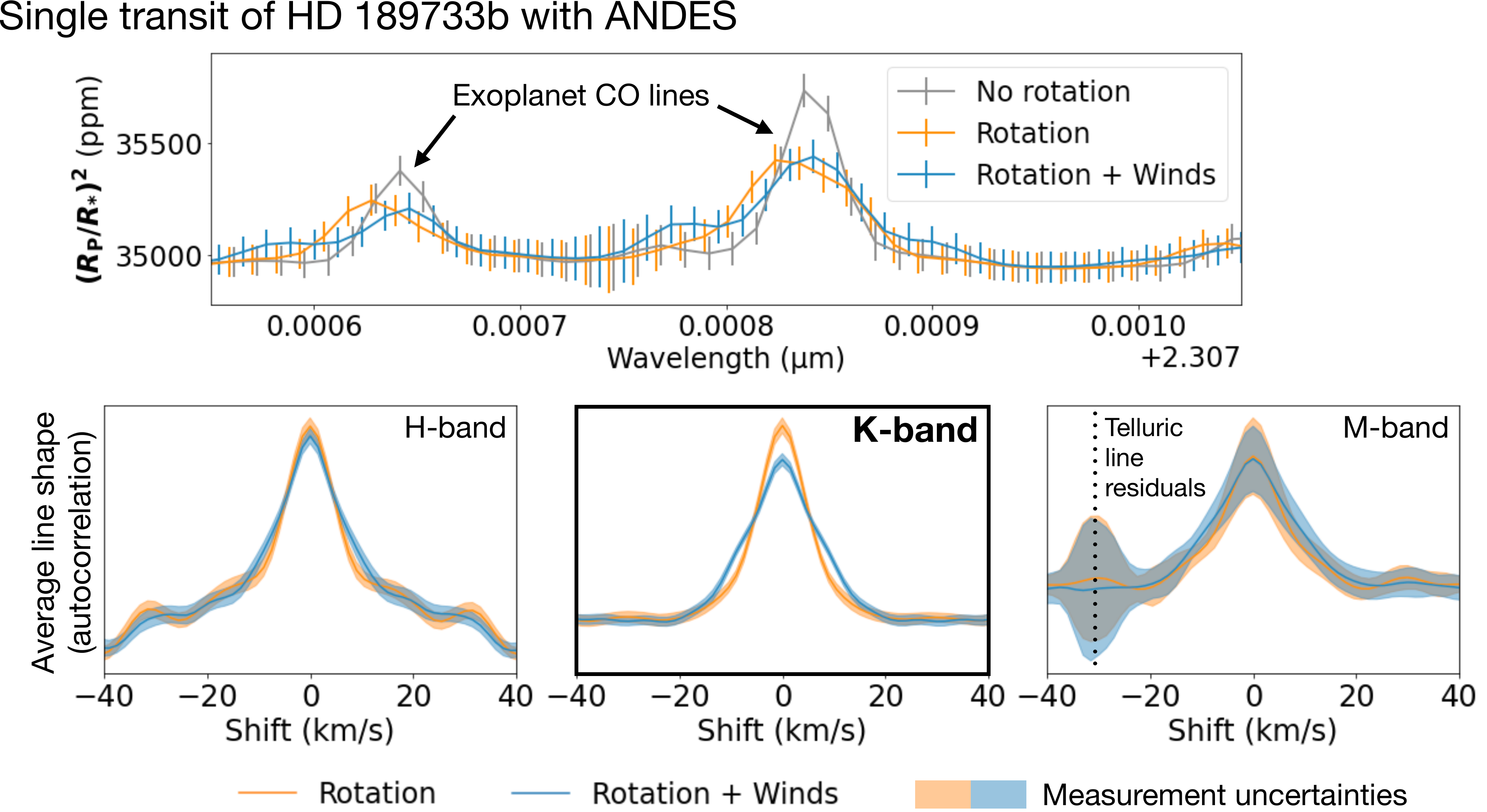}
  \caption{Simulated transit of HD 189733b with ELT ANDES, which will reveal the wind patterns in the atmospheres of hot Jupiters. {\it Top panel:} transit depth in the K-band, assuming a single transit observed at $R=100,000$. {\it Bottom panel:}  average spectral line shapes extracted from simulated observations of the CO band. For each band, the same number of detector pixels was assumed, using a width of 0.003, 0.006, and 0.012~$\mu$m for H-, K-, and M-band, respectively.}
  \label{fig:HD189}
\end{figure*}

\subsubsection{Observing the direct impact of forming giant planets on the PPD}
The torque of the protoplanet acting on the gas will push back the gas in the outer disc, resulting in carving a gap into the PPD of $\sim\,\mathrm{R}_\mathrm{Hill}$ \citep{Papaloizou1986,Crida2006}. Since giant planets are favored to form outside the water ice line, the gas there is too cold for significant VIS or NIR emission. However, taking into account the high luminosity of the accreting protoplanet, the gas in the vicinity of the protoplanet may be heated up to $\sim 1\,000\,\mathrm{K}$ \citep{Montesinos2015,Garate2021}. Both high spatial and spectral resolution might allow to observe the modified spectral emission of the disc with a peak emission at ($\sim 2.5$\,\textmu{}m) with the low end of the K-band ($\sim2.4$\,\textmu{}m). Further, the accretion flow of from the PPD onto the CPD is expected to be at least partially supersonic, which would shift emission lines of embedded molecules like CO. Tracing the velocity field of ongoing gas accretion onto a giant planet would put observational constraints on the accretion structure which would provide important observational evidence for the construction of advanced hydro-dynamical models.


\section{Science cases in the K-band}
\label{sec:Kband}

The baseline design of ANDES does not include the K-band spectrograph, which is currently a goal. However, several science cases described above already discussed the important, at times even essential, role that the K-band plays in enabling certain kinds of studies (see the various mentions in sections \ref{sec:gasgiants} and \ref{sec:PPDs}).
This is also reflected by the fact that the single most important band for high-resolution spectroscopy of exoplanets to date has been the K-band. It enabled the first molecular detection in an exoplanet at high-resolution, the first measurement of atmospheric wind speeds \citep[both][]{snellendekok2010}, the first measurement of planetary spin rates \citep{snellenbrandl2014}, the first brightness map of a brown dwarf \citep{Crossfield2014}, and the first detection of an isotopologue in an exoplanet atmosphere \citep{zhangsnellen2021}. All of these successes used the fact that the K-band is ideal to observe the strong, regularly spaced lines of the abundant and chemically stable CO molecule. The property of CO as an excellent dynamical tracer \citep{savelkempton2023} and the sensitivity increase of the ELT will enable us to pursue several novel, groundbreaking science cases. While K-band is a goal, not in ANDES' baseline design, we propose making ANDES a ``CO machine'', ensuring that the opportunities of K-band high-resolution spectroscopy will not remain untapped on the ELT. The transformational power of ANDES K-band lies in the study of atmospheric dynamics in gas giant planets, as already alluded to in Section~\ref{sec:gasgiants}. We will also discuss the role of CO as an antibiomarker in terrestrial planets. However, due to the tens of nights that would have to be spent on a single target, we deem the small planet science case less compelling for the K-band.

\subsection{Close-in gas giant planets}
\label{sec:kband_close_in_dynamics}
Figure~\ref{fig:HD189} (upper panel) shows how high-resolution K-band transmission spectra obtained with ANDES will reveal the wind patterns in the atmospheres of hot Jupiters. For this estimate, we post-processed the three-dimensional temperature and velocity field obtained from General Circulation Models in \citet{drummondmayne2018} for the prototypical hot Jupiter HD 189733b with the pRT-Orange transmission code (Molli\`ere et al., in prep.). We then studied the average line shape by calculating the auto-correlation function of the spectrum. Even for a single transit, the effect of winds on the transmission spectrum can be extracted with high significance in the K-band (lower panel). H-band is significantly worse because the CO lines are weaker, while in the M-band, the CO line measurements are affected by thermal background and low telluric transmission. Therefore, the K-band is the best prospect for accurately tracing the dynamics.\footnote{Other molecules that may be more readily detected in other bands, such as $H_2O$, are less direct tracers of atmospheric dynamics, as their abundances are also shaped by variations in temperature around the planet. CO is ideal because of its thermal stability and strong, widely separated lines.}

The predictions in Figure~\ref{fig:HD189} were made assuming a spectral width of 0.006 in the K-band. However, CO’s full wavelength range of interest in the K-band extends from 2.3--2.4\,$\mu$m. Expanding the wavelength range corresponds to an increase in the S/N by a factor of four. Considering this full spectral range of CO, we find that there are 64 transiting planets for which one can infer the wind properties, as shown in Figure~\ref{fig:HD189}. For this estimate, we restricted the sample of known transiting exoplanets to equilibrium temperatures $> 1000$\,K (to ensure CO visibility in gas-dominated planets), mass errors $<30$\%, and made use of the so-called transit spectroscopy metric \citep{kemptonbean2018}. The stars were not checked for visibility from Cerro Amazones. Therefore, while the detailed analysis method to invert high-resolution spectra to obtain information on the wind patterns still needs to be worked out in the coming decade, we find that high-resolution K-band observations with ANDES will enable studying the wind patterns of many tens of transiting planets with high precision. This data set will be revolutionary for understanding the circulation in gas giant planets and the underlying physical phenomena that control them, potentially also shedding light on their internal evolution \citep{tremblinchabrier2017,kollkomacek2018}. We again note that the numbers presented here (many tens of planets) are comparable to those presented in Section \ref{sec:gasgiants} ($\sim$100 planets), but not directly comparable, since the sensitivity analysis was carried out differently. Moreover, the number quoted in this section relied on K-band alone, while in Section \ref{sec:gasgiants} the K-band was included in addition to other bands. These results both point to K-band playing a crucial role for undestanding the atmospheric dynamics of gas giants. At the same time, studying how the results in Section~\ref{sec:gasgiants} depend on the inclusion of K-band would be beneficial.

In addition to transmission spectra, we can place powerful and complementary constraints on the atmospheres of these planets from the measurement of their emission spectra. These spectra probe a higher pressure regime than transmission, can be obtained for both transiting and non-transiting planets ---greatly expanding the number of planets we can characterize--- and allow us to measure both the day- and night-sides of the planets, given sufficient signal-to-noise. Current high-resolution spectroscopic characterization measurements of hot Jupiters have demonstrated sensitivity to the three-dimensional thermal structure of these planets \citep{beltzrauscher2021,hermandemooij2022,vansluijsbirkby2023}, and spatial variations in winds, probed as a function of longitude \citep{pinobrogi2022} or through different molecular species \citep{brogiemeka-okafor2023}. Since the Doppler shifts in high-resolution spectra are a function of both the thermal and wind spatial structures \citep{zhangkempton2017}, the robust properties of CO, as described above, again make it a powerful tracer for dynamics in emission spectra.

\subsection{Mapping directly imaged planets and brown dwarfs}
Characterizing the two-dimensional brightness distribution of brown dwarfs and directly imaged exoplanets would allow a fundamental advance for the theory of clouds, convection and magnetic fields of substellar hydrogen-dominated celestial bodies and globally in our understanding of planetary formation and evolution. The latter is because improving our understanding of convection would allow for better planetary structure models, which are used for planet evolution models, that ultimately depend on the heat that planets retain from formation \citep[e.g.,][]{baraffechabrier2003,mordasinimarleau2017}.

\begin{figure*}[t!]
  \centering
  \includegraphics[width=1.\linewidth]{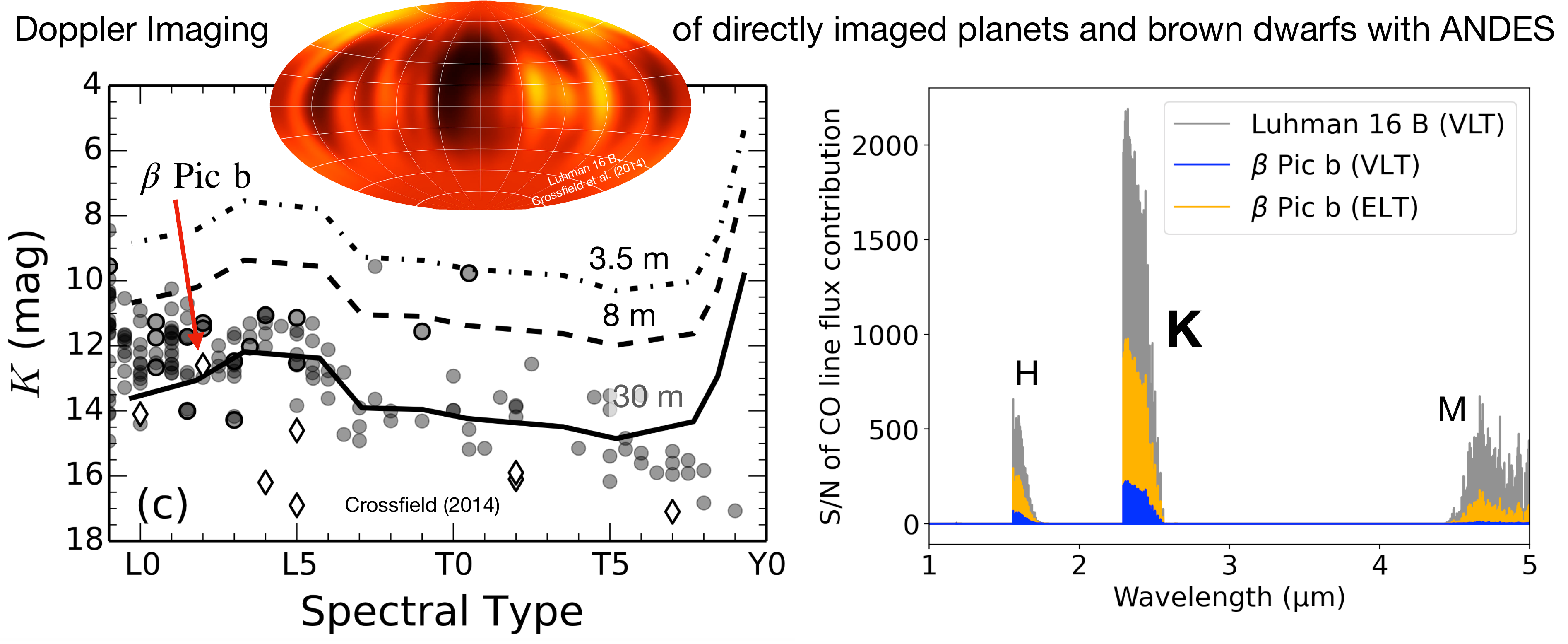}
  \caption{K-band feasibility of Doppler imaging. {\it Left panel:} Limiting magnitude for Doppler Imaging in the K-band, taken from \citet{Crossfield2014} The three horizontal lines show the limits for 3.5 (dot-dashed), 8 (dashed) and 30 m (solid) telescopes. Diamonds are known directly imaged exoplanets as of 2014. {\it Right panel:} modeled S/N per pixel of the negative flux (i.e., absorption) contribution of CO lines in the spectra of Luhman 16~B and $\beta$~Pic~b, observed for 5 h with the VLT or ELT (ANDES/METIS) at $R = 100,000$.}
  \label{fig:mapping}
\end{figure*}

Due to its strong lines in the K-band, carbon monoxide will enable us to map the surface brightnesses of directly imaged planets and brown dwarfs. This is done by employing the so-called Doppler imaging method, which tracks how temperature heterogeneities (spots or clouds) rotate in and out of view during observation. Brown dwarfs and directly imaged planets rotate with equatorial speeds of up to 20--30\,km\,s$^{-1}$, so resolving the motion of spots across the targets’ visible disks requires high spectral resolution. Effectively, the maps are constructed from time-dependent variations in the CO line shapes. This was first demonstrated by \citep{Crossfield2014} for Luhman 16\,B, an L-T transition object in a brown dwarf binary system. The two brown dwarfs are the closest known to date (2 pc), which made the mapping possible already with CRIRES at the VLT. As demonstrated in a follow-up paper, K-band is the band of choice when mapping the surfaces of brown dwarfs and exoplanets, and this will only really be possible in the era of 30-m (and above) class telescopes \citep[][also see the left panel in Figure~\ref{fig:mapping}]{crossfield2014b}. More specifically, the ANDES K-band spectrometer would unlock 10s of brown dwarfs for surface mapping and a few directly imaged exoplanets, the most accessible being $\beta$~Pic~b. For further illustration, we show the S/N of the flux contribution of CO lines (which cause absorption features) in $\beta$~Pic~b and Luhman 16\,B in the right panel of Figure~\ref{fig:mapping}. K-band clearly dominates over the CO lines accessible in the H- and M-bands.

\subsection{Ancillary gas giant science in the K-band}
In addition to tracing dynamics, ANDES K-band will enable a population study of the CO and $^{13}$CO content of gaseous exoplanets, both of which are important formation tracers \citep[e.g.,][]{oebergmurrayclay2011,madhusudhanamin2014,mordasinivanboekel2016,zhangsnellen2021,mollieremolyarova2022}. CO is particularly interesting because it is the major carbon-carrying volatile species once chemically favored \citep[between $\sim$1000-3500 K, see][]{loddersfegley2002}, and highly stable. For formation constraints, elemental abundance ratios are of interest (e.g., C/O, N/O), for which the abundance of other species must be inferred ($H_2O$, $CH_4$, $HCN$, $C_2H_2$, $NH_3$, $PH_3$, $CO_2$, ...). Detecting these species would also be crucial for informing chemical disequilibrium models \citep[e.g.,][]{agundezparmentier2014,zahnlemarley2014,baeyensdecin2021}. Some of these are accessible in the K-band or can be seen when using the bluer near-IR bands of ANDES (YJH), which would be observed simultaneously with the K-band. As mentioned above, planetary emission spectra of close-in planets will be accessible at any time in the ELT era (so no eclipse scheduling is necessary) as long as the planet moves fast enough to distinguish it from telluric and stellar lines \citep[as already used in, e.g.,][]{brogidekok2014}. This will allow ANDES to quickly build up a large library of planetary emission spectra to be analyzed with high-resolution retrieval methods \citep[e.g.,][]{brogiline2019,gibsonnugroho2022}. Such ground-based observations are, therefore, highly complementary even to current space-based telescopes such as JWST, which require scheduling for transits and eclipses or long continuous staring for phase curves.

\subsection{Atmospheric characterization of small HZ planets}

We also tested whether planetary CO can be detected in the K-band with stellar light reflected off terrestrial, habitable zone planets. While the classical biomarker molecule $O_2$ is not visible in the K-band, the CO is an important molecule to rule out false-positive detections of bioactivity. Other molecular species that could contextualize an $O_2$ detection are better observed in other bands, at least for systems with M-dwarf hosts, and considering transmission spectra \citep{curriemeadows2023}. CO is a useful anti-biomarker since it traces the abiotic production of $O_2$, driven by the photolysis of $O_2$, which most dominantly occurs for M-dwarfs \citep[e.g.,][]{harmanschwieterman2015,meadows2017}. We studied the case of Proxima Cen~b, the closest habitable zone planet outside the solar system, and assumed a $CO_2$-dominated atmosphere with clouds that increase the reflected light visible from the planet. To estimate the detectability of CO, we first computed 3-Dimensional Global Climate Model simulations \citep{Turbet2016}, then used a reflected light line-by-line radiative transfer model to compute high-resolution spectra at maximum elongation, and eventually estimated the $S/N$ of a CO detection using a cross-correlation approach. We found that about 15 nights of observations are sufficient to detect CO at 5$\sigma$ if a CO/$CO_2$ ratio between 10$^{-2}$ and 10$^{-4}$ is assumed. In principle, CO has lines in the H-, K-, and M-bands\footnote{As we argued above, the K-band is the best band for giant planets.}. However, for CO/$CO_2$$ < 10^{-2}$, CO is not visible in the H band. For CO/$CO_2$ $ > 0.01$, however, which corresponds to a higher $O_2$ mixing ratio in the $CO_2$ photolysis scenario, we find that the K-band’s CO lines become optically thick and saturate, and are thus no longer a viable pathway for detecting CO. In this high CO mixing ratio case, the H-band allows us to detect CO weak lines in about 35 nights. In fact, the lower diffraction limit in the H-band allows to probe Proxima Cen~b at shorter angular separations and thus probe the planet at more favorable orbital phase angles (with an expected contrast gain of a factor 2-3), which brings back the required number of nights to 12-17, a number somewhat comparable to K-band estimates. Moreover, the detection of CO in the H-band can be confirmed in the K-band using non-detection of the planet, given that CO at high mixing ratios saturates the band.

We therefore argue that to ensure a CO detection, if the molecule is present in the planet’s atmosphere, we require K- in addition to ANDES’ H-band because the detectability of CO is highly dependent on the atmospheric properties such as the CO/$CO_2$ ratio, the surface pressure, the cloud coverage, cloud albedo, and cloud top pressure, which all control the depth of the CO lines in the H- and K-band reflected light spectrum. We thus conclude that K-band may be a viable option to detect CO as an anti-biomarker in the reflected light of exoplanets but that an in-depth study of the various conceivable atmospheric cases would help to fully assess this goal's feasibility.

Also, the emitted light case in the M-band should be studied. However, our first tests show that it may be unfeasible due to lower angular resolution, stellar flux suppression, and increased telluric thermal background and absorption. Due to the broad (partially saturated) line cores, We also note that a resolution of R = 20,000 is sufficient to detect CO in the K-band. Thus, observations at the intended resolution of ANDES may only be necessary if the high-resolution is required to distinguish planet and telluric lines. We note that the line core absorption also depends on the atmospheric CO abundance. Lastly, while Proxima Cen~b is the best case in terms of S/N, we argue that at least five planets may ultimately be amenable to K-band characterization: Proxima Cen~b, Ross 128~b, GJ 273b, Wolf 1061~c, and GJ 682~b, with exposure times between 2-5 times longer than Proxima Cen. However, several of these targets are very close to the diffraction limit of ANDES in the K band at maximum elongation, which will make efficient stellar light suppression challenging. As studying these planets is a considerable time investment, one option would be first to explore whether the planets can be detected in reflected light, which requires up to 10 times less exposure time. After that, these habitable zone planets may be observed for 10s of nights (likely over multiple years) to establish the presence of (anti-)biomarkers in their atmospheres.

We also assessed the detectability of (anti-)biomarker molecules in habitable-zone terrestrial planets with transmission spectroscopy. For this, planets orbiting M-dwarfs are the most feasible candidates, with favorable transit depths and short orbital periods. The most promising planetary system for this is TRAPPIST-1, which is a 7-planet system around an M8V host star at 12.5 pc distance from the solar system, with up to four of its planets in the present-day habitable zone \citep{gillontriaud2017}. However, as reported in \citep{curriemeadows2023}, CO is best detected in the H-band in transmission of habitable zone planets, requiring $\sim$100 transits. $CO_2$ may be detectable in the bluest part of the K-band ($\sim$2.1 µm), but also here \citet{curriemeadows2023} argue for H-band being a better option.

We also studied whether $PH_3$ may be detectable in the K-band, assuming a $CO_2$-dominated atmosphere. This is motivated by the claimed $PH_3$ detection in Venus, where it was interpreted as a potential biomarker \citep{greavesrichards2021}. For TRAPPIST-1b we found that $PH_3$ is only detectable (in 40 nights) when increasing the $PH_3$ abundances by a hundredfold when compared to the value of 10 ppb reported in \citet{greavesrichards2021}, making this an unlikely scenario, although according to \cite{Sousa2020} $PH_3$ could enter a runaway buildup and reach such abundances.

\begin{figure*} 
  \centering
  \includegraphics[width=0.9\linewidth]{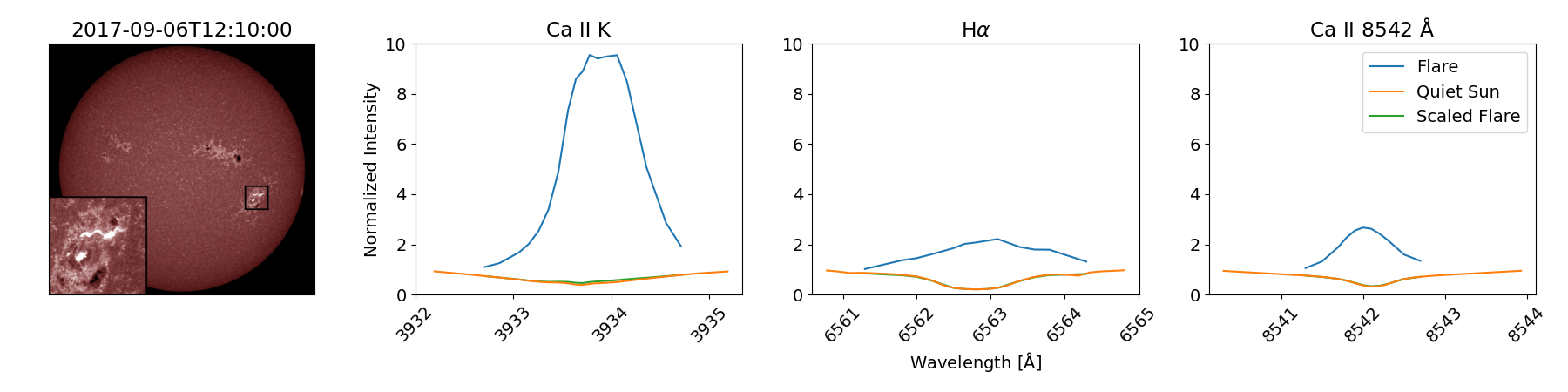}
  \caption{A comparison of the spectral response of the Ca II K, H$\alpha$, and Ca II 8542~\AA\ lines to changes in solar activity, demonstrating the sensitivity of the Ca II K line over the other chromospherically active lines. \textbf{First panel:} An SDO/AIA \citep{2012SoPh..275...17L} observation of the Sun at 2017-09-06T12:10:00 in the 1700~\AA\ continuum showing one of the largest solar flares of the last solar cycle ($\approx 10^{32}$ erg). An enlarged view of the same image is shown on the lower left side of this figure. \textbf{Second-fourth panel:} Flare spectrum in blue, quiet Sun spectrum in orange, and a scaled flare spectrum with a 1\% fill factor in green for each of the three aforementioned lines. The line core of the Ca II K line is reacting much more strongly to the flare than the line cores of the H$\alpha$ line and the lines of the Ca infrared triplet.}
  \label{fig:activity_tracers_flare}
\end{figure*}

\subsection{Protoplanetary Disks}
\label{sec:PPDs_Kband}

As discussed in Section~\ref{sec:PPDs}, various science cases are of interest to the protoplanetary disk community in the K-band, such as the characterization of young stellar objects via emission lines from the inner disk and weak photospheric features or measuring the magnetic field strengths of T~Tauri stars via Zeeman splitting (e.g., \citealt{johns-krull2007}, and references therein). Some of the most transformative science cases, which the K-band uniquely enables, concern the role of the angular momentum loss in protoplanetary disks via molecular winds and the characterization of embedded protostars and their accretion disks in different environments (see Sect.\,\ref{sec:PPDs}). These studies are highly relevant in defining the timescales for accretion and disk dispersal during the early evolution and, more in general, for planet formation, as protoplanetary disks are the birthplaces of exoplanetary systems.


\section{Exoplanet Science cases in the U-band}

Currently, the ANDES baseline design extends to 4000\,\AA, but with the goal of extending the spectral coverage to 3500\,\AA, covering the U-band. The U-band strongly benefits the core exoplanet science case of characterizing atmospheres and detecting biosignatures for habitable-zone exoplanets and makes several additional exoplanet science cases possible to study. 

The core exoplanet science case of studying habitable-zone exoplanets relies on the different planetary and stellar rest frames, moving at different radial velocities, to tell planetary and stellar absorption and emission signatures apart (see for example \cite{2019AJ....157..114B, 2018arXiv180604617B}). The atmospheres of hot Jupiters are successfully studied using this technique. However, the difference between the planetary and stellar rest frame velocity is much smaller for temperate planets, pushing the planetary and stellar signatures in time-series of spectra closer together, see Fig.\ref{fig:example_restframes}. If the surface of the host star could be expected not to change during ANDES observations, this would be unproblematic. However, especially the low-mass stars that ANDES will target for temperate-planet observations display stellar activity features even at old ages, which manifest themselves as changing spots and faculae as well as multitudes of small and sometimes also larger flares \citep{2005LRSP....2....8B, 2011ApJ...743...48W, 2020AJ....159...60G,Pietras2022}. In fact, a recent analysis of spectral lines in low-mass stars showed that most of those lines show at least a minor response to changes in stellar activity \citep{2023A&A...674A..61L}. For ANDES this means that stellar activity changes can be expected to produce a non-negligible signal in the stellar rest frame, which needs to be quantified and separated from the planetary signal in the adjacent planetary rest frame.

The most sensitive tracers of stellar activity accessible to ANDES are the Ca\,II H and K lines at 3933 and 3968\,\AA, i.e.\ in the ANDES U-band. We illustrate this in Fig.\ref{fig:activity_tracers_flare} by comparing the spectral response of the Ca II K, H$\alpha$, and Ca II 8542~\AA\ lines to a solar flare with an energy of $\approx 10^{32}$ erg as observed by the Swedish~1-m~Solar~Telescope \citep[SST, ][]{Scharmer03} and described in \cite{2023arXiv230903373P}, with quiet Sun observations from \cite{2023A&A...671A.130P}. The former is an average taken over an area of approximately 2x2 arcseconds inside of the lower brightened area marked in the first panel of the figure. Additionally, a scaled flare response is generated using the NESSI code \citep{2023arXiv231106200P} where the same profile is assumed to take up 1\% of the solar area, in this case the Ca II K line is the only line where a response is clearly visible. We note that in the stellar context, where one would observe the spectrum of the full stellar disk, this would correspond to a rather weak flare. Average observed flare energies of cool stars tend to be an order of magnitude more energetic \citep[e.g. ][]{Maehara15,Pietras2022}, however the ratios between the three aforementioned lines are expected to stay largely the same.


In the stellar context, several studies have shown that the sensitivity of Ca\,II H\&K to even small activity changes cannot be matched by other chromospherically sensitive lines such as H$\alpha$, Ca~IRT or the Na~D lines \citep{2017A&A...605A.113M, 2022AN....34323996D, 2023A&A...670A..71F}. Additionally, it has been shown that these lines are uniquely suitable for deriving the filling-factor and distribution of different types of active regions \citep{2023MNRAS.tmp.3174C}. Having strictly simultaneous access to CaII H\&K data with ANDES will strongly help isolate planetary absorption features and benefit the core exoplanetary science case.

In addition, the U-band will enable additional exoplanet science cases:

Due to Rayleigh scattering, clear exoplanet atmospheres will produce an upward slope towards the blue end of the spectrum. Techniques like the chromatic Rossiter-McLaughlin effect \citep{2015A&A...580A..84D} or chromatic Doppler Tomography \citep{2022A&A...657A..23E} can be used to test for such blue skies with data from the ANDES U-band. The presence of clear skies can be used as a consistency check for the interpretation of biosignature detections; most biosignatures should not produce a detectable signal in cases of cloudy/hazy skies.

Ozone, which is a possible biosignature in exoplanet atmospheres \citep{2018AsBio..18..630M}, produces a broad absorption band that intensifies sharply towards near-ultraviolet wavelengths. In the ANDES U-band, the wavelength region shorter than 3900\,\AA\ is the most relevant one for this. Since ozone does not produce sharp absorption features here but rather a broad absorption band, a possible detection method with ANDES is the chromatic Rossiter-McLaughlin effect or chromatic Doppler Tomography.

Lava worlds, i.e.,\ hot rocky exoplanets, may produce absorption in the CaII H\&K lines themselves. The expected mechanism is exospheres absorbing at these wavelengths, for example, due to sputtered species arising from the rocky surface, or material degassed from a magma ocean\citep{2022SSRv..218...15L}. So far, searches for this effect have mainly focused on 55~Cnc~e, but have not yet been successful \citep{2016A&A...593A.129R}. ANDES U-band spectra would efficiently probe this scenario.

\section{Additional Exoplanet Science Cases}

\subsection{Radial Velocity Mass Measurements}

\begin{figure*}
\centering
\includegraphics[width=0.99\linewidth]{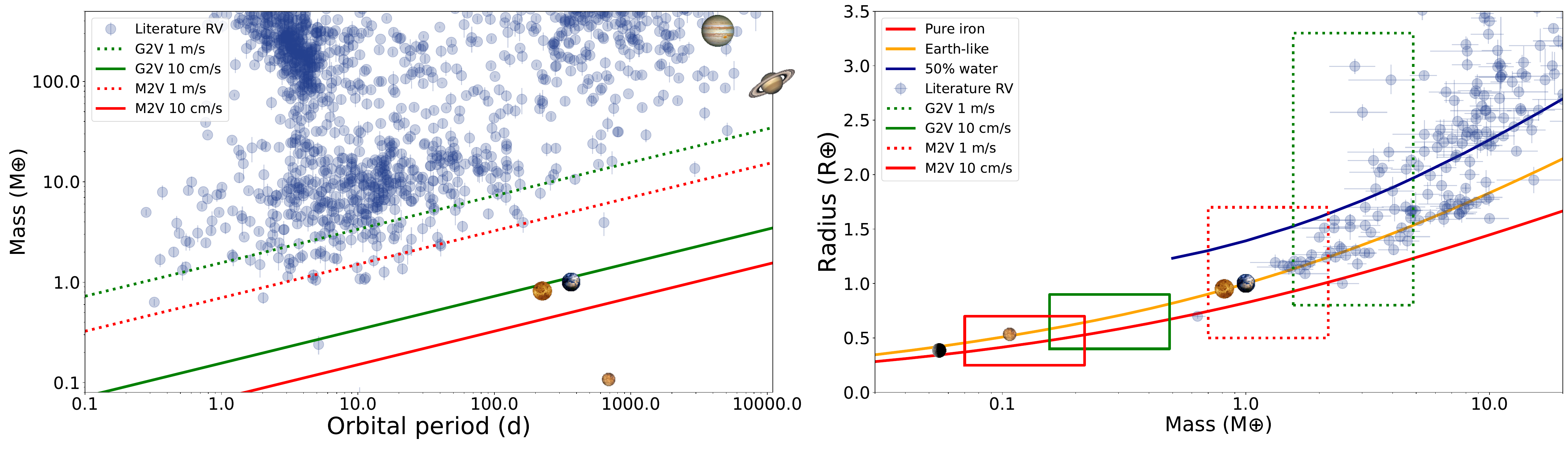}
\caption{\label{fig:rvsection} Left panel: shows the distribution of known exoplanets with masses measured via RV, along with some of the solar system's planets. The dotted and thick lines show the detection limits assuming 1 m/s and 10 cm/s precision for solar and M2-type stars, respectively. Right panel: shows the mass-radius diagram of known low-mass planets with masses and radii measured with uncertainties lower than 33\%. The dotted and thick squares show the regions of the parameter-space accessible assuming 1\,m\,s$^{-1}$ and 10\,cm\,s$^{-1}$ precision, respectively.}
\end{figure*}

One of the scientific objectives of ANDES is to detect and characterize exoplanets by measuring their induced radial velocity signals. ANDES will achieve a minimum radial velocity precision of 1\,m\,s$^{-1}$, with a goal of 10\,cm\,s$^{-1}$. A precision of 1\,m\,s$^{-1}$ is sufficient to detect the signals of super-Earths and mini-Neptunes orbiting Sun-like stars and Earth-mass planets orbiting the habitable zone of low-mass stars \citep{Anglada2016}.  Reaching the goal of 10\,cm\,s$^{-1}$ would make it capable of detecting the signals created by Earth-mass planets orbiting in the habitable zones of Sun-like stars or Mars-like planets orbiting around low-mass stars \citep{Pepe2021}. 

Figure~\ref{fig:rvsection}, left panel, shows the expected detection limits of ANDES for the case of a G2-type star (1\,M$_{\odot}$) and an M2-type star (0.3\,M$_{\odot}$). With 1\,m\,s$^{-1}$ precision, ANDES will be able to detect planets at the low end of the mass distribution of known exoplanets. A precision of 10\,cm\,s$^{-1}$ would make it capable of investigating a still unexplored region of the parameter space. 
 
It is known that by improving the determination of the masses of the exoplanets, it is possible to constrain their interior structure and formation history \citep{Luque2022}. Currently, the precision of the masses of many transiting exoplanets is limited by the RV precision of the data, which is limited by the apparent brightness of the host stars. Paired with the massive collecting area of the ELT, ANDES will be able to measure the masses of transiting exoplanets with enough precision to constrain their interior structure, even those orbiting faint stars. It will be capable of measuring the masses of Earth-like exoplanet candidates detected by space missions such as PLATO \citep{Rauer2014}, which are expected to be located in systems at a moderate distance to the Sun. ANDES will be able to confirm candidate signals of Earth-like exoplanets detected by ground-based surveys such as the Terra Hunting Experiment \citep{Thompson2016} by exploiting its superior data quality. It will also be capable of characterizing the population of low-mass exoplanets of the solar neighborhood, even those orbiting very faint stars. 

With a precision of 1\,m\,s$^{-1}$, ANDES will be able to fully populate the current low-mass regime by studying fainter targets. Scaling from the capabilities of ESPRESSO@VLT \citep{Pepe2021}, ANDES@ELT will be able to reach 1\,m\,s$^{-1}$ precision in solar-type stars up to $V=15$\,mag and in M-dwarfs up to $V=16$\,mag\footnote{According to the SN relationships presented in \citep{Pepe2021}, and scaling by the collected area of the ELT. }. This difference already provides more than 200 known candidate systems in which ESPRESSO cannot achieve a precision of 1\,m\,s$^{-1}$, including 24 systems hosting Earth-size planets\footnote{Based on data of the Nasa exoplanet archive; 2023/09/28}. With a precision of 10\,cm\,s$^{-1}$, ANDES would be able to explore an empty region of the parameter space up to magnitude $V=10$\,mag for solar-type stars and $V=10.8$\,mag for M-dwarfs (Figure~\ref{fig:rvsection}, right panel). Currently, there are 28 known systems for which ANDES would be the only instrument that could achieve a precision of 10\,cm\,s$^{-1}$.

\subsection{Stellar pulsations}

The physical properties of the exoplanets' host stars are fundamental to understanding their nature, origin, and evolution. To start with, planetary radii derived from transit observations are expressed in terms of the host star radius, while planetary masses derived from radial velocity measurements depend on the host star's mass. In addition to mass and radius, the abundance of C, O, and rock-forming elements like Mg, Si, and Fe determines the mineralogy, structure, geodynamics, and potential habitability of rocky-like planets. Furthermore, models of planet formation have inferred that the frequency of certain types of planets does show a dependence on the host star's metallicity and mass. Also, planetary periods and eccentricities might be related to the presence of heavy elements in the stellar host.

This is especially critical for M dwarfs \citep[e.g.][]{2015A&A...577A.132M}. First, they are faint in the optical, so high signal-to-noise spectra tend to be limited by telescope size. Furthermore, the temperatures of M dwarf atmospheres are cool enough to form diatomic and triatomic molecules that create thousands of spectral lines that are poorly known and, moreover, many of them blend with each other. Therefore, many of the spectral lines traditionally used in the spectral analysis of solar-type stars are blended or not present, while the spectral synthesis of M dwarfs suffers from the incomplete knowledge of many molecular data as well as from the accuracy of the atmospheric models used.

Asteroseismology, which makes use of the radial and non-radial pulsations of stars permeating throughout the stellar atmosphere, is a powerful tool to provide independent constraints on the fundamental parameters of the stars along with the internal structure. Asteroseismology is based on comparing of patterns of observed oscillation frequencies with theoretical predictions based on stellar evolution models. Many applications of asteroseismology of relevance to exoplanet host stars have been reported. For example, using 29 months of Kepler data, \cite{2014ApJ...781...67F} yield the radius of Kepler-10\,b with an unprecedented uncertainty of only 125\,km. The authors also confirm that Kepler-10 was (at that moment) the oldest known rocky-planet-harbouring system (10.41\,Gyr). If the planet transits, asteroseismology gives the orbital eccentricity \citep[e.g.,\ Kepler-419][]{2014ApJ...791...89D}. Furthermore, the stellar rotation axis derived from asteroseismology can be combined with the inclination of the planetary orbit plane to yield the host star’s obliquity directly \citep[e.g.\ Kepler 56; the first spin-orbit misalignment in a multiplanet system,][]{2013Sci...342..331H}. See \cite{2018ASSP...49..119H} for more examples and a recent review.

Stellar oscillations appear all across the Hertzsprung-Russell diagram. Theoretical studies \citep{2012MNRAS.419L..44R,2014MNRAS.438.2371R} predict that low-mass M-dwarf stars have the potential to pulsate. These studies predict short periodicities ranging from 20 minutes up to 3 hours and empirically estimated amplitudes of just a few $\mu$mag or a few tens of cm\,s$^{-1}$. So far, the analysis of HARPS data has shown that no signals below 0.5\,m\,s$^{-1}$ can be detected with a confidence level better than 90\% on M dwarfs \citep[see][]{2017MNRAS.469.4268B}. Nonetheless, \cite{2017MNRAS.469.4268B} report some power excess for the two most long-term stable M dwarfs in their sample, one of them with a 2 hours period and an amplitude as low as 0.36\,m\,s$^{-1}$, compatible with the theoretical prediction for low-radial, low-order g-modes. \cite{2019hsax.conf..266B} acquired more observations for the pulsating candidate but could not confirm the oscillations. 

The better sensitivity and precision offered by ANDES will allow a better sampling of the predicted short periods and reduce the noise level, thus increasing the chances of detecting the stellar pulsations in M stars for the very first time. Since the stellar pulsation signatures are like a fingerprint of the star, the combination of optical and near-infrared spectra will allow a complete characterization of the pulsation modes and strengths and, therefore, a more detailed analysis of the internal structure and physics of M-stars. Based on the results by \cite{2017MNRAS.469.4268B}, and considering the lower limit for the expected pulsation frequencies (3 hours), we estimate that 50 measurements per star should be enough for the detection of these modes with an SNR of 5. Assuming a total integration time per target of 180 seconds and setting a V magnitude limit of 11, we estimate that a sample of 30--65 M-dwarfs can be targeted in 100--200 hours.  

It is also worth noticing the stellar atmospheric studies that can be done at high-resolution with ANDES, which is important for assessing if metallicity correlations are present at the stellar/sub-stellar boundary. For this work, the near-infrared band is particularly useful. For example, \cite{2012ApJ...748...93R} used the equivalent width of the Na~I can Ca~I triplet and the H$_{\rm 2}$0-K2 index in the K-band of high-resolution spectra for calculating stellar basic parameters and metallicities. This technique has been widely extended to provide calibrations and near-infrared indexes to characterize large samples of M dwarfs. In this way, we can move forward with modeling the atmospheres where the transition from a hydrogen-fusing object to a dusty object is difficult to model accurately.

\subsection{RM planet characterization/Obliquity measurements}

Spectra taken before, during, and after transit can be used to probe not only the constituents of and physical conditions in extrasolar planet atmospheres but also to measure the projection of the spin-orbit angle on the sky plane, i.e.,\ the projected obliquity of the host star. When a transiting planet hides a portion of the stellar disk, the corresponding radial-velocity component is diminished in the disk-integrated stellar spectrum, leading to line-profile distortions known as the Rossiter–McLaughlin (RM) effect (see \cite{Albrecht2022} and references therein). While the Sun's obliquity is low ($\sim 7^\circ$\ relative to the ecliptic) this is not necessarily true for extrasolar planet systems. See, for example, the TEPCAT catalogue\footnote{\url{https://www.astro.keele.ac.uk/jkt/tepcat/}} for an up-to-date listing. Spin-orbit misalignment can be created during the epoch of planet formation as well as later during the main-sequence lifetime of the system. Therefore, the degree of alignment between the stellar spin and orbital spin contains important information about the system's past history. Unfortunately, no RM measurements have been conducted for Earth-sized planets orbiting solar-like stars as these are out of reach of current instruments \footnote{The RM effect in the Trappist system has been measured.  Asteroseismic measurements have led to inclination measurements of solar-like stars with Earth-sized planets in two systems (Kepler-50 and Kepler-65)}. This leaves a major gap in our knowledge. ANDES will be able to determine the obliquities of just such systems. There is an additional advantage for transiting Earth-mass planets on long-period orbits around solar-like stars. Their RM amplitude is larger - of the order of 30\,cm\,s$^{-1}$ - relative to the radial velocity reflex semi-amplitude, 9\,cm\,s$^{-1}$ for Earth \citep{Gaudi2007}. The signal occurs also on a timescale of a few hours and not months or years. Therefore, confirming such a transiting planet from a transit survey might be more readily achievable through measurements of the RM effect. Although the host stars' projected obliquity will be known, no mass estimate of the planet is gained.

In summary, the RM effect can be observed and analyzed for "free" if transmission spectroscopy is conducted in a transiting exoplanet system. The quantity measured, projected stellar obliquity, is unknown for the types of systems ANDES will target. For other classes of exoplanet systems spin-orbit misalignments have defied theoretical expectations and have been proven useful in understanding the histories of these systems. In addition, for low-mass planets on long-period orbits, confirmation of transiting exoplanet candidates through the RM effect will be more readily achievable than via traditional radial velocity measurements obtained throughout the orbit.

\subsection{The Chromatic Rossiter-Mclaughlin effect}

The Chromatic Rossiter-Mclaughlin (CRM) effect can be exploited to gain additional atmospheric information on exoplanet atmospheres from high-resolution spectra which normally are best accessible through analysis of additional low-resolution spectra \citep[e.g.,][]{Snellen2004,Dreizler2009,Santos2020,Borsa2016,Esparza-Borges2022}. Its measurement principle relies on the amplitude of the RM effect at different wavelengths and depends on the square of the star-to-planet radius ratio at the respective wavelength. Determining the amplitudes in the different wavelength regions allows us to measure the apparent star-to-planet radii ratios over the complete wavelength range of high-resolution echelle spectra. By measuring the CRM, we will obtain access to low-res features (e.g., Rayleigh scattering, clouds) that probe different atmospheric pressure ranges and are inaccessible to high-res transmission spectroscopy \citep{Yan2015}.

It comes with several disadvantages, chiefly only planetary atmospheres in wavelength regions where stellar lines are present can be probed. Because the signal in the stellar lines is analyzed, the Signal-to-Noise ratio (SNR) is reduced relative to the case of standard low-resolution observations. Some advantages make an analysis of the CRM in particular systems desirable. The same spectrum as the high-resolution atmospheric study can be used; no additional observations are required. Therefore, narrow and broad spectral features are measured simultaneously. Its dependence on astrophysical stellar noise differs from that of low-resolution observations. For faster-rotating stars, its SNR decreases linearly with planet radius instead of the square of this quantity, potentially offsetting some of the disadvantages in specific systems \citep[e.g.,][]{Albrecht2022}.



\subsection{Stellar science cases addressed by exoplanet observations}

Questions such as the nature of the stellar dynamo and stellar surface features are actively investigated in ANDES' working group 2 (see respective White Paper). Observations of exoplanet transits yield additional information about the host star that can be used to address those stellar science goals and do not necessarily require separate observations. Instead, relevant stellar observables can be extracted from the same observational data collected for exoplanet transmission or emission spectra. 

One example is the differential rotation of cool stars with convective envelopes, which is an important parameter in stellar dynamo studies. In cases where the exoplanetary orbit is even only slightly misaligned with the stellar rotation axis, transit observations sample different latitudes on the host star. The collected spectra can be used to reconstruct the spectrum of the occulted patch of the stellar surface \citep{2016A&A...588A.127C}, yielding sensitive measurements of stellar differential rotation and stellar limb darkening \citep[e.g. ][]{2015A&A...573A..90M,2023arXiv231006681C,2023A&A...674A.174C,2023MNRAS.522.4499D,2023A&A...673A..71R} or stellar granulation \citep{2017A&A...597A..94C}.

Another illustration pertains to the spectral line profile linked to the granulation pattern, which, in turn, significantly relies on the stellar fundamental parameters. Each spectral line has unique finger-prints in the spectrum that depend on line strength, depth, shift, width, and asymmetry across the granulation pattern depending on their height of formation and sensitivity to the atmospheric conditions \citep{2000A&A...359..743A}. In this context, it is possible to measure and characterize the granulation pattern of the hosting star performing (spatially-) resolved spectroscopy during planet transits \citep{2017A&A...605A..91D,2017A&A...605A..90D,2018A&A...616A.144D,2019A&A...631A.100C,2021A&A...649A..17D}.

Other examples of stellar science cases being addressed by exoplanet observations with ANDES are starspot properties encoded in spot occultations during transits \citep[e.g. ][]{2021arXiv210406173B} or more generally stellar magnetic field \citep[e.g. ][]{2019ApJ...879...55C}, short-term stellar variability \citep[e.g. ][]{2023A&A...670A..24S}, and serendipitous observations of stellar flares, particularly in the case of M dwarf stars for which 
 ANDES will be crucial in terms of shaping the behavior of the complex observed atmospheres \citep[e.g. ][]{2018A&A...615A..14F,2020MNRAS.499.5047M} and determine the impact of flares on the chemical composition of transmission spectra \citep{2022A&A...667A..15K}.


\section{Solar System Science with ANDES}

\subsection{Doppler velocimetry technique in planetary atmospheres}

The Doppler effect measurement has been previously used to derive wind estimates in the atmospheres of Venus \citep{Machado:2012,Goncalves:2020}, Mars \citep{Machado:2020}, Jupiter \citep{Goncalves:2019,Silva:2022} or Saturn \citep{Machado:2022}. 

The technique consists in measuring the Doppler shift of solar lines reflected by the planetary disk \citep{Gaulme:2018} and using it to directly derive the zonal wind across a large range of latitudes and local times, in the dayside. The Doppler velocimetry method, used in the previously referred works, takes advantage of the backscattered solar radiation at the planetary atmosphere. Doppler wind speeds can be retrieved from observing stellar lines (e.g., Fraunhofer lines) at very high spectral resolution. This method was previously applied for determining radial velocities of stars \citep{Baranne:1996} and in asteroseismology \citep{Martic:1999}.

The technique was also used in recent planetary measurements in the visible range \citep{Civeit:2005}. The measurements of the wind velocities in planetary atmospheres are retrieved using an adaptaded form of the absolute accelerometry (AA) technique (Connes, 1985), applied to high-resolution spectra obtained at the Very Large Telescope (ESO). Its applicability to planetary atmospheric winds retrieval has also been demonstrated for the case of Venus \citep{Machado:2012,Goncalves:2020}, Mars \citep{Machado:2020}, Jupiter \citep{Goncalves:2019,Silva:2022}, Saturn \citep{Machado:2022} and even for Titan atmosphere's dynamical studies in the scope of an exploratory study \citep{Luz:2005,Luz:2006}.

The method is based on an optimum correlation between a measured spectrum and a reference spectrum. A detailed analysis of the particular method used for outer planets has been made by \citet{Civeit:2005}. Instead of determining the Doppler shifts on individual lines the AA technique takes the ensemble of lines and achieves a statistical precision in the velocity determination which varies inversely with the spectral line density of the source. Hence, observations at solar wavelengths with a high resolution spectrometer are particularly suited to this technique, since the solar spectrum scattered off by the planetary atmospheres carries the signatures of some 4000 Fraunhofer solar absorption lines in the wavelength domain we use, yielding a theoretical precision of a few m$s^{-1}$. 
In a Titan's atmosphere dynamical study, as a consequence of the observations’ geometry and of the moon's rotation, the line of sight Doppler shift at each pixel on the image of the planetary disk is affected by the combination of the relative orbital motion, the planetary motion and the cloud particles’ motion.

In the AA technique the velocity information is extracted from the variation of the spectral intensity in a given measurement relative to an idealized spectral intensity, which can be an average of many observations, or even a model. 
Let us denote I ($\lambda$)  and $I_{0}$($\lambda$) respectively the Doppler-shifted (measured) and the idealized spectral intensities of the target at a given wavelength. Then, assuming Doppler shifts smaller than the typical line width, combining the first order expansion $I_{0}$ - I $\simeq$ $\delta$$\lambda$$\partial$$I_{0}$/$\partial$$\lambda$ and $\delta$$V_{m}$/c = $\delta$$\lambda$/$\lambda$, yields -$\delta$$V_{m}$($\lambda$)/c = (I-$I_{0}$)/($\lambda$$\partial$$I_{0}$$/$$\partial$$\lambda$), where $\delta$$V_{m}$ is the measured velocity signal projected on the line of sight. Computing a weighted average over a given spectral domain yields

\begin{equation}
  \Delta V =  \frac{\int{\frac{\delta V_{m}}{\sigma^2(\delta V_{m})}}d\lambda}{\int{\frac{1}{\sigma^2(\delta V_{m})}}d\lambda}
\end{equation}

where $\sigma^{2}$($\delta$V$_{m}$($\lambda$))  is the variance of the velocity measurement at wavelength $\lambda$ . The rule for optimum weighting is that each weight be proportional to the inverse square of the RMS error. Under the assumption of pure photon noise $\sigma^{2}$(I($\lambda$)) = I ($\lambda$) with:

\begin{equation}
 - \frac{\Delta V}{c} =  \frac{\int{(I(\lambda)- I_{0}(\lambda))}M(\lambda)d\lambda}{\int{\lambda \frac{\partial I_{0}}{\partial \lambda}} M(\lambda) d\lambda}
\end{equation}

Where M($\lambda$) = ($\lambda$/I)$\partial$I$_{0}$/$\partial$$\lambda$ is called the mask function \citep{Connes:1985}. The mask function can be interpreted as a weighting function which amplifies the differences between the reference and the shifted spectrum within absorption lines, and nullifies them where the spectrum is flat. The technique takes into consideration the whole spectrum, independently of individual spectral lines, and taking the variance it can be shown that the standard deviation is inversely proportional to the packing of lines in the spectrum.
The constant Doppler shift arising from the relative motion of the planet and the observer, and systematic shifts from instrumental effects, are eliminated by considering the differential measurements between a spectrum taken at a given point of the disk and the reference spectrum, taken at the center of the disk where the phase angle is null, which are both observed almost simultaneously.

ANDES's combined high spectral and spatial resolutions in IFU mode will allow us to push the Doppler velocimetry technique at unprecedented levels to derive wind maps at very high spatial resolution. This is, however, at the cost of being able to close the Adaptive Optics loop, which requires targeting planetary objects with a relatively small angular size, such as Titan. The typical angular diameter of Titan is 0.8". With a diffraction limit (at 2~$\lambda$/D, 0.75\,$\mu$m) of ANDES at about 8\,mas, we expect to reach a typical horizontal resolution of 50\,km~$\times$~50\,km. Using the radial velocity photon noise limits \citep{Reiners:2020}\footnote{We used the online tool \url{http://www.astro.physik.uni-goettingen.de/research/rvprecision/}}, we evaluate that a 1~hour integration time is sufficient to reach a 2\,m\,s$^{-1}$ precision per element of resolution for Titan. However, this approach neglects the effect of instrument stability, stellar activity, calibration, and others that can significantly increase noise level \citep{Goncalves:2019} and deserve to be quantified in detail.

Interestingly, Titan has a forest of strong CH$_4$ lines from the visible (as seen in previous ESPRESSO observations of Titan; Turbet et al., in preparation) to the infrared \citep{Campargue:2012}; these lines could be used to improve further the SNR of wind measurements (in addition to using stellar lines), as well as to extend wind measurements at altitudes above the haze layer. However, doing so requires an experimental effort to accurately constrain CH$_4$ line parameters in the visible (see an example in \citealt{Campargue:2023}).

Moreover, high-resolution spectroscopy presents a unique opportunity to observe a planetary target with a CH$_4$-rich atmosphere, from which CH$_4$ optical proprieties can be studied and retrieve related isotopic ratios. It also showcases the use of a close planetary target to test new methods for chemical retrieval of minor atmospheric compounds (some of them with relevant astrobiological interest as shown in R. Silva et al., in revision in Planetary and Space Science), in preparation for upcoming studies of cold terrestrial exoplanet atmospheres.

\subsection{Probing tenuous planetary atmospheres}

ANDES high spectral resolution and sensitivity will make it a powerful instrument to study tenuous planetary atmospheres, where due to the low pressures ($\mu$bar to nbar range), line profiles are defined by thermal broadening, yielding intrinsic FWHM of $\sim$0.01 cm$^{-1}$ at 2 $\mu$m. Such atmospheres include the sublimation-driven atmospheres of Pluto, Triton and Io, the latter including also a volcanic contribution. A common feature is the existence of an orbital pressure cycle, related to changing heliocentric distance and/or subsolar latitude. On Pluto and Triton, the pressure seasonal cycle is best monitored through stellar occultations \citep[e.g.,][]{2021ApJ...923L..31S}, but infrared spectroscopy can also contribute by measuring the N$_2$ ice temperature \citep{2018Icar..314..274M}, which determines the total pressure. Moreover, CH$_4$ and CO are present at the surface and in the atmosphere. In particular, Pluto's northern latitudes are covered by methane deposits, which are a major source of atmospheric methane \citep{2019Icar..329..148B}. But how such deposits will react as this region will reach Northern Solstice by 2029 is unknown and global climate model predictions can be tested against temporally-sampled and spatially-resolved (if the AO loop can be closed on a 0.1" target) observations with ANDES. VLT/CRIRES observations \citep{2015Icar..246..268L} have demonstrated the possibility to monitor Pluto's atmospheric methane from ground-based IR spectroscopy but the ability to map it on Pluto's and Triton's disks will be unique to ANDES.

Other Transneptunian objects (TNOs) such as Quaoar, Gongdong, Makemake, and Eris are known to have volatile-covered surfaces \citep[N$_2$, CH$_4$, see review in][]{2020tnss.book..109B}, and recent JWST observations have shown that CO$_2$ and CO are also common on the surfaces of Kuiper Belt objects \citep{2023PSJ.....4..130B}. This is generally consistent with volatile escape models \citep[e.g.][]{2011ApJ...738L..26B, 2020tnss.book..127Y} that show that large KBO are able to retain their volatiles over the age of the solar system. This leaves the possibility that other TNOs besides Pluto and Triton may exhibit atmospheres at least along some parts of their orbits. From an observational perspective, the target object should be warm enough when observed -- typically, a 21 K (resp. 29 K) N$_2$ (resp. CH$_4$) atmosphere is at the transition between collisional and ballistic -- and global atmospheres are more promising for detection than atmospheres restricted to the subsolar region. Stellar occultations (e.g. \citet{2012Natur.491..566O} for Makemake) provide complementary searches for atmospheres, but by nature can constrain them only near the terminator, a priori not the most favourable region.

Although most of the knowledge of Io's nanobar-class atmosphere -- see review in \citet{2023ASSL..468..233D} -- is based on UV and sub-mm observations, it is also accessible to near-IR spectroscopy, with the spatially-resolved detection of SO$_2$ gas at 4\ $\mu$m in sunlight and of SO emission at 1.7 $\mu$m in Jupiter eclipse. The latter has been known for over 20 years but its interpretation remains uncertain. The most plausible scenario \citep{2023JGRE..12807872D} is that it represents emission of hot SO gas directly ejected from volcanic vents at a high quenching temperature. However, studies of the correlation between this emission and the distribution of hot spots have given contradictory results, and the overall spectral shape (so far restricted to the disk-average spectrum due to S/N limitations) is still defying modelling. ANDES observations combining both high spectral and high spatial resolution should shed light on the origin of this emission. Finally, plume atmospheres, such as Enceladus', recently observed by JWST \citep{2023NatAs...7.1056V}, and Europa's elusive one \citep{2023Sci...381.1305V}, may also be explored with ANDES. Although the major gas H$_2$O (and CO$_2$) will be inaccessible from the ground, other species diagnostic of sub-surface ocean chemical conditions, e.g. CO, CH$_4$, C$_2$H$_6$, CH$_3$OH, may be searched for.

\subsection{Isotopic ratios in comets}

Isotopic ratios represent a powerful tracer of the origin of planetary bodies, especially for small bodies, that suffered less chemical evolution during their lifetime, compared to planets. The comets are among the most primitive materials in the solar system and can provide unique constraints on the composition of the pre-solar nebula. Their main advantage for measuring their chemical composition is the presence of a coma with species (molecules, radicals, ions, and atoms) in gas phase. It permits a detailed chemical analysis of the cometary ices by means of spectroscopy. In the spectral  range covered by ANDES many emission lines due to different radicals can be observed. The main source of information -- that implies a high signal-to-noise ratio -- coming from these emission lines is the isotopic ratios.

The two main isotopic ratios that can be measured in the spectral range covered by ANDES are $^{12}$C/$^{13}$C and  $^{14}$N/$^{15}$N that can be measured in the following species: CN (for both ratios, in the U band), NH$_2$ (for nitrogen, in the V band), C$_2$ (in the B band), CO$^+$ (for carbon, in the B band) and N$_2^+$ (in the U band). These last two species are not always detected in the inner coma. The results obtained so far, mainly with 8-m class telescopes \citep{wyckoff:2000, manfroid:2009, rousselot:2012, rousselot:2014, shinnaka:2016, opitom:2019, rousselot:2023}  indicate a $^{12}$C/$^{13}$C ratio similar to the terrestrial value of 89 (but with large errorbars) for C$_2$, CN and CO$^+$ (only one measurement in this ion, rarely observed in the inner coma) and a significant enrichment of $^{15}$N compared to $^{14}$N in both CN and NH$_2$ radicals compared to terrestrial value (by a factor of about 2) and solar bulk \citep{marty:2011}  and Jupiter ammonia \citep{fouchet:2004} (by a factor of about 3), only one lower limit being so far obtained for the $^{14}$N/$^{15}$N ratio for N$_2$ with ground-based observations. 

The significant improvement in the signal-to-noise ratio expected with ANDES at ELT, compared to UVES at VLT (the main spectrograph used so far for these observations) opens very interesting possibilities, both for comets belonging to our solar system and interstellar comets expected to be observed in the near future.  A significant reduction of the errorbars is expected, leading to more constraints on the origin of the comets (e.g. according to their dynamical type) and first measurements in interstellar comets -- never done so far -- will provide unique insight in extrasolar planetary systems.

\section{Synergies with other instruments}

As a next-generation instrument, ANDES will be preceded by a suite of ground- and space-based facilities for exoplanet science, namely the ELTs, JWST, PLATO, ARIEL, ALMA and SKAO, and will be a vital stepping stone on the way to the study of Earth analogues with the possible future great observatories under study like the Habitable Worlds Observatory (HabWorlds) and the Large Interferometer For Exoplanets (LIFE). It therefore plays a crucial role in further advancing our knowledge of exoplanet atmospheres through synergy and complementarity with other instruments, some of which we discuss below.

\subsection{Ground-based}
ANDES will follow on the exoplanet characterization efforts of the first generation of spectroscopic workhorses for the ELT, i.e. METIS \citep{Quanz2015}, HARMONI \citep{Houlle2021}, and MICADO \citep{Trippe2010}. For exoplanet atmospheric characterization at high spectral resolution, each has its own niche, complementary to ANDES. 

MICADO offers single slit ($0.016^{\prime\prime}\times3^{\prime\prime}$) mode, AO-assisted, high contrast spectroscopy at $R < 20,000$ for non-instantaneous wavelengths between $0.84-2.46 \mu$m. At similar wavelengths, HARMONI hosts an AO-assisted, image slicer IFU, with a high contrast mode covering the H \& K band (non-simultaneously) at $R=18,000$, with excellent spatial sampling (spaxel size) of 3.88 mas. Both have the potential to search for the reflected light spectra of nearby and widely-separated exoplanets. HARMONI's IFU in particular could observe Proxima b (separation $<37$ mas), though the design of its focal plane mask and apodizer may require some modification from its current design or use non-standard observing modes to enable this (Vaughan et al. submitted). Thanks to its $R=100,000$ resolution, the ANDES infrared IFU avoids the issue of stellar PSF saturation that necessitates the HARMONI focal plane mask. Thus, ANDES will be able to observe reflected light spectra of exoplanet atmospheres at much higher spectral resolution, leading to not only tighter abundance constraints, but more comprehensive chemical analyses of rocky exoplanets, thanks to its increased sensitivity to weaker lines from less abundant species in these wavelength regimes, including the key biosignature species oxygen.

METIS will offer an IFU in the L \& M bands ($3-5\mu$m) at $R=100,000$. It is the perfect complement to ANDES. METIS will be primarily sensitive to the thermal emission of temperate rocky worlds, providing detailed analyses of their composition and thermal structure \citep{Bowens2021}. Combined, ANDES and METIS will obtain a holistic, detailed view of these planets across the full optical-NIR regime, probing both reflected and thermal emission resulting in measurements of the planetary albedo, its heat budget and ultimately establishing the main parameters of its global climate. METIS and ANDES will also use their high spectral resolution to probe the exact shape of spectral lines in the atmospheres of gas giants, providing complementary wavelength information that has the potential to examine dynamic and chemical effects, particularly in disequilibrium, at different altitude/pressure levels in greater detail than ever before.

ANDES will also have significant synergies with METIS regarding proto-planetary disk science. METIS will be able to image and spectrally map the inner dusty disk structure, it will study the excitation and dynamics of CO fundamental emission in the inner gaseous disk, and it will allow a study of the gas chemistry in the hot region of the disk through observations of key species such as $H_2$, H$_2$O, and organic molecules. On the other hand, ANDES will provide dynamical information and full characterization in terms of physical parameters of the atomic component of the inner gas, it will study the accretion process and provide measurements of the magnetic field, and it will understand the influence of jets in the star-disk interaction.

In addition, observations of molecular lines at higher energies with respect to those covered by METIS will give important information for a correct interpretation of the excitation mechanism. The science performed by ANDES in the field of proto-planetary disks will be, therefore, highly complementary to that of METIS and the combination of observations from the two instruments will allow a complete picture of the inner disk region and its interaction with the still accreting star.

\subsection{Space-based}

JWST \citep{Gardner2006}, which is already a reality, and ARIEL \citep{Tinetti2018}, with a foreseen launch in 2029, are both expected to be operational within a similar time frame (early 2030's) as ANDES. Their combined observations are expected to yield crucial synergies, enabling the exploration of a wide parameter space of exoplanet atmospheres. JWST is a more powerful observatory than ARIEL for single observations, but ARIEL holds a great advantage which is that it covers the entire spectral region from 0.5 to 8 microns simultaneously and that it is a dedicated mission for exoplanet population statistics. Both missions together will measure global planetary atmospheric parameters, including temperature-pressure profiles, cloud coverage, and bulk molecular abundances of more than 1000 planets \citep{Edwards2022}. These space-based observations offer very complementary information as they can access parts of the electromagnetic spectrum that are inaccessible to ground-based observatories due to absorption and scattering in Earth's atmosphere. 

High contrast imaging instruments at ELT will examine a volume-limited population of young, non-transiting exoplanets formed with the same initial composition as the ARIEL and JWST sample but at considerably colder temperatures (beyond 100 AU). On the other hand, high-dispersion spectroscopy instruments are tailored to study the atmospheres of exoplanets overlapping with ARIEL's sample and those accesible to JWST.

With a resolving power of $R > 100,000$, ELTs can resolve atomic and molecular bands in exoplanet spectra into hundreds or thousands of individual lines. These signals can be combined for detection, offering astrophysical information over small wavelength scales that JWST or ARIEL won't be able to obtain. In particular ANDES will measure planet metallicities  \citep{Hoeijmakers2018_k9, CasasayasBarris2019, Yan2019}, inhomogeneities and dynamics in the planetary atmosphere through time-resolved transmission measurements \citep{Ehrenreich2020}, individual lines that can be used as tracers to understand evaporation and atmospheric evolution processes \citep{Yan2018, Nortmann2018}, planetary albedos through the detection of reflected light at different wavelengths \citep{Martins2018} and detailed information about the properties of stellar active regions, augmenting JWST and ARIEL's capability to mitigate stellar activity-induced noise in retrieved transmission spectra \citep{Rackham2018, Boldt2020}. More importantly, the high resolution spectroscopic observations can be combined with the bulk compositions, T-P profiles, and thermal inversions derived from ARIEL and JWST data in a single analysis framework \citep{Brogi2017}. These combined analysis of high and low dispersal data analysis will provide a holistic picture of planetary atmospheres. 

Finally, ANDES radial velocity measurements can be used to determine accurate planet masses for small planets or faint host stars, which might be crucial for some of the best earth-sized targets PLATO \citep{Rauer2014} might find, and also to provide information on planetary rotation and high-altitude wind speeds for the most favorable targets. Accurate mass measurements are necessary to interpret the atmospheric
characterization data, whether it is to understand the lack of an atmosphere or to perform retrievals on molecular features \citep{batalha2019precision, Changeat2019}.

\section{Conclusions}
\label{sec:Concl}

In this work we have presented detailed simulations of some of the ground-breaking exoplanet (and solar system) science that will be attainable with ANDES at the ELT. ANDEs will feature a seeing-limited spectrograph with a minimum spectral resolution of $R=100,000$, covering simultaneously the wavelength range of 0.5--1.8\,\textmu m, and with the goal for an extended coverage into the U and the K bands. It will also feature an AO-assisted mode with an IFU covering simultaneously the Y, J, and H bands.

The main objective of the ANDES project is to characterize the atmosphere of rocky planets in the habitable zone, mostly around M-dwarf stars. This is particularly relevant as the characterization of these accessible and abundant planets is crucial to understanding their habitability and determining if they could support life. ANDES will also focus on the detection of reflected light from the planetary atmosphere, allowing to explore their structure, composition, and surface conditions. This will provide an unprecedented detailed understanding beyond basic parameters such as mass, radius, and bulk composition. A golden sample of at least five nearby non-transiting earth-sized habitable zone planets will be accessible to ANDES for atmospheric characterization and search for biomarkers within a few nights. This would be a major scientific milestone for exoplanets  and astrobiology, an objective that no other currently approved astronomical facility will be able to reach. 

Furthermore, ANDES will also be dedicated to studying the initial conditions under which planets form and the physical mechanisms that lead to the different observed planet types populations and system architectures. This involves investigating the composition and spatial distribution of atomic and molecular gas in the inner regions of young circumstellar disks, which is needed for a better understanding of the initial formation and later evolution of planetary systems.

Since ANDES will be operational at the same time as NASA's JWST and ESA's ARIEL missions, it will provide enormous synergies in the characterization of planetary atmospheres at high and low spectral resolution. This will open a true golden era for full comprehensive understanding of the diversity of exoplanet atmospheres and how they form and evolve.

\bmhead{Acknowledgments}

EP and HP acknowledge financial support from the Spanish Ministry of Science and Innovation (MICINN) project PID2021-125627OB-C32 and PGC2018-098153-B-C31. JIGH and ASM acknowledge financial support from the Spanish Ministry of Science and Innovation (MICINN) project PID2020-117493GB-I00. FD thanks Hayley Beltz for valuable support with the WASP-76 b ANDES simulations. JLB acknowledges funding from the European Research Council (ERC) under the European Union’s Horizon 2020 research and innovation program under grant agreement No 805445. BLCM acknowledges continuous grants from the Brazilian funding agencies CNPq and Print/CAPES/UFRN. This study was financed in part by the Coordenaçao de Aperfeiçoamento de Pessoal de Nivel Superior - Brasil (CAPES) - Finance Code 001






\clearpage

\begin{appendices}



\section{ANDES AO-assisted performance modelling}
\label{App:one}

\subsection{ANDES AO expected performance}

The current ANDES-SCAO design includes a modulated pyramid WFS sensor inspired on the HARMONI-SCAO in what respect to the main system parameters (wavelength range, number of sub-apertures, and control strategy).
We obtained a first estimation of the achievable contrast, starting from the end-to-end AO simulations from \cite{Charlotte2022} performed for HARMONI-SCAO, and adding error contributions that had been neglected so far.
In particular, we summed up to the HARMONI-SCAO wavefront residuals the effects of: 1) $SP$: static petalling; 2) $DP$: dynamic petalling; 3) $J$: residual PSF jitter.
As petalling error we refer here to the differential piston between the 6 segments of ELT-M4.
Then, we accumulated 2000 instantaneous PSFs (corresponding to 4s of integration) obtained from these residuals and we computed the contrast achieved by the SCAO system. This approach provides an optimistic estimation of the performance because the mentioned errors  will impact the AO loop itself, increasing the AO residuals. Here, we are neglecting this aspect that will be taken into account in the near future with dedicated end-to-end simulations.

The contrast is considered as the ratio between the radial profile and the peak of the PSF. For the contrasts reported in Section~\ref{sec:aogeneral}, we considered: AO reference star $I\leq8$; median seeing conditions ($0.71asec$, "$L_0 = 50m$); $DP = 50nm$ RMS and  $J=100nm$ RMS, accounting for vibration and wind shake residuals.
About $SP$, we envisioned two possible scenarios: high piston - M4 is delivered to the instrument with $SP=200nm$ RMS and SCAO system is not able to correct for this static error that persists during the observation; low piston - $SP=25nm$ thanks to a good phasing performed by ELT before the handover to ANDES, or thanks to the SCAO capability to correct the petalling error down to this level.

The low piston scenario will also enable the use of a coronagraph. The simulations show that even a classic Lyot coronagraph will provide a contrast gain $\geq10$ in the region around $20mas$ at $1600nm$, reaching values $<1.0 \cdot 10^{-3}$.

\subsection{Planet observations simulations}

We have simulated ANDES-SCAO observations of known exoplanets to explore the target sample that can be probed in this mode. We base the simulation on the stellar and planetary data available in the “Planetary Systems Composite Data” table\footnote{\url{https://exoplanetarchive.ipac.caltech.edu/cgi-bin/TblView/nph-tblView?app=ExoTbls&config=PSCompPars}} in the NASA Exoplanet Archive\footnote{\url{https://exoplanetarchive.ipac.caltech.edu/index.html}}.
We note that a small subset of the systems must include some quantities required for the simulation. Instead of excluding these systems, we choose to replace these missing values either with sensible default values or with values based on the rest of the quantities.
\begin{itemize}
    \item For stars, some stellar radius, mass, or effective temperature estimates need to be included. We set the missing effective temperatures to 5000\,K, and then we fill in the missing mass and radius values using either mass-radius or temperature-mass-radius relations.
    \item For planets, the estimates for some of the orbital parameters, radii, or masses are missing. We default the missing eccentricities, arguments of periastron, and inclinations to $0, 90^\circ$, and $90^\circ$, respectively. Next, we calculate the missing semi-major axes using the planet's orbital period and stellar mass (and assume circular orbits). Finally, for planets with either a mass or radius estimate but not both, we set the missing value to match Jupiter's if the existing value indicates the planet to be Jupiter-sized.
\end{itemize}

We first calculate the maximum projected planet-star distance for each planet based on the planet's orbital parameters (semi-major axis, inclination, eccentricity, and argument of periastron), after which we use the maximum projected distance and the distance to the system to calculate the planet's maximum angular separation from its host star. Next, we calculate the planet-star radius and area ratios based on the planet and star radii, and then the planetary equilibrium temperature based on the stellar effective temperature and the planet's semi-major axis.

The planet-star flux ratios are calculated as a sum of the planet's reflected and emitted light, $F_\mathrm{r}$ and $F_\mathrm{e}$, respectively, relative to the flux from its host star,
\begin{equation}
    f = \frac{F_\mathrm{r} + F_\mathrm{e}}{F_\star}.
\end{equation}
The reflected flux ratio is calculated using the maximum projected planet-star distance, assuming an earth-like albedo of 0.3 and a phase angle of $\pi$/2. In contrast, the emitted flux ratio is calculated using the planet-star area ratio, planet equilibrium temperature, and the star's effective temperature, approximating both the planet and the star as black bodies.

We calculate the number of photons received by a resolution element per second as a function of magnitude using a first-order polynomial in log10(photons/s) fitted to photons/s estimates created by the ANDES ETC v1.1 for a compact source. 

We use the ETC to calculate the S/N ratios (SNRs) for two magnitudes (5 and 10) with a 6000\,s exposure time for each passband. The number of photons/s is obtained as SNR$^2$ / 6000, after which a polynomial is fit to the two $\log_{10}$(photons/s) estimates as a function of AB magnitude.\!\footnote{We also tested that the logarithm of the number of photons per second obtained using the ETC is linear in AB magnitude, as it should.} We set the exposure time to 6000\,s, the sky background to 30 counts, the telescope and optics temperature to 200\,K, and the read-out noise and dark current to 0 to ensure that $\approx100\%$ of the noise comes from photon noise. 
We convert the V, J, H, and K mags into the AB magnitude system\footnote{\url{https://en.wikipedia.org/wiki/AB_magnitude}} to calculate the number of photons received by a resolution element for each star. We first convert the magnitude into a flux density using 
\texttt{PyAstronomy.magToFluxDensity\_bessel98}. The method calculates the flux density in Jy, which we then convert to AB magnitude as
\begin{equation}
    m_\mathrm{AB} = -2.5 \log_{10} \left(\frac{f_\nu}{3621~\mathrm{Jy}}\right)
\end{equation}
After this, the AB magnitude can be used with the photons/second model to estimate the number of photons arriving at a resolution element in a given time.

We use the contrast curves for the low (750, 1000, 1600, 2200 nm) and high (1000, 1600, 2200 nm) piston scenarios discussed above and provided by Anne-Laure Cheffot and Enrico Pinna (priv. Comm.). The final SNR is calculated following \citet{Lovis2017} and \citet{Snellen2015}, as
\begin{equation}
    \mathrm{SNR} = f \sqrt{t_\mathrm{e} F_\star n_\mathrm{l} K},
\end{equation}
and the time to a given SNR as
\begin{equation}
    T_\mathrm{SNR} = \frac{\mathrm{SNR}^2}{f^2 F_\star n_\mathrm{l} K},
\end{equation}
where $f$ is the planet-star flux ratio, $t_\mathrm{e}$ is the exposure time, $F_\star$ is the flux from the star, $n_\mathrm{l}$ the number of lines (assumed here arbitrarily to be 1000, which is a typical number for other ground-based spectrographs), and K is 1/contrast. The SNR considers only photon noise and ignores other white noise sources (readout) and systematics (in particular telluric and stellar lines are assumed to be perfectly removed).

\begin{figure*}
  \centering
  \includegraphics[width=0.49\linewidth]{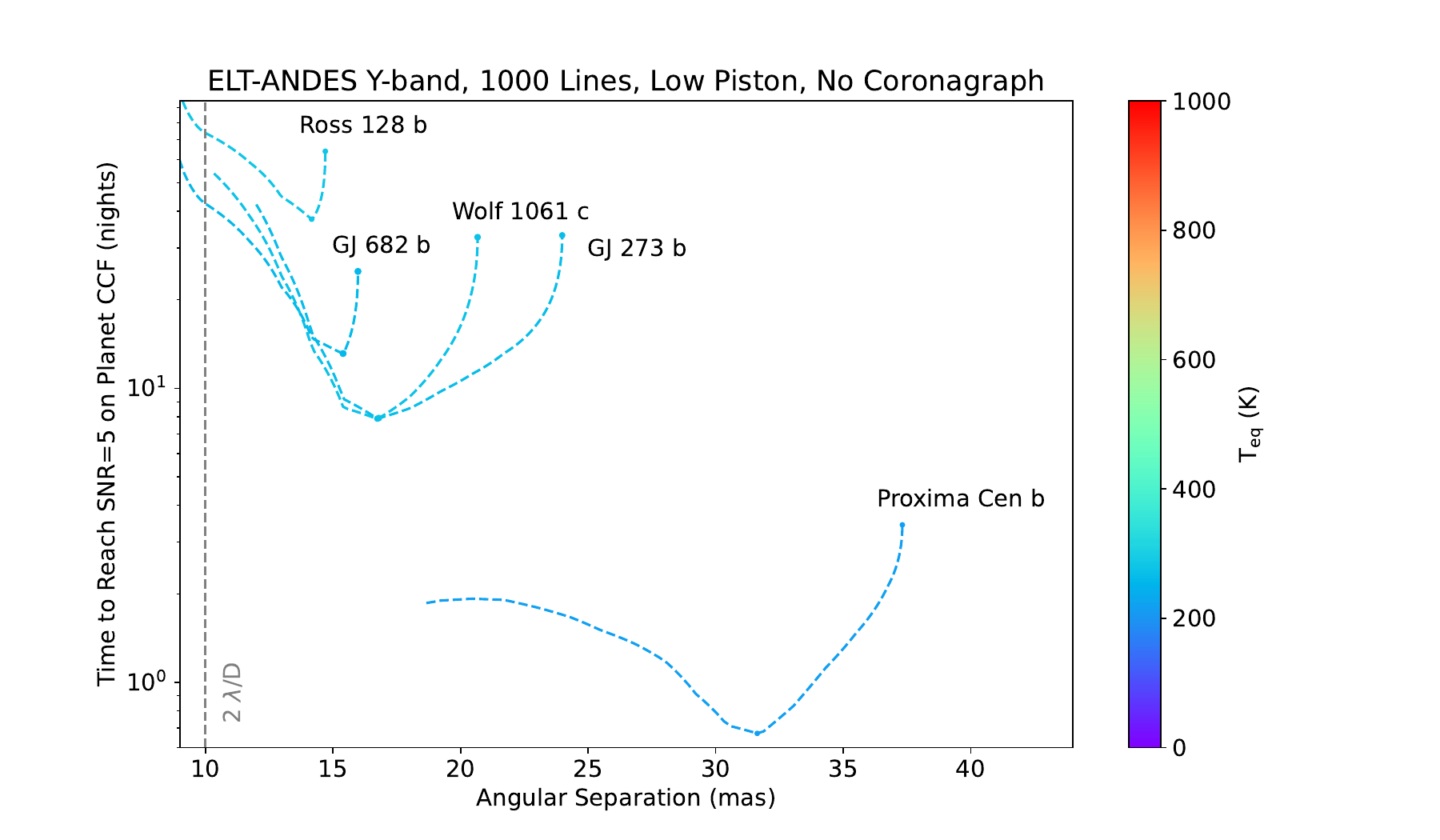}
  \includegraphics[width=0.49\linewidth]{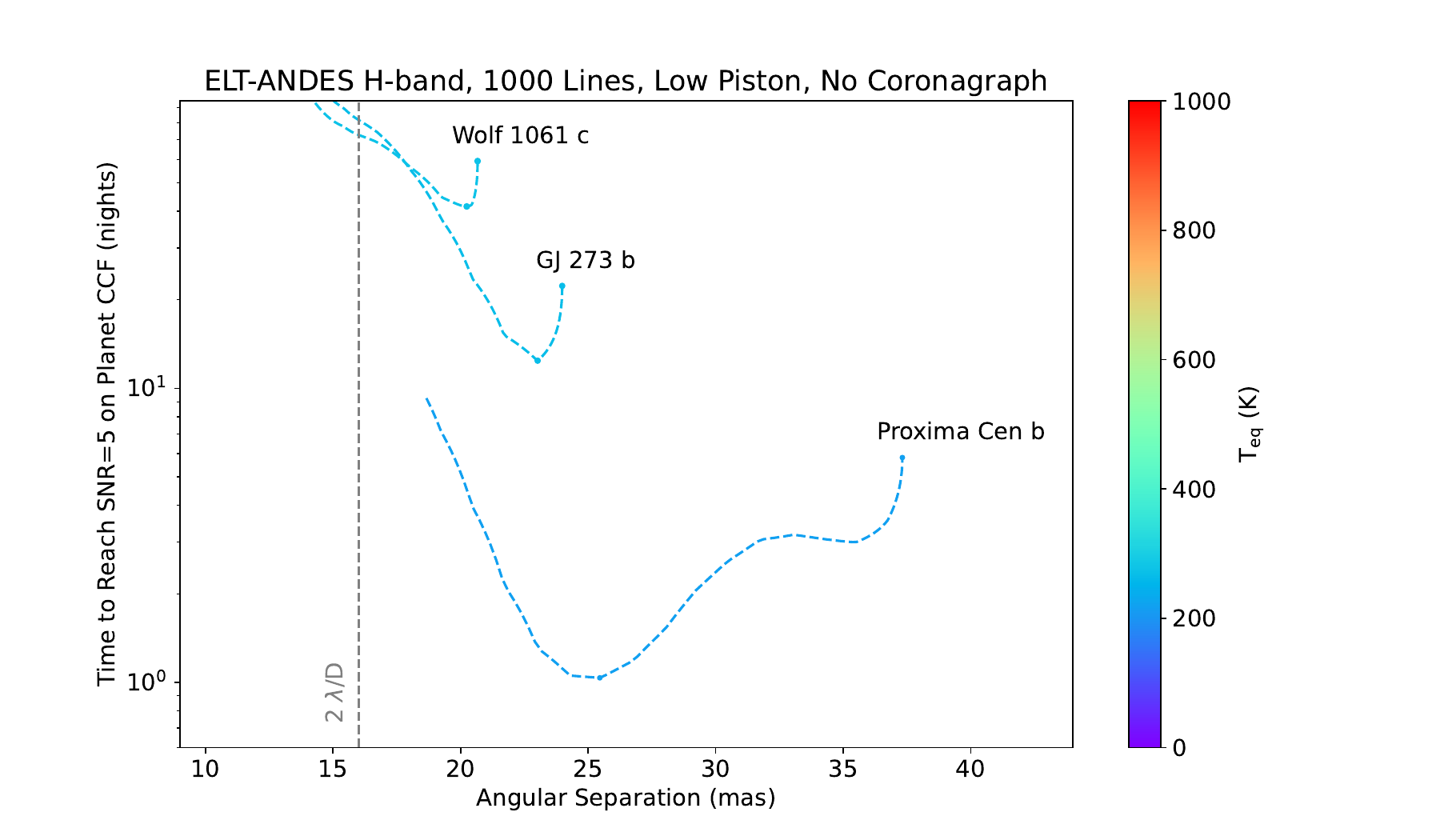}
  \caption{ Same as Figure~\ref{fig:lovis2}, but here the contrast curves assume a low-piston scenario and no coronagraph.}
  \label{fig:lovis_appendix}
\end{figure*}




\end{appendices}


\bibliography{sn-bibliography}

\end{document}